\documentclass[10pt]{article}
\usepackage{amsmath,amssymb}

\usepackage{empheq}
\usepackage{graphicx}
\usepackage{soul}
\usepackage{float}
\usepackage{slashed}
\usepackage{amsthm}
\usepackage{amsfonts}
\usepackage{amsthm}
\usepackage{amsmath}
\usepackage{amscd}
\usepackage{stmaryrd}
\usepackage{bbm}
\newtheorem{theorem}{Theorem}[section]
\newtheorem*{theorem*}{Theorem}
\graphicspath{ {Writings/} }
\usepackage{geometry}
\usepackage{enumerate}
\usepackage{pgf,tikz}
\usepackage{csquotes}
\usepackage[bottom]{footmisc}

\usepackage{caption}
\usepackage{subcaption}

\usepackage{color}
\usepackage{xcolor}
\newcommand{\bsy}{\ensuremath{\boldsymbol}}

\DeclareMathOperator{\chih}{\widehat{\chi}}
\DeclareMathOperator{\omegah}{\hat{\omega}}

\DeclareMathOperator{\nablasl}{\slashed{\nabla}}

\DeclareMathOperator{\chibh}{\widehat{\underline{\chi}}}
\DeclareMathOperator{\omegabh}{\hat{\underline{\omega}}}
\DeclareMathOperator{\etab}{\underline{\eta}}
\DeclareMathOperator{\omegab}{\underline{\omega}}
\DeclareMathOperator{\yb}{\underline{\xi}}

\DeclareMathOperator{\alphab}{\underline{\alpha}}
\DeclareMathOperator{\betab}{\underline{\beta}}
\DeclareMathOperator{\chib}{\underline{\chi}}

\DeclareMathOperator{\etao}{\overset{\text{\scalebox{.6}{$(1)$}}}{\eta}}
\DeclareMathOperator{\zetao}{\overset{\text{\scalebox{.6}{$(1)$}}}{\zeta}}

\DeclareMathOperator{\omegabo}{\overset{\text{\scalebox{.6}{$(1)$}}}{\hat{\omegab}}}
\DeclareMathOperator{\trchio}{\overset{\text{\scalebox{.6}{$(1)$}}}{\text{tr}\chi}}
\DeclareMathOperator{\trchibo}{\overset{\text{\scalebox{.6}{$(1)$}}}{\text{tr}\chib}}
\DeclareMathOperator{\chiho}{\overset{\text{\scalebox{.6}{$(1)$}}}{\chih}}
\DeclareMathOperator{\chibho}{\overset{\text{\scalebox{.6}{$(1)$}}}{\chibh}}

\DeclareMathOperator{\alphao}{\overset{\text{\scalebox{.6}{$(1)$}}}{\alpha}}
\DeclareMathOperator{\alphabo}{\overset{\text{\scalebox{.6}{$(1)$}}}{\alphab}}
\DeclareMathOperator{\betao}{\overset{\text{\scalebox{.6}{$(1)$}}}{\beta}}
\DeclareMathOperator{\betabo}{\overset{\text{\scalebox{.6}{$(1)$}}}{\betab}}
\DeclareMathOperator{\rhoo}{\overset{\text{\scalebox{.6}{$(1)$}}}{\rho}}
\DeclareMathOperator{\sigmao}{\overset{\text{\scalebox{.6}{$(1)$}}}{\sigma}}
\DeclareMathOperator{\ybo}{\overset{\text{\scalebox{.6}{$(1)$}}}{\yb}}

\newtheorem{corollary}[theorem]{Corollary}

\newtheorem{prop}[theorem]{Proposition}
\newtheorem{remark}[theorem]{Remark}
\newtheorem{definition}[theorem]{Definition}

\usepackage[affil-it]{authblk}
\usepackage[linktoc=all]{hyperref}
\hypersetup{
    colorlinks = true,
    linkcolor = {black},
    citecolor = {black}
}

\usepackage{wasysym}
\usepackage{verbatim}
\usepackage{cleveref}
\usepackage{mathtools}
\usepackage{stmaryrd}

\begin{document}

\title{A new gauge for gravitational perturbations of Kerr spacetimes II:\\The linear stability of Schwarzschild revisited}
\author{Gabriele Benomio}
\affil{\small Princeton Gravity Initiative, Princeton University\\Washington Road, Princeton NJ 08544, United States of America}

\maketitle

\begin{abstract}
We present a new proof of linear stability of the Schwarzschild solution to gravitational perturbations.~Our approach employs the system of linearised gravity in the new geometric gauge of \cite{benomio_kerr_system}, specialised to the $|a|=0$ case.~The proof fundamentally relies on the novel structure of the transport equations in the system.~Indeed, while exploiting the well-known decoupling of two gauge invariant linearised quantities into spin $\pm 2$ Teukolsky equations, we make enhanced use of the \emph{red-shifted} transport equations and their stabilising properties to control the gauge dependent part of the system.~As a result, an \emph{initial-data} gauge normalisation suffices to establish both orbital and \emph{asymptotic} stability for \emph{all} the linearised quantities in the system.

The absence of future gauge normalisations is a novel element in the linear stability analysis of black hole spacetimes in geometric gauges governed by transport equations.~In particular, our approach simplifies the proof of \cite{DHR}, which requires a \emph{future} normalised (double-null) gauge to establish asymptotic stability for the full system.
\end{abstract}

\tableofcontents

\section{Introduction}

In the realm of black hole stability problems, the Schwarzschild solution $$g_M=-\left( 1-\frac{2M}{r} \right)dt^2+\left(1-\frac{2M}{r} \right)^{-1} dr^2 +r^2 (d\theta^2+\sin^2\theta\,d\phi^2)$$ has played a particularly prominent role.~Being the simplest black hole solution to the vacuum Einstein equations, it has been the object of most of the works pioneering the mathematical study of both scalar and gravitational perturbations of black hole spacetimes \cite{Dafermos_Rodnianski_decay_waves_schwarzschild, DHR, DHRT}.~Alongside its intrinsic interest, a solid understanding of the Schwarzschild solution and its stability properties has proven fundamental to address the more general, and complicated, Kerr geometry \cite{FullKerr, Andersson_Blue_Lin_Kerr, Hintz_Vasy_Kerr, Klainerman_Szeftel_Kerr_small_a_1, Giorgi_Klainerman_Szeftel_wave_estimates}.~See the introduction in \cite{benomio_kerr_system} for an overview of the vast literature on the subject.

\medskip

In this work, we present a new proof of linear stability of the Schwarzschild solution to gravitational perturbations.~The proof employs the system of linearised gravity in the new gauge of \cite{benomio_kerr_system}, specialised to the $|a|=0$ case.~The new structural properties of the system lead to interesting conceptual simplifications in the stability analysis compared to previous works.

\medskip

As we shall explain in the present introduction, the nature and significance of the simplifications concern the treatment of so-called \emph{gauge normalisations}.~After briefly discussing the meaning of gauge normalisations, we illustrate the new features of our analysis in relation to previous work \cite{DHR}.~Some more detailed (but still introductory) comments are deferred to the overview in Section \ref{sec_overview}.

\subsection{Gauge normalisations in linear theory}

The formulation of the vacuum Einstein equations as a system of partial differential equations requires a choice of gauge.~Typically, a choice of gauge only \emph{partially} fixes the gauge freedom in the problem.~At the level of the linearised vacuum Einstein equations, a manifestation of the residual freedom is the existence of a class of special solutions to the linearised system, called \emph{pure gauge solutions}, corresponding to infinitesimal diffeomorphisms preserving the choice of gauge.~Adding a particular pure gauge solution to a general solution to the linearised system is viewed as a choice of gauge normalisation and fixes the residual freedom in linear theory.~A convenient choice of gauge normalisation usually depends on the general solution to the linearised system considered.

\medskip

\emph{Geometric gauges} governed by transport equations have been \emph{future} normalised when adopted in linear stability problems for black hole spacetimes (e.g.~\cite{DHR, Giorgi_full_RN}).~This means that, given a general solution to the linearised system, the choice of gauge normalisation is made depending on the evaluation at (all) \emph{future times} of certain (linearised) dynamical quantities associated to the solution.~As a result of the future gauge normalisation, the transport equations in the system, \emph{which would not in general yield decay}, could be exploited to obtain an \emph{asymptotic} stability result for the gauge dependent part of the system.~On the other hand, partial stability results, as for instance \emph{orbital} stability in \cite{DHR} (see Section \ref{sec_intro_DHR}), could already be achieved by choosing gauge normalisations which \emph{only} depend on the evaluation of \emph{initial-data} (linearised) quantities.~This latter type of normalisation, referred to as \emph{initial-data gauge normalisation}, is, in general, easier to treat than a future gauge normalisation, in that it is entirely performed at initial time, instead of being dynamically determined.

\medskip

When the linear stability result is employed as a direct building block in a proof of nonlinear stability, the use of future gauge normalisations directly transfers from the linear to the nonlinear problem.~One such instance is the nonlinear result \cite{DHRT} in a \emph{double-null gauge}, where a key future gauge normalisation adopted in the proof manifestly inherits the future gauge normalisation introduced in the linear result \cite{DHR}.~We also point out that, even if a proof of nonlinear stability is not directly constructed around a linear stability result, as for instance in \cite{StabMink}, the future gauge normalisations possibly entering the nonlinear problem can be viewed as arising from an approach to the linearised problem employing future gauge normalisations. 

\medskip

The new proof of linear stability of the Schwarzschild solution presented in this work contributes towards a better understanding of the role of gauge normalisations in black hole stability problems.~Indeed, our proof employes a geometric gauge governed by transport equations (i.e.~the new gauge of \cite{benomio_kerr_system}) but, unlike previous works, \emph{both} orbital and \emph{asymptotic} stability are established in an \textbf{initial-data gauge normalisation}.~The gauge normalisation is performed once and for all, and no future gauge normalisations are, at any point, required.~Future gauge normalisations can be circumvented in virtue of the new structure of the transport equations in the system.~See Section \ref{sec_intro_new_result}.

\medskip

In the following two sections, we put the proof of \cite{DHR} and our work in closer comparison to elucidate the novelties of our approach.

\subsection{Linear stability in double-null gauge} \label{sec_intro_DHR}

The seminal work \cite{DHR} is the first to establish the linear stability of the Schwarzschild solution to gravitational perturbations.~The result is formulated in a \emph{double-null gauge} and requires a \emph{future} gauge normalisation.

\medskip

\begin{theorem}[Linear stability of Schwarzschild in double-null gauge \cite{DHR}] \label{th_DHR}
Consider the system of linearised gravity in a double-null gauge of \cite{DHR}.~Then, all appropriately initial-data normalised solutions are uniformly bounded in terms of the size of their initial data and, upon adding a suitable pure gauge solution, decay in time to a linearised Kerr solution.~The additional pure gauge solution is itself uniformly bounded in terms of the size of the initial data and effectively re-normalise the gauge to the \ul{future}.
\end{theorem}

\medskip

The proof of Theorem \ref{th_DHR} proceeds as follows:
\begin{enumerate}
\item \textbf{Decay for the gauge invariant linearised quantities}

In a double-null gauge, two gauge invariant linearised quantities in the system exactly decouple into two spin $\pm 2$ Teukolsky equations.~By implementing a transformation theory for general solutions to the Teukolsky equations, work \cite{DHR} derives higher-order gauge invariant linearised quantities which satisfy the Regge--Wheeler equation and are proven to decay.~The transformation theory allows to eventually retrieve decay for the original gauge invariant linearised quantities, which can thus be controlled independently from the rest of the system.
\item \textbf{Initial-data gauge normalisation}

As a first step towards proving stability for the gauge dependent linearised quantities, work \cite{DHR} chooses an appropriate initial-data gauge normalisation, i.e.~an appropriate notion of \emph{initial-data normalised solutions}.

Crucially, some of the normalisation identities are \emph{propagated} globally-in-time by the system of equations.~In particular, the initial location of the event horizon is fixed by the initial-data gauge normalisation of \cite{DHR} and is propagated so as to remain fixed for all times.
\item \textbf{Boundedness for the gauge dependent linearised quantities}

The gauge dependent linearised quantities relative to initial-data normalised solutions are shown to be uniformly bounded by the size of their initial data.~The uniform boundedness statement does not lose derivatives, thus providing a true orbital stability result.

At this stage, \emph{decay} can only be obtained for \emph{some} of the gauge dependent linearised quantities.~For instance, the linearised ingoing shear of an initial-data normalised solution cannot be shown to decay.
\item \textbf{Future gauge normalisation}

To prove decay for \emph{all} the gauge dependent linearised quantities, one needs an additional gauge re-normalisation, performed along the entire event horizon.~The \emph{future} gauge re-normalisation is obtained by adding yet another pure gauge solution to an initial-data normalised solution. 
\item \textbf{Decay for the gauge dependent linearised quantities}

All the gauge dependent linearised quantities relative to \emph{future} re-normalised solutions are proven to decay.~Importantly for nonlinear applications, the pure gauge solution necessary for the future gauge re-normalisation is shown to be itself uniformly bounded by the size of the initial data.
\end{enumerate}

\subsection{The main result:~Linear stability in the new gauge}  \label{sec_intro_new_result}

The present work proves the linear stability of the Schwarzschild solution to gravitational perturbations in the new geometric gauge of \cite{benomio_kerr_system}.~In contrast with Theorem \ref{th_DHR}, the asymptotic stability result does \emph{not} require a future gauge re-normalisation.

\medskip

\begin{theorem}[Linear stability of Schwarzschild in the new gauge of \cite{benomio_kerr_system}] \label{th_new_gauge}
Consider the system of linearised gravity in the new gauge of \cite{benomio_kerr_system}, specialised to the $|a|=0$ case.~Then, all appropriately \ul{initial-data} normalised solutions are uniformly bounded in terms of the size of their initial data and \ul{decay} in time to a linearised Kerr solution.
\end{theorem}

\medskip

See Theorem \ref{intro_main_theorem} in the overview for a more detailed (but still informal) version of Theorem \ref{th_new_gauge}.

\medskip

The absence of future gauge normalisations simplifies our proof of linear stability, allowing to eliminate Step 4 of the proof of Theorem \ref{th_DHR} and obtain Steps 3 and 5 simultaneously.~Our proof is otherwise conceptually similar to the one of Theorem \ref{th_DHR} and can be summarised as follows:
\begin{enumerate}
\item \textbf{Decay for the gauge invariant linearised quantities}

In the new gauge of \cite{benomio_kerr_system}, two gauge invariant linearised quantities in the system exactly decouple into two spin $\pm 2$ Teukolsky equations.~These linearised quantities are the analogue of the decoupling quantities of \cite{DHR}.~The proof of decay for our gauge invariant linearised quantities proceeds exactly as in Step 1 of the proof of Theorem \ref{th_DHR}. 
\item \textbf{Initial-data gauge normalisation}

As in Step 2 of the proof of Theorem \ref{th_DHR}, we introduce an \emph{initial-data} gauge normalisation.~Since the gauge dependent part of our system is larger than the one in \cite{DHR} (see Section \ref{sec_overview_system}), our initial-data gauge normalisation necessarily prescribes a larger set of normalisation identities, including some identities for the new linearised quantities (see Section \ref{sec_overview_renormalised_solutions}).~A related new feature of our gauge normalisation is the presence of a wider class of pure gauge solutions that one can employ to normalise general solutions to the system.~In particular, we identify two independent classes of pure gauge solutions, namely \emph{coordinate} pure gauge solutions and \emph{frame} pure gauge solutions (see Section \ref{sec_overview_pure_gauge_solutions}).~These new technical ingredients in the implementation of Step 2 may be viewed as a manifestation of the use of non-integrable frames in the derivation of the system in \cite{benomio_kerr_system}.

It is crucial that our system propagates some of the normalisation identities.~In analogy with the proof of Theorem \ref{th_DHR}, the location of the event horizon is fixed at initial time by the gauge normalisation and is propagated so as to remain fixed for all times.
\item \textbf{Boundedness and decay for the gauge dependent linearised quantities}

In contrast with the proof of Theorem \ref{th_DHR}, our initial-data gauge normalisation of Step 2 suffices to establish \emph{both} orbital stability (i.e.~uniform boundedness without loss of derivatives) and \emph{asymptotic} stability for \emph{all} the gauge dependent linearised quantities.

Our proof of decay for the gauge dependent linearised quantities fundamentally relies on the novel structure of the \emph{outgoing} transport equations in the system (see Section \ref{sec_overview_system}).~Indeed, the choice of initial-data gauge normalisation of Step 2 allows an enhanced use of the \emph{red-shifted} transport equations available in the system.~These are equations which, \emph{along the event horizon}, take the schematic form
\begin{equation*}
\nablasl_4 \overset{\text{\scalebox{.6}{$(1)$}}}{\Gamma}+2\,k\,\kappa_M \overset{\text{\scalebox{.6}{$(1)$}}}{\Gamma}=\ldots \, ,
\end{equation*}
with $\overset{\text{\scalebox{.6}{$(1)$}}}{\Gamma}$ denoting a gauge dependent linearised quantity, $k\in \left\lbrace 1,2 \right\rbrace$ and $\kappa_M=1/4M$ the \emph{positive} surface gravity of the Schwarzschild event horizon.~Red-shifted transport equations can be employed to prove integrated energy \emph{decay} by direct, forward-in-time integration from the initial data,\footnote{One also needs to exploit that the right hand side of the equation suitably decays.} without introducing any future gauge re-normalisation (see Section \ref{sec_overview_analysis_gauge_dependent_quantities}).~The absence of future gauge re-normalisations in our analysis is, in this sense, intimately tied to the presence of the \textbf{red-shift effect} at the event horizon and how \textbf{the stabilising properties of the red-shift effect are effectively captured by the new structure of our system of linearised gravity}.\footnote{It is interesting to note that if one were to employ the system of linearised gravity in the gauge of \cite{benomio_kerr_system} to address the (linear) asymptotic stability of Minkowski space (i.e.~$|a|=M=0$, and no red-shift effect), a \emph{future} gauge normalisation would be needed.}

The main novel element of our proof of decay is the treatment of the linearised \emph{ingoing} shear.~As part of our scheme, we establish integrated energy decay for the linearised ingoing shear by direct, forward-in-time integration of the red-shifted transport equation 
\begin{equation} \label{intro_lin_4_chibh_a_0}
\nablasl_4 \chibho+\left(\frac{1}{r}\left(1-\frac{2M}{r}\right)+\frac{2M}{r^2}\right)\chibho = \frac{1}{r}\,\chiho \, ,
\end{equation}
for which the hierarchical structure of the system allows prior control on the right hand side.~Our approach may be contrasted to the one taken in the proof of Theorem \ref{th_DHR} to estimate the corresponding linearised quantity.~Indeed, the analogue of the equation \eqref{intro_lin_4_chibh_a_0} in \cite{DHR} possesses an additional angular derivative of a linearised connection coefficient on the right hand side, which one does not a-priori control and thus prevents from directly exploiting the equation.~The scheme to prove decay for the linearised ingoing shear of \cite{DHR} has to take an alternative route, which involves integrating an \emph{ingoing} transport equation (in fact, backwards from the event horizon) and ultimately requires, as outlined in Step 4 of Section \ref{sec_intro_DHR}, a \emph{future} gauge re-normalisation.~See \cite{DHR} and Sections II.3.3 and II.3.4 in \cite{DHRT} for a more detailed description of the proof in double-null gauge.
\end{enumerate}

\subsection{The Kerr problem}

Beyond its independent interest, the present work sets the stage for a future stability analysis of the system of linearised gravity in the gauge of \cite{benomio_kerr_system} in the general $|a|<M$ case.~In fact, some of the essential aspects which are expected to characterise the future analysis, such as the central role of the red-shifted transport equations, are already manifest (and easier to appreciate) in the present work.~In analogy to the main result of this work, the analysis will produce both a true orbital stability result (i.e.~without loss of derivatives) and an asymptotic stability result, thus indicating that the framework developed in \cite{benomio_kerr_system} may be well suited to address the nonlinear stability problem in the full sub-extremal range $|a|<M$.

\subsection{Acknowledgements}

The present work builds upon the framework developed by the author in his doctoral thesis \cite{benomio_thesis}.~The author is indebted to Gustav Holzegel for numerous discussions over the years and for fundamentally contributing to this work with insightful ideas.~The author is also grateful to Mihalis Dafermos, Igor Rodnianski and Martin Taylor for fruitful conversations.~The author's work has been supported by a Roth Scholarship at Imperial College London, by the Royal Society through the Royal Society Tata University Research Fellowship R1-191409 and by the Princeton Gravity Initiative at Princeton University.~The author also acknowledges the University of M\"{u}nster for hospitality during a visit as Young Research Fellow.

\section{Overview}  \label{sec_overview}

We present an overview of the main elements of the proof.

\subsection{The Schwarzschild exterior manifold}

In Section \ref{sec_The_Schwarzschild_exterior_manifold}, we recall the definition of the Schwarzschild exterior manifold $(\mathcal{M},g_{M})$ and make a choice of differentiable structure and null frame on $\mathcal{M}$.

\medskip

To facilitate the comparison with previous work \cite{DHR}, we consider the Eddington--Finkelstein (double-null) differentiable structure $$(u,v,\theta^A)$$ on $\mathcal{M}\setminus\mathcal{H}^+$, where $\mathcal{H}^+:=\partial\mathcal{M}$ denotes the future event horizon.~In Eddington--Finkelstein coordinates, the metric $g_M$ reads
\begin{equation} \label{intro_schw_m}
g_M=-4\,\Omega^2_M(u,v)\,du\,dv+r^2(u,v)\,\slashed{\gamma}_{\theta^A\theta^B}d\theta^Ad\theta^B \, ,
\end{equation}
with $r(u,v)$ the standard Schwarzschild radius on $\mathcal{M}\setminus\mathcal{H}^+$, $$\Omega^2_M(u,v):=1-\frac{2\,M}{r(u,v)}$$ and $\slashed{\gamma}$ the round metric on the unit two-sphere.~Our choice of coordinates differs from the star-normalised, outgoing principal differentiable structure from Section 5.7 of \cite{benomio_kerr_system} in the $|a|=0$ case, for which the ingoing coordinate is a spacelike coordinate.

\medskip

To the metric $g_M$, we associate the pair of null vector fields
\begin{align}  \label{intro_e4_e3}
e_4^{\text{as}}&=\partial_v \, , & e_3^{\text{as}}&=\Omega^{-2}_M\,\partial_u 
\end{align}
on $\mathcal{M}\setminus\mathcal{H}^+$.~The vector fields \eqref{intro_e4_e3} smoothly extend to global, \emph{regular}, non-degenerate vector fields on the whole $\mathcal{M}$, including on $\mathcal{H}^+$.~The vector fields \eqref{intro_e4_e3} can be completed to a local null frame $\mathcal{N}_{\text{as}}$.~Cf.~\cite{DHR}, where a different (non-regular) choice of scaling for the null frame vector fields is made.

\medskip

The frame $\mathcal{N}_{\text{as}}$ is the algebraically special frame of $(\mathcal{M},g_M)$.~The frame $\mathcal{N}_{\text{as}}$ coincides with the algebraically special frame introduced in \cite{benomio_kerr_system}, specialised to the $|a|=0$ case.~The frame $\mathcal{N}_{\text{as}}$ is integrable and adapted to the double-null foliation of $(\mathcal{M},g_M)$ induced by the Eddington--Finkelstein differentiable structure, i.e.~we have $$(\mathfrak{D}_{\mathcal{N}_{\text{as}}})_p = T_p\mathbb{S}^2_{u(p),v(p)}$$ for any $p\in\mathcal{M}$.~In the present paper, the $\mathfrak{D}_{\mathcal{N}_{\text{as}}}$ tensors of \cite{benomio_kerr_system} are referred to as \emph{$\mathbb{S}^2_{u,v}$-tensors}. 

\medskip

The connection coefficients and curvature components of $g_M$ relative to $\mathcal{N}_{\text{as}}$ are $\mathbb{S}^2_{u,v}$-tensors on $\mathcal{M}\setminus \mathcal{H}^+$ which extend \emph{regularly} to the whole $\mathcal{M}$, including $\mathcal{H}^+$.~The non-vanishing connection coefficients and curvature components are
\begin{align*}
\omegah_M&=\frac{2M}{r^2} \, , & 
(\text{tr}\chi)_M &=\frac{2\Omega_M^2}{r}  \, ,  &  (\text{tr}\chib)_M &=-\frac{2}{r} \, , & \rho_M &=-\frac{2M}{r^3} \, .
\end{align*}

\subsection{The system of linearised gravity} \label{sec_overview_system}

The system of linearise gravity on $(\mathcal{M},g_M)$ of Section \ref{sec_system_linearised_gravity} is obtained by setting $|a|=0$ in the system of linearised gravity of Section 10 in \cite{benomio_kerr_system}.~A \emph{solution} to the linearised system of equations is intended as a collection of all the linearised quantities appearing in the system.\footnote{By linearity, the sum of solutions to the linearised system of equations is a solution to the linearised system of equations.}~All the linearised quantities in the system extend to \emph{regular} smooth scalar functions and $\mathbb{S}^2_{u,v}$-tensors on the whole $\mathcal{M}$, including on $\mathcal{H}^+$.~No additional regularising weights are needed.

\subsubsection{Structural properties}

For a discussion of the key structural properties of the system, see Section 1.4 and Appendix B in \cite{benomio_kerr_system}.~For the sake of the present overview, we recall that the linearised second variational formula
\begin{equation} \label{intro_4_chih}
\nablasl_4 \chiho +(\text{tr}\chi)\chiho-\omegah \chiho =  -\alphao
\end{equation}
is \emph{blue-shifted} and preserves the structure of the corresponding linearised equation from \cite{DHR}.~On the other hand, the linearised ingoing shear satisfies the \emph{red-shifted} transport equation
\begin{equation} \label{intro_4_chibh}
\nablasl_4 \chibho+\frac{1}{2}\,(\text{tr}\chi)\chibho+\omegah\chibho = -\frac{1}{2}\,(\text{tr}\chib)\chiho \, .
\end{equation}
In contrast with \cite{DHR}, where the corresponding linearised equation couples with an additional angular derivative of a linearised connection coefficient, \emph{our equation \eqref{intro_4_chibh} only couples with the linearised outgoing shear}.~As another remarkable feature of our linearised system of equations, the linearised Raychaudhuri equation
\begin{equation} \label{intro_lin_4_trchi}
\nablasl_4 (\trchio)+ (\text{tr}\chi)(\trchio)-\omegah \, (\trchio) =  0 
\end{equation}
and the transport equation for the linearised antitrace of the outgoing second fundamental form\footnote{See Section 1.5.2 of \cite{benomio_kerr_system} for an introductory discussion on the appearance of the antitraces of the second fundamental forms in the equations.~We recall that this is intimately tied to the use of non-integrable frames in the derivation of the system of linearised gravity in \cite{benomio_kerr_system}, which persists in the $|a|=0$ case.}  
\begin{equation} \label{intro_lin_4_atrchi}
\nablasl_4 (\overset{\text{\scalebox{.6}{$(1)$}}}{\slashed{\varepsilon}\cdot\chi})+(\text{tr}\chi)(\overset{\text{\scalebox{.6}{$(1)$}}}{\slashed{\varepsilon}\cdot\chi})-\omegah \, (\overset{\text{\scalebox{.6}{$(1)$}}}{\slashed{\varepsilon}\cdot\chi})   = 0 \, ,
\end{equation}
both completely decouple from the rest of the system.~In \cite{DHR}, the linearised equation corresponding to \eqref{intro_lin_4_trchi} only decouples along the event horizon.

\medskip

Our linearised system loses the symmetry between the equations for the barred and unbarred quantities present in \cite{DHR}.\footnote{Here we refer to the symmetry of the equations under the exchange of the barred and unbarred quantities and, in the case of the transport equations, $\nablasl_4$ and $\nablasl_3$-derivatives.~Note that such a symmetry is already lost in our \emph{nonlinear} system of renormalised vacuum Einstein equations of \cite{benomio_kerr_system} (see the schematic form (14)-(16) of the renormalised null structure equations in Section 1.4.3).}~Indeed, the simplification of the outgoing transport equations \eqref{intro_4_chih}-\eqref{intro_lin_4_atrchi} corresponds to a complication (if compared to \cite{DHR}) in the form of the ingoing transport equations and elliptic equations.~These latter equations are the most affected by the appearance of the new (if compared to \cite{DHR}) linearised quantities 
\begin{align*}
&(\overset{\text{\scalebox{.6}{$(1)$}}}{\slashed{\varepsilon}\cdot\chi}) \, , &   &(\overset{\text{\scalebox{.6}{$(1)$}}}{\slashed{\varepsilon}\cdot\chib}) \, , & &\ybo
\end{align*}
in the system, as one can see from the linearised second variational formula
\begin{equation} \label{intro_3_chib}
\nablasl_3 \chibho+(\text{tr}\chib)\chibho = -2\,\slashed{\mathcal{D}}_2^{\star} \ybo    -\alphabo
\end{equation}
and the Codazzi equation
\begin{equation}
\slashed{\text{div}}\chiho = \frac{1}{2}\nablasl (\trchio) -\frac{1}{2}\,{}^{\star}\nablasl (\overset{\text{\scalebox{.6}{$(1)$}}}{\slashed{\varepsilon}\cdot\chi})  +\frac{1}{2}\,(\text{tr}\chi) \zetao  -\betao +\frac{1}{2}\,(\nablasl_3(\text{tr}\chi))\overset{\text{\scalebox{.6}{$(1)$}}}{\mathfrak{\slashed{\mathfrak{f}}}}_{4}+\frac{1}{2}\,(\nablasl_4(\text{tr}\chi))\overset{\text{\scalebox{.6}{$(1)$}}}{\mathfrak{\slashed{\mathfrak{f}}}}_{3}   \, .  \label{intro_lin_codazzi}
\end{equation}
In fact, all the ingoing transport equations appearing in our linearised system of null structure equations couple with a derivative of a linearised connection coefficient on the right hand side.~We note that, in constrast, the linearised equation in \cite{DHR} corresponding to \eqref{intro_3_chib} does \emph{not} couple with any linearised connection coefficient on right hand side.

\medskip

The linearised null structure and Bianchi equations in our system are complemented by the equations for the linearised induced metric and frame coefficients.~The equations for the linearised frame coefficients may be thought as replacing the equations for the linearised lapse function and shift vector field in \cite{DHR}.~For comments on the derivation and structure of these equations, see Sections 1.4.3 and 1.5.2 in \cite{benomio_kerr_system}.

\medskip

We remark that the linearised frame coefficients appear on the right hand side of the linearised null structure and Bianchi equations more frequently than the linearised lapse function and shift vector field in \cite{DHR}, yielding a stronger coupling between the equations for the linearised frame coefficients and the rest of the system.\footnote{This fact arises from the use of variable horizontal distributions (which persists for $|a|=0$) in the derivation of the system of linearised gravity in \cite{benomio_kerr_system}.~The stronger coupling is already apparent from the Codazzi equation \eqref{intro_lin_codazzi}, which in \cite{DHR} does not couple with the linearised lapse function or shift vector field.}~Nonetheless, the outgoing transport equations in our system are only partially affected by this stronger coupling.\footnote{The linearised frame coefficients only appear in the outgoing transport equations when one linearises the angular derivative of a connection coefficient or curvature component which is non-vanishing on the background.~See, for instance, the equations \eqref{4_omegabo} and \eqref{4_betabo}.}~In this regard, one may compare the completely decoupled Raychaudhuri equation \eqref{intro_lin_4_trchi} with the corresponding linearised Raychaudhuri equation in the ingoing direction
\begin{equation*}
\nablasl_3 (\trchibo)+ (\text{tr}\chib)(\trchibo) =  (\text{tr}\chib)\omegabo +2\, \slashed{\text{div}} \ybo  -\nablasl_3(\text{tr}\chib)\overset{\text{\scalebox{.6}{$(1)$}}}{\mathfrak{\underline{\mathfrak{f}}}}_4  -\nablasl_4(\text{tr}\chib)\overset{\text{\scalebox{.6}{$(1)$}}}{\mathfrak{\underline{\mathfrak{f}}}}_3   \, ,
\end{equation*}
where some linearised frame coefficients appear on the right hand side.

\subsubsection{Initial data, well-posedness and asymptotic flatness}

The system of linearised gravity will be treated in a self-contained fashion, without any reference to the nonlinear theory and its derivation in \cite{benomio_kerr_system}.~In Section \ref{sec_initial_data_well_posedness}, we define smooth \emph{seed initial data} for the system of linearised gravity as a set of linearised quantities which can be freely prescribed on the union of null cones $$\mathcal{S}_{u_0,v_0}:= C_{u_0}\cup \underline{C}_{v_0} \, ,$$ with $C_{u_0}:=\left\lbrace u=u_0 , v\geq v_0\right\rbrace$ an outgoing null cone and $\underline{C}_{v_0}:=\left\lbrace u\geq u_0 ,  v=v_0 \right\rbrace$ an ingoing null cone.~The initial value formulation of the system of linearised gravity is globally well-posed.~See Proposition \ref{prop_well_posedness} and Appendix \ref{appendix_well_posedness}.~We also introduce a notion of \emph{(pointwise) asymptotic flatness} for the seed initial data, which is needed for the formulation of our main result.

\subsection{Classes of special solutions} \label{sec_special_solutions}

The linear stability result of the present work is formulated as a decay statement for solutions to the linearised system of equations (see already Theorem \ref{intro_main_theorem}).~As expected, decay does not hold for all solutions to the linearised system of equations unconditionally.~Indeed, one expects dynamical solutions to the linearised system arising from smooth, asymptotically flat seed initial data to converge, in the evolution, to a suitable notion of \emph{linearised Kerr solution}, which should be itself a stationary, thus \emph{non-decaying}, solution to the linearised system.~Moreover, the linearised equations are manifestly compatible with the existence of solutions for which some of the linearised quantities \emph{grow} for all times (equation \eqref{intro_lin_4_trchi} is, for instance, compatible with exponential growth along the event horizon).

\medskip

It turns out that the obstructions to formulating an unconditional decay statement can be fully understood by identifying two classes of special solutions to the linearised system, namely the classes of \emph{pure gauge solutions} and \emph{reference linearised solutions}, and by implementing a spherical-mode analysis of solutions relative to the $Y^{\ell}_m$-spherical harmonics of the $\mathbb{S}^2_{u,v}$-spheres.

\medskip

This section discusses the two classes of special solutions, whereas Section \ref{sec_overview_renormalised_solutions} will address the so-called \emph{initial-data normalised solutions}.~These notions are presented in a different order in Section \ref{sec_normalised_solutions} of the body of the paper.~The exposition adopted in this overview gives a better account of the logic behind the formulation of these notions, although when proving rigorous statements one has to follow a different path.

\subsubsection{Pure gauge solutions} \label{sec_overview_pure_gauge_solutions}

In Section \ref{sec_pure_gauge_solutions}, we discuss the class of \emph{pure gauge solutions}, which may be viewed as divided into two sub-families: 
\begin{itemize}
\item \textbf{Coordinate pure gauge solutions}

The first sub-family of pure gauge solutions, called \emph{coordinate pure gauge solutions}, is presented in Section \ref{sec_coord_pure_gauge_solns}.~This family arises from the residual freedom available in choosing the fixed differentiable structure on the ambient manifold $\mathcal{M}$ in the derivation of the system in \cite{benomio_kerr_system} (see Section 6.1 of \cite{benomio_kerr_system}).~In fact, one can prescribe infinite one-parameter families of local coordinates $$(\bsy{u}(\epsilon),\bsy{v}(\epsilon),\bsy{\theta^A}(\epsilon))$$ on $\mathcal{M}$ such that $(\bsy{u}(0),\bsy{v}(0),\bsy{\theta^A}(0))=(u,v,\theta^A)$ and relative to which the metric $g_M$ is in the outgoing principal gauge of \cite{benomio_kerr_system}\footnote{See Section 6.2 in \cite{benomio_kerr_system}.} \emph{to linear order in $\epsilon$}.~Such one-parameter families of coordinates can be written in their most general form as
\begin{align*}
\bsy{u}(\epsilon)&=u+\epsilon\cdot h_2(u,v,\theta^A) \, , & \bsy{v}(\epsilon)&=v+\epsilon\cdot h_1(u,v,\theta^A)  \, ,  & \bsy{\theta^A}(\epsilon)&=\theta^A+\epsilon\cdot h^{\theta^A}(u,v,\theta^B)
\end{align*}
and thus be generated by the four scalar functions $(h_1,h_2,h^{\theta^A})$.~The outgoing principal gauge conditions impose a system of four ODEs in the $e_4^{\text{as}}$-direction for the four scalar functions (see the equations \eqref{ODE_pure_gauge_1}-\eqref{ODE_pure_gauge_3}), which thus one only has the freedom to prescribe on the initial ingoing null cone $\underline{C}_{v_0}$.

The coordinate pure gauge solutions can be explicitly read off from the linearisation of the transformed metric components of $g_M$ under the $\epsilon$-change of coordinates above.~In particular, the linearised frame coefficients generated by the change of coordinates can be derived from the linearised metric components using Proposition 9.4 in \cite{benomio_kerr_system}.~For instance, one has $$\bsy{g}_{uu}(\epsilon)=0+\epsilon\cdot (-4\,\Omega^2_M\partial_uh_1)+\mathcal{O}(\epsilon^2)$$ and 
\begin{equation*}
\overset{\text{\scalebox{.6}{$(1)$}}}{\underline{\mathfrak{f}}}_3  =(4\Omega^4_M)^{-1}\overset{\text{\scalebox{.6}{$(1)$}}}{g}_{uu}  \, , 
\end{equation*}
which yields $$\overset{\text{\scalebox{.6}{$(1)$}}}{\underline{\mathfrak{f}}}_3=-\nablasl_3 h_1 \, .$$ All the linearised connection coefficients and curvature components can then be deduced by solving the full linearised system of equations.~See Definition \ref{def_coordinate_pure_gauge_soln}.

\item \textbf{Frame pure gauge solutions}

The second sub-family of pure gauge solutions, called \emph{frame pure gauge solutions}, is presented in Section \ref{sec_frame_pure_gauge_solns}.~This family arises from the residual freedom available in choosing the fixed frame on the ambient manifold $\mathcal{M}$ in the derivation of the system in \cite{benomio_kerr_system} (see Section 6.3 in \cite{benomio_kerr_system}).~In fact, one can prescribe infinite one-parameter families of local frames $$\widetilde{\bsy{\mathcal{N}}}(\epsilon)$$ on $\mathcal{M}$ such that $\widetilde{\bsy{\mathcal{N}}}(0)=\mathcal{N}_{\text{as}}$ and which, \emph{to linear order in $\epsilon$}, are null relative to $g_M$ and satisfy all the properties listed in Section 6.3 of \cite{benomio_kerr_system} relative to $g_M$.~Such one-parameter families can be written in their most general form as
\begin{align*}
\widetilde{\bsy{e}}_{\bsy{4}}(\epsilon)&=e_4^{\text{as}} \, , &  \widetilde{\bsy{e}}_{\bsy{3}}(\epsilon)&= e_3^{\text{as}}+ \epsilon\cdot k^A\, e_A^{\text{as}}  \, , & \widetilde{\bsy{e}}_{\bsy{A}}(\epsilon)&= e_A^{\text{as}}+\frac{\epsilon}{2} \cdot k_A \, e_4^{\text{as}} 
\end{align*}
and thus be generated by the $\mathbb{S}^2_{u,v}$ vector field $k$.~The scalar functions $k^A$ are required to satisfy a system of two ODEs in the $e_4^{\text{as}}$-direction (see the equation \eqref{ODE_k}), and thus can only be freely prescribed on the initial ingoing null cone $\underline{C}_{v_0}$.

The frame pure gauge solutions can be explicitly read off from the linearisation of the transformed frame coefficients of $\mathcal{N}_{\text{as}}$ under the $\epsilon$-frame transformation above.~For instance, one has
\begin{align*}
\bsy{\underline{\mathfrak{f}}}_A(\epsilon): & =g_M(\widetilde{\bsy{e}}_{\bsy{3}}(\epsilon),e_A^{\text{as}}) \\ &=\epsilon\cdot k^B g_M(e_B^{\text{as}},e_A^{\text{as}}) \\
&=\epsilon\cdot k_A \, ,
\end{align*}
which yields
\begin{equation*}
\overset{\text{\scalebox{.6}{$(1)$}}}{\underline{\mathfrak{f}}}_A=k_A \, .
\end{equation*}
We note that frame transformations do not generate linearised metric components, whereas all the linearised connection coefficients and curvature components can be deduced by solving the full linearised system of equations.~See Definition \ref{def_frame_pure_gauge_soln}.
\end{itemize}

We remark that the two sub-families of pure gauge solutions are \emph{not} equivalent, meaning that one can sum an identically vanishing coordinate pure gauge solution with a non-trivial frame pure gauge solution, and vice-versa.~It is also true that there exist coordinate pure gauge solutions that cannot be realised as a linear combination of frame pure gauge solutions, and vice-versa.~The emergence of two distinct sub-families of pure gauge solutions may be seen as the linear manifestation of the disentanglement of frames and spacetime foliation in the derivation of the system in \cite{benomio_kerr_system}, which persists in the $|a|=0$ case.

\medskip

By looking at the complete Definitions \ref{def_coordinate_pure_gauge_soln} and \ref{def_frame_pure_gauge_soln} of pure gauge solutions, the reader should note that any pure gauge solution satisfies
\begin{align} \label{intro_alpha_alphab_vanish_gauge}
\alphao&= 0 \, , & \alphabo&=0 \, .
\end{align}
We say that the linearised quantities \eqref{intro_alpha_alphab_vanish_gauge} are \emph{gauge invariant}.~The linearised quantities which are not gauge invariant are said to be \emph{gauge dependent}.

\subsubsection{Reference linearised solutions} \label{sec_overview_reference_linearised_solutions}

In Section \ref{sec_reference_linearised_solutions}, we present the \emph{reference linearised solutions}.~To obtain this class of solutions, we consider the Schwarzschild metric $g_M$ and Kerr metric $g_{a,M}$ in suitable coordinates and with a suitably associated null frame and then linearise the metric components and frame coefficients relative to the mass and angular momentum parameters $M$ and $a$.~We remark that suitable choices of coordinates and frame are such that the outgoing principal gauge conditions and frame properties of Sections 6.1 and 6.3 in \cite{benomio_kerr_system} are satisfied \emph{to linear order in the spacetime parameter relative to which one linearises}.~Such choices are \emph{not} unique, and thus there is no unique way to define the families of reference linearised solutions.~Reference linearised solutions arising from different (admissible) choices of coordinates and (or) frames differ by a pure gauge solution.
\begin{itemize}
\item \textbf{Reference linearised Schwarzschild solutions}

By linearising the Schwarzschild metric \eqref{intro_schw_m} relative to the mass parameter $M$ around a non-vanishing mass $M_0$,\footnote{The metric \eqref{intro_schw_m} is in outgoing principal gauge.} one obtains a one-parameter family of linearised metric components, parametrised by $M$.~As pointed out in \cite{DHR}, some care is needed when linearising the metric \eqref{intro_schw_m}, in that the null coordinates $(u,v)$, as well as the radial function $r$, depend themselves on the mass parameter $M$.~To circumvent this issue, one can define rescaled null coordinates $(\hat{u},\hat{v})$ and the scalar function $x$ which are \emph{independent} of $M$ and such that
\begin{equation*}
g_M=4M^2\left(-4\left(1-\frac{1}{x}\right)d\hat{u}\,d\hat{v}+x^2 d\sigma_2\right) \, . 
\end{equation*}
The linearised induced metric can be immediately read off, namely
\begin{align*}
\overset{\text{\scalebox{.6}{$(1)$}}}{\widehat{\slashed{g}}}&=0 \, , & (\text{tr}\overset{\text{\scalebox{.6}{$(1)$}}}{\slashed{g}})&=-2\,\mathfrak{m} \, , 
\end{align*}
where $\mathfrak{m}:=-2(M-M_0)/M_0$.~To obtain the linearised frame coefficients, we consider the null frame vector fields $\widehat{e}_4^{\text{as}}:=2M\,e_4^{\text{as}}$ and $\widehat{e}_{3,M}^{\text{as}}:=(2M)^{-1}\,e_3^{\text{as}}$.~The rescaling of the null vector fields is necessary for the vector field $\widehat{e}_4^{\text{as}}$ to remain \emph{fixed} relative to $M$ (as to satisfy the frame property (197) of Section 6.3 in \cite{benomio_kerr_system} relative to $M$).\footnote{We note that the rescaling of the null frame vector fields also automatically fixes the frame property (199) of Section 6.3 in \cite{benomio_kerr_system} relative to $M$.~Indeed, we have $\widehat{\omegah}_M=x^{-2}$ relative to the frame vector fields $(\widehat{e}_4^{\text{as}},\widehat{e}_{3,M}^{\text{as}})$, which is independent of $M$.}~Relative to the $(\hat{u},\hat{v})$ coordinates, the null frame vector fields read
\begin{align*}
\widehat{e}_4^{\text{as}}&=\partial_{\hat{v}} \, ,  & \widehat{e}_{3,M}^{\text{as}}&=\frac{1}{4M}\, \left(1-\frac{1}{x}\right)^{-1}\partial_{\hat{u}} \, .
\end{align*}
To give an example of linearisation of a frame coefficients, one can Taylor-expand in $M$ (around $M_0$) the following scalar function
\begin{align*}
g_{M_0}(\widehat{e}_{3,M}^{\text{as}},\widehat{e}_{4}^{\text{as}})&= -2+4\,\frac{(M-M_0)}{M_0}+\mathcal{O}((M-M_0)^2)
\end{align*}
and obtain the linearised frame coefficient
\begin{equation*}
\overset{\text{\scalebox{.6}{$(1)$}}}{\underline{\mathfrak{f}}}_4= \mathfrak{m} \, .
\end{equation*}
All the linearised connection coefficient and curvature components are obtained by solving the full linearised system of equations.~The collection of all the linearised quantities obtained form the one-parameter family of \emph{reference linearised Schwarzschild solutions} $\mathsf{S}(\mathfrak{m})$, with $\mathfrak{m}\in\mathbb{R}$ (see Definition \ref{def_ref_schw}).~As one can see from the complete Definition \ref{def_ref_schw}, \emph{the family of reference linearised Schwarzschild solutions is supported on $\ell=0$-spherical modes}.

\item \textbf{$\bsy{\ell=1}$-reference linearised Kerr solutions}

The derivation of the reference linearised Kerr solutions is more subtle.~One can consider the Kerr metric in coordinates $(\bar{t},r,\theta,\bar{\phi})$, obtained from the standard Boyer--Lindquist coordinates $(t,r,\theta,\phi)$ via the transformation\footnote{We introduce the standard scalar functions $\Delta(r)=r^2-2Mr+a^2$ and $\Sigma(r,\theta)=r^2+a^2\cos^2\theta$ of the Boyer--Lindquist coordinates.}
\begin{align*}
d\bar{t}&=dt-\frac{r^2+a^2}{\Delta}\,dr \, , & d\bar{\phi}&=d\phi-\frac{a}{\Delta}\,dr \, .
\end{align*}
It is an easy check that, in the coordinates $(\bar{t},r,\theta,\bar{\phi})$, the Kerr metric $g_{a,M}$ takes the form
\begin{align*}
g_{a,M}=& \, -\left(1-\frac{2Mr}{\Sigma}\right) d\bar{t}^2-2\,d\bar{t}\,dr-\frac{4aMr}{\Sigma}\,\sin^2\theta\,d\bar{t}\,d\bar{\phi}+2 a \sin^2\theta\,dr\,d\bar{\phi} \\ &+\Sigma \, d\theta^2+\frac{(r^2+a^2)^2-a^2\Delta\sin^2\theta}{\Sigma}\sin^2\theta\,d\bar{\phi}^2
\end{align*}
and is in outgoing principal gauge (to any order of $a$).~We then consider the frame $\widehat{\mathcal{N}}_{\text{as}}$ with frame vector fields
\begin{align*}
\widehat{e}_{4,a}&=\frac{\Delta}{\Sigma}\,\partial_r  \, , \\
\widehat{e}_{3,a}&=\left(2+\frac{4Mr}{\Delta}+\frac{a^2r\sin^2\theta}{M\Sigma}\right) \partial_{\bar{t}}-\partial_r-\frac{a^2\sin(2\theta)}{2M\Sigma}\,\partial_{\theta}+a\left( \frac{2}{\Delta}+\frac{r}{M\Sigma}\right)\partial_{\bar{\phi}} \, , \\
\widehat{e}_{1,a}&= \frac{a^2\sin(2\theta)}{2\Sigma}\,\partial_{\bar{t}}+\frac{r}{\Sigma}\,\partial_{\theta}+\frac{a\cot\theta}{\Sigma}\,\partial_{\bar{\phi}} \, , \\
\widehat{e}_{2,a}&=\frac{ar\sin\theta}{\Sigma}\,\partial_{\bar{t}}+\frac{a\Delta\sin\theta}{2M\Sigma}\,\partial_r-\frac{a\cos\theta}{\Sigma}\,\partial_{\theta}+\frac{r\csc\theta}{\Sigma}\,\partial_{\bar{\phi}} \, .
\end{align*}
The frame $\widehat{\mathcal{N}}_{\text{as}}$ is obtained by performing the $\mathcal{O}(a)$-rotation of the algebraically special frame $\mathcal{N}_{\text{as}}$ of Kerr of Section 5.3 in \cite{benomio_kerr_system} such that
\begin{align*}
\widehat{e}_{4,a}&=e_4^{\text{as}}  \, , &
\widehat{e}_{3,a}&=e_3^{\text{as}}+\frac{a}{M}\,\sin\theta \,e_2^{\text{as}} \, , \\
\widehat{e}_{1,a}&= e_1^{\text{as}} \, , &
\widehat{e}_{2,a}&=e_2^{\text{as}}+\frac{a}{2M}\,\sin\theta \,e_4^{\text{as}} \, ,
\end{align*}
with $(e_1^{\text{as}},e_2^{\text{as}})$ as in footnote 57 of \cite{benomio_kerr_system}.~The frame $\widehat{\mathcal{N}}_{\text{as}}$ satisfies all the properties listed in Section 6.3 of \cite{benomio_kerr_system} \emph{to linear order in $a$} (around $a=0$), whereas the frame $\mathcal{N}_{\text{as}}$ fails at satisfying the property (201) in Section 6.3 of \cite{benomio_kerr_system} to linear order in $a$ (note that $\etab_{a,M}=\mathcal{O}(a)$ from (159) in Section 5.4 of \cite{benomio_kerr_system}) and thus cannot be employed to correctly derive the reference linearised Kerr solutions.~We linearise the frame coefficients of $\widehat{\mathcal{N}}_{\text{as}}$ relative to the angular parameter $a$ (around $a=0$), keeping the mass parameter $M$ fixed.~All the linearised connection coefficients and curvature components are obtained by solving the full linearised system of equations, e.g.
\begin{align*}
  (\overset{\text{\scalebox{.6}{$(1)$}}}{\slashed{\varepsilon}\cdot\chi})_{m=0}&=\frac{2 \mathfrak{a} \Omega^2_M}{r^2} \, Y_{m=0}^{\ell=1} \, , &
(\overset{\text{\scalebox{.6}{$(1)$}}}{\slashed{\varepsilon}\cdot\chib})_{m=0}&=\frac{2 \mathfrak{a}}{M r^2}\left(r+M\right) Y_{m=0}^{\ell=1} \, , & \betabo_{m=0}&=\frac{3 \mathfrak{a}}{r^3}\,r\,{}^{\star}\nablasl Y_{m=0}^{\ell=1} \, .
\end{align*}
The collection of all the linearised quantities obtained form three linearly independent one-parameter families of \emph{$\ell=1$-reference linearised Kerr solutions} $\mathsf{K}(\mathfrak{a},m)$, with $\mathfrak{a}\in\mathbb{R}$ and $m=-1,0,1$ (see Definition \ref{def_ref_kerr}).\footnote{Strictly speaking, the procedure that we described extracts the $\ell=1$-reference linearised Kerr solutions $\mathsf{K}(\mathfrak{a},m)$ with $m=0$.~The $\ell=1$-reference linearised Kerr solutions with $m=\pm 1$ are then trivial to obtain.}~As one can see from the complete Definition \ref{def_ref_kerr}, \emph{the family of $\ell=1$-reference linearised Kerr solutions is supported on $\ell=1$-spherical modes}.
\end{itemize}

The family of reference linearised Schwarzschild solutions and $\ell=1$-reference linearised Kerr solutions generate the four-parameter family of \emph{reference linearised Kerr solutions $\mathcal{K}_{\mathfrak{m},\mathfrak{s}_m}$}, with $\mathfrak{s}_m\in\mathbb{R}$, defined as the sum of the reference linearised solution $\mathsf{S}(\mathfrak{m})$ and the reference linearised solutions $\mathsf{K}(\mathfrak{a},m)$ such that
\begin{equation*}
\sigmao[\mathcal{K}_{\mathfrak{m},\mathfrak{s}_{m}}]=\sum_m \mathfrak{s}_{m}\cdot  \sigmao_{m}[\mathsf{K}(\mathfrak{a},m)] \, .
\end{equation*}
\emph{The family of reference linearised Kerr solutions is supported on $\ell=0,1$-spherical modes}.~In particular, the associated gauge invariant linearised quantities \eqref{intro_alpha_alphab_vanish_gauge} identically vanish.

\subsection{Initial-data normalised solutions} \label{sec_overview_renormalised_solutions}

In Section \ref{sec_normalised_solutions}, we define \emph{initial-data normalised solutions} to the system of linearised gravity.~These are the solutions for which the linear stability result is formulated (see already Theorem \ref{intro_main_theorem}).

\medskip

The definition of initial-data normalised solutions states two sets of identities, namely the normalisation identities for the $\ell=0,1$ and $\ell\geq 2$-spherical projections of initial-data normalised solutions (see Definition \ref{def_initial_gauge_norm}).~All the identities are formulated on the initial ingoing null cone $\underline{C}_{v_0}$.~In particular, the identities for the higher modes are formulated on the \emph{whole} initial ingoing null cone $\underline{C}_{v_0}$ in the case of the linearised frame coefficients, i.e.
\begin{align} \label{intro_normalisation_frame_comps}
\overset{\text{\scalebox{.6}{$(1)$}}}{\mathfrak{\underline{\mathfrak{f}}}}_4|_{\ell\geq 2} &= 0 \, , & \overset{\text{\scalebox{.6}{$(1)$}}}{\mathfrak{\slashed{\mathfrak{f}}}}_{4}|_{\ell\geq 2} &=0  \, , 
\end{align}
and on the \emph{initial horizon sphere} $\mathbb{S}^2_{\infty,v_0}$ for the remaining linearised quantities appearing in the definition, i.e. 
\begin{align} \label{intro_normalisation_metric}
(\textup{tr}\overset{\text{\scalebox{.6}{$(1)$}}}{\slashed{g}})|_{\ell\geq  2}&=0 \, , & \slashed{\textup{curl}}\,\slashed{\textup{div}}\overset{\text{\scalebox{.6}{$(1)$}}}{\widehat{\slashed{g}}}&= 2 \sigmao|_{\ell\geq 2} 
\end{align}
and\footnote{The form of the second of the identities \eqref{intro_normalisation_trchi} differs from how the identity is stated in Definition \ref{def_initial_gauge_norm}, where it reads $(\slashed{\text{div}}\etao+\rhoo)|_{\ell\geq 2}=0$.~The form in \eqref{intro_normalisation_trchi} is more convenient for the discussion of the present overview.}
\begin{align} \label{intro_normalisation_trchi}
(\trchio)|_{\ell\geq 2}&=0 \, , & \nablasl_3(\trchio)|_{\ell\geq 2}&=0  \, .
\end{align}
We note that the identities \eqref{intro_normalisation_metric} have no analogue in the definition of initial-data normalised solutions in \cite{DHR} (see Definition 9.1 therein).~We shall explain later in the overview why these are needed.~We also note that we do not state any normalisation identities on the initial outgoing null cone $C_{u_0}$.

\medskip

Crucially, the normalisation identities \eqref{intro_normalisation_frame_comps}-\eqref{intro_normalisation_trchi} are \emph{propagated} by the linearised system of equations.~The identities \eqref{intro_normalisation_frame_comps} thus hold globally on spacetime, whereas the identities \eqref{intro_normalisation_metric} and \eqref{intro_normalisation_trchi} hold globally along the event horizon.~We remark that the propagation of the full set of normalisation identities \eqref{intro_normalisation_frame_comps}-\eqref{intro_normalisation_trchi} is a special feature of \emph{linear} theory and plays a key role in the problem.~For instance, the first of the normalisation identities \eqref{intro_normalisation_trchi}, which fixes the initial location of the event horizon, is propagated by virtue of the homogeneous structure of the linearised transport equation \eqref{intro_lin_4_trchi}.~As a result, the initial location of the event horizon fixes the location of the event horizon for all future times.~An analogous property does not survive in \emph{nonlinear} theory, where the corresponding transport equation possesses nonlinear terms on the right hand side and the location of the event horizon can only be determined teleologically.

\subsubsection{Spherical projection and generality}

Initial-data normalised solutions satisfy a key property:~The $\ell=0,1$-spherical projection of any such solution coincides with a suitable reference linearised Kerr solution,\footnote{Consistently, the reference linearised Kerr solutions are initial-data normalised solutions.} and is thus non-dynamical (see Proposition \ref{prop_proj_kernel_kerr}).~Remarkably, the \emph{initial data} of an initial-data normalised solution completely identify the reference linearised Kerr solution coinciding with its $\ell=0,1$-spherical projection.~Such a reference linearised solution is interpreted as the reference linearised Kerr solution to which the initial-data normalised solution converges (as we shall prove) in the evolution.

\medskip

The linear stability result is formulated as an integrated energy decay statement for the dynamical $\ell\geq 2$-spherical projection of initial-data normalised solutions (see already Theorem \ref{intro_main_theorem}).\footnote{We note that the statement of Theorem \ref{intro_main_theorem} further assumes that the initial-data normalised solution arises from \emph{asymptotically flat} seed initial data.}~The fact that the result only considers initial-data normalised solutions does not determine any loss of generality.~Indeed, \emph{any} general solution to the linearised system of equations can be transformed into an initial-data normalised solution by adding an appropriate pure gauge solution (see Proposition \ref{prop_initial_data_normalisation}).~The pure gauge solution needed to achieve the initial-data normalisation can be entirely read off from the initial data of the given solution.

\subsection{The main theorem}

The main result of the paper is an integrated energy decay statement for angular derivatives of the linearised induced metric, frame coefficients, connection coefficients and curvature components of any \emph{initial-data} normalised solution to the linearised system of equations.~We state a first version of the main result in Theorem \ref{intro_main_theorem} below.\footnote{The combination of angular operators applied to the linearised curvature components in \eqref{intro_iled_alpha}-\eqref{intro_iled_rho_sigma} guarantees that all the quantities appearing in the estimates are \emph{exactly} subtracted of their $\ell=0,1$-spherical projection, and are thus supported on $\ell\geq 2$-spherical modes (see Section \ref{sec_The_Schwarzschild_exterior_manifold} for remarks on these operators).~Since, as previously observed, the $\ell=0,1$-spherical projection of an initial-data normalised solution coincides with a reference linearised Kerr solution, the theorem characterises the full dynamical part of the solution.}~The precise statement of the result is Theorem \ref{th_main_theorem} of Section \ref{sec_main_theorem}.~Our main theorem may be compared to Theorem 4 in \cite{DHR}, where the linear stability result is stated for \emph{future} normalised solutions.

\medskip

\begin{theorem}[Linear stability of the Schwarzschild solution, first version] \label{intro_main_theorem}
Consider an \ul{initial-data} normalised solution to the system of linearised equations arising from smooth, asymptotically flat seed initial data on $\mathcal{S}_{u_0,v_0}$.~Assume that the initial energy $\mathbb{E}_0$ defined in \eqref{def_initial_energy_norm} is finite.~Then, 
\begin{itemize}
\item The energy $\mathbb{E}_0$ uniformly bounds weighted $L^2$-fluxes on null cones for five angular derivatives of all linearised curvature components.~Moreover, for any $\delta>0$, $v\geq v_0$ and $u\geq u_0$, the degenerate (at trapping, $r=3M$) integrated energy decay estimates for five angular derivatives of the linearised curvature components 
\begin{align}
\int_{v_0}^v d\bar{v} \int_{u_0}^{u} d\bar{u}\int_{\mathbb{S}^2_{\bar{u},\bar{v}}} \sin\theta \,d\theta \, d\phi \,\frac{\Omega^2}{r^{1+\delta}} \left( 1-\frac{3M}{r}\right)^2  |\mathcal{A}^{[5]}(r^3\alphao)| ^2 &\lesssim \mathbb{E}_0 \, , \label{intro_iled_alpha}\\
\int_{v_0}^v d\bar{v} \int_{u_0}^{u} d\bar{u}\int_{\mathbb{S}^2_{\bar{u},\bar{v}}} \sin\theta \,d\theta \, d\phi \,\frac{\Omega^2}{r^{1+\delta}} \left( 1-\frac{3M}{r}\right)^2  |\mathcal{A}^{[5]}(r \alphabo)| ^2  &\lesssim \mathbb{E}_0 \, , \label{intro_iled_alphab}\\
\int_{v_0}^v d\bar{v} \int_{u_0}^{u} d\bar{u}\int_{\mathbb{S}^2_{\bar{u},\bar{v}}} \sin\theta \,d\theta \, d\phi \,\frac{\Omega^2}{r^{1+\delta}}\left( 1-\frac{3M}{r}\right)^2  |\mathcal{A}^{[4]}r\slashed{\mathcal{D}}_2^{\star}(r^3\betao)| ^2  &\lesssim \mathbb{E}_0 \, , \label{intro_iled_beta}\\
\int_{v_0}^v d\bar{v} \int_{u_0}^{u} d\bar{u}\int_{\mathbb{S}^2_{\bar{u},\bar{v}}} \sin\theta \,d\theta \, d\phi \,\frac{\Omega^2}{r^{1+\delta}}\left( 1-\frac{3M}{r}\right)^2  |\mathcal{A}^{[4]}r\slashed{\mathcal{D}}_2^{\star}(r^2\betabo)| ^2  &\lesssim \mathbb{E}_0 \, , \label{intro_iled_betab}\\
\int_{v_0}^v d\bar{v} \int_{u_0}^{u} d\bar{u}\int_{\mathbb{S}^2_{\bar{u},\bar{v}}} \sin\theta \,d\theta \, d\phi \,\frac{\Omega^2}{r^{1+\delta}} \left( 1-\frac{3M}{r}\right)^2  |\mathcal{A}^{[3]}r^2\slashed{\mathcal{D}}_2^{\star}\slashed{\mathcal{D}}_1^{\star}(r^3\rhoo,r^3\sigmao)| ^2 &\lesssim \mathbb{E}_0 \label{intro_iled_rho_sigma} 
\end{align}
hold.~In particular, analogous integrated energy decay estimates, but without degenerating factor, hold for one less angular derivative of all linearised curvature components.
\item Analogous uniform boundedness and (possibly non-degenerate) integrated energy decay estimates (in terms of $\mathbb{E}_0$) for up to five angular derivatives of the linearised connection coefficients, frame coefficients and induced metric hold.
\end{itemize}
The pointwise norm $|\cdot|$ and all angular operators are defined (in the standard way) in Section \ref{sec_The_Schwarzschild_exterior_manifold}.
\end{theorem}

\medskip

The top-order terms in the initial energy $\mathbb{E}_0$ involve five derivatives of linearised curvature components.~Thus, the uniform boundedness and (degenerate) integrated energy decay estimates for the linearised curvature components do \emph{not} lose derivatives.

\medskip

The proof of Theorem \ref{intro_main_theorem} is divided into two parts, the first one addressing the \emph{gauge invariant} linearised quantities in the system of equations and the second one treating the \emph{gauge dependent} linearised quantities.~The first part transfers the gauge invariant analysis of \cite{DHR}.~The second part implements a \emph{new scheme}.

\subsection{Analysis of the gauge invariant linearised quantities}  \label{sec_overview_analysis_gauge_invariant}

The first part of our proof is carried out in Section \ref{sec_analysis_gauge_invariant} and relies on the exact decoupling of the gauge invariant linearised quantities
\begin{align} \label{intro_alphas_gauge_inv_quantities}
&\alphao \, , & &\alphabo 
\end{align}
from the rest of the linearised system of equations.~The linearised quantities \eqref{intro_alphas_gauge_inv_quantities} satisfy two decoupled \emph{spin $\pm 2$ Teukolsky equations}.\footnote{See Section 1.4.3 in \cite{benomio_kerr_system}.}~One can define the (regular) derived gauge invariant linearised quantities
\begin{align*}
\overset{\text{\scalebox{.6}{$(1)$}}}{P}&:=-\frac{1}{2\,r^3}\nablasl_3(r^2\nablasl_3(r\alphao)) \, , &  \overset{\text{\scalebox{.6}{$(1)$}}}{\underline{P}}&:=-\frac{1}{2\,r^3\Omega^2}\nablasl_4(r^2\,\Omega^{-3}\nablasl_4(r\,\Omega^4\alphabo)) \, ,
\end{align*} 
which, after suitable radial rescaling, satisfy the so-called \emph{Regge--Wheeler equation}.~Both the Teukolsky and Regge-Wheeler equations are discussed at length in \cite{DHR, benomio_kerr_system}.~The analogue of the linearised quantities \eqref{intro_alphas_gauge_inv_quantities} in \cite{DHR} are still gauge invariant, exactly decouple and satisfy the same equations as the ones satisfied by \eqref{intro_alphas_gauge_inv_quantities}.

\medskip

One can analyse the linearised quantities \eqref{intro_alphas_gauge_inv_quantities} independnetly from the rest of the linearised system of equations.~In particular, one can apply Theorems 1 and 2 of \cite{DHR} as blackbox results to our problem and obtain the integrated energy decay estimates \eqref{intro_iled_alpha} and \eqref{intro_iled_alphab} of Theorem \ref{intro_main_theorem}.~Analogous estimates can be formulated for the derived gauge invariant linearised quantities.

\medskip

The control over the gauge invariant linearised quantities is crucial for the second part of the proof, where integrated energy decay estimates for the gauge dependent quantities are addressed.~In particular, we will make essential use of certain identities relating the (derived) gauge invariant linearised quantities and the gauge dependent linearised quantities.~Examples of such identities are the following
\begin{align}
\overset{\text{\scalebox{.6}{$(1)$}}}{P}&= \slashed{\mathcal{D}}_2^{\star}\slashed{\mathcal{D}}_1^{\star}(-\rhoo,\sigmao) -\frac{3}{4}\,\rho\,(\textup{tr}\chib)\chiho-\frac{3}{4}\,\rho\,(\textup{tr}\chi)\chibho  +(\nablasl_3\rho)\,\slashed{\mathcal{D}}_2^{\star}\overset{\text{\scalebox{.6}{$(1)$}}}{\mathfrak{\slashed{\mathfrak{f}}}}_{4}+(\nablasl_4\rho)\,\slashed{\mathcal{D}}_2^{\star}\overset{\text{\scalebox{.6}{$(1)$}}}{\mathfrak{\slashed{\mathfrak{f}}}}_{3}  \, ,  \label{intro_rel_P}\\ 
\overset{\text{\scalebox{.6}{$(1)$}}}{\underline{P}}&=\slashed{\mathcal{D}}_2^{\star}\slashed{\mathcal{D}}_1^{\star}(-\rhoo,-\sigmao) -\frac{3}{4}\,\rho\,(\textup{tr}\chib)\chiho-\frac{3}{4}\,\rho\,(\textup{tr}\chi)\chibho  +(\nablasl_3\rho)\,\slashed{\mathcal{D}}_2^{\star}\overset{\text{\scalebox{.6}{$(1)$}}}{\mathfrak{\slashed{\mathfrak{f}}}}_{4}+(\nablasl_4\rho)\,\slashed{\mathcal{D}}_2^{\star}\overset{\text{\scalebox{.6}{$(1)$}}}{\mathfrak{\slashed{\mathfrak{f}}}}_{3} \, .  \label{intro_rel_Pb}
\end{align}
Compared to \cite{DHR}, the right hand sides of \eqref{intro_rel_P} and \eqref{intro_rel_Pb} possess additional terms, which depend on the linearised frame coefficients.\footnote{Initial-data normalised solutions satisfy a simplified version of these identities, but some of the additional terms still appear.~See Remark \ref{rmk_derived_gauge_inv_quantities_rels}.}

\subsection{Analysis of the gauge dependent linearised quantities}

The analysis of the gauge dependent linearised quantities is the core of our new proof of linear stability of the Schwarzschild solution and is carried out in Section \ref{sec_analysis_gauge_dependent}.~The key new feature of the analysis is that integrated energy decay for \emph{all} the gauge dependent linearised quantities is established for initial-data normalised solutions, without introducing any future gauge re-normalisation.~Such a novelty relies on the new structure of the linearised \emph{outgoing} transport equations, which present the features described in Section \ref{sec_overview_system}.~In particular, the the \emph{red-shifted} transport equations play a fundamental role in our new scheme.

\subsubsection{Horizon energy fluxes} \label{sec_overview_horizon_fluxes}

To prove integrated energy decay estimates, one first needs to control the \emph{horizon} energy fluxes of some of the gauge dependent linearised quantities (see Section \ref{sec_horizon_fluxes}).~To achieve that, the structure of the linearised elliptic equations in the system and the identities for (and previous control over) the gauge invariant linearised quantities are crucial.~For initial-data normalised solutions, these equations and identities coincide with their correspondending ones in \cite{DHR} \emph{when restricted to the event horizon}.~We note, for instance, the identities \eqref{intro_rel_P} and \eqref{intro_rel_Pb} reducing to
\begin{align*}
\overset{\text{\scalebox{.6}{$(1)$}}}{P}&= \slashed{\mathcal{D}}_2^{\star}\slashed{\mathcal{D}}_1^{\star}(-\rhoo,\sigmao) -\frac{3}{4}\,\rho\,(\textup{tr}\chib)\chiho    \, , &
\overset{\text{\scalebox{.6}{$(1)$}}}{\underline{P}}&=\slashed{\mathcal{D}}_2^{\star}\slashed{\mathcal{D}}_1^{\star}(-\rhoo,-\sigmao) -\frac{3}{4}\,\rho\,(\textup{tr}\chib)\chiho    
\end{align*} 
and the linearised Codazzi equation \eqref{intro_lin_codazzi} reducing to 
\begin{align*}
\slashed{\text{div}}\chiho =  -\betao 
\end{align*} 
along the event horizon.~This convenient feature allows to closely follow the arguments of \cite{DHR} and achieve control over all the horizon energy fluxes controlled therein.~One, for instance, controls the following energy fluxes for the linearised outgoing shear
\begin{equation} \label{intro_fluxes_outgoing_shear}
\int_{\mathcal{H}^+(v_0,\infty)}dv \sin\theta \,d\theta\, d\phi\,(|\chiho|^2+|\nablasl_3\chiho|^2 ) \lesssim \mathbb{E}_0 \, .
\end{equation}

\medskip

Importantly, in our proof, we need control over additional horizon energy fluxes to the ones obtained in \cite{DHR}.~In particular, for the linearised induced metric, we derive the estimate
\begin{equation}   \label{intro_fluxes_metric}
\int_{\mathcal{H}^+(v_0,\infty)}dv \sin\theta \,d\theta\, d\phi\, ( |\slashed{\textup{div}}\,\slashed{\textup{div}}\overset{\text{\scalebox{.6}{$(1)$}}}{\widehat{\slashed{g}}} |^2+|\slashed{\textup{curl}}\,\slashed{\textup{div}}\overset{\text{\scalebox{.6}{$(1)$}}}{\widehat{\slashed{g}}} |^2) \lesssim \mathbb{E}_0  \, , 
\end{equation} 
which can be achieved by exploiting the linearised Gauss equation, the intrinsic expression for the linearised Gauss curvature in terms of the linearised induced metric (see \eqref{aux_expression_K}) and the (global, along the event horizon) normalisation identities \eqref{intro_normalisation_metric}.~The control over the energy flux \eqref{intro_fluxes_metric} is where the novel normalisation identities \eqref{intro_normalisation_metric} enter crucially in our proof.

\subsubsection{Integrated decay for the gauge dependent linearised quantities} \label{sec_overview_analysis_gauge_dependent_quantities}

To elucidate the central role of the red-shifted transport equations in the gauge dependent part of the analysis, we consider the red-shifted linearised null structure equations in schematic form\footnote{Analogous considerations apply to the red-shifted transport equations for the linearised frame coefficients.}
\begin{equation} \label{intro_model_transport_eqn}
\nablasl_4\overset{\text{\scalebox{.6}{$(1)$}}}{\Gamma}+\frac{p}{2}\,(\text{tr}\chi) \,\overset{\text{\scalebox{.6}{$(1)$}}}{\Gamma}+k \omegah \overset{\text{\scalebox{.6}{$(1)$}}}{\Gamma}=\overset{\text{\scalebox{.6}{$(1)$}}}{F} \, ,
\end{equation} 
with $p\in\mathbb{Z}_0$ and $k\in\mathbb{N}$.~We recall that the connection coefficient $\omegah$ is an everywhere \emph{positive} (including \emph{on} the event horizon) smooth scalar function.~We think of the linearised quantities in the equation \eqref{intro_model_transport_eqn} as symmetric traceless $\mathbb{S}^2_{u,v}$ two-tensors (and thus supported on $\ell\geq 2$-spherical modes), possibly obtained by taking a suitable combination of angular derivatives, e.g.
\begin{equation*}
\overset{\text{\scalebox{.6}{$(1)$}}}{\Gamma}=\left\lbrace  \chibho \, , \, \slashed{\mathcal{D}}_2^{\star}\zetao \, , \, \slashed{\mathcal{D}}_2^{\star}\nablasl(\trchibo) \, ,  \, \ldots  \right\rbrace \, .
\end{equation*}
For a linearised connection coefficient satisfying an equation of the form \eqref{intro_model_transport_eqn}, we can establish a global integrated energy decay estimate of the form
\begin{equation} \label{intro_model_iled}
\int_{v_0}^v d\bar{v} \int_{u_0}^u d\bar{u} \int_{\mathbb{S}^2_{\bar{u},\bar{v}}} \sin\theta\, d\theta \, d\phi \, \frac{\Omega^2}{r^{1+\delta}} \, |r^p\overset{\text{\scalebox{.6}{$(1)$}}}{\Gamma}|^2 \lesssim \mathbb{E}_0  
\end{equation}
for any $\delta>0$, $v\geq v_0$ and $u\geq u_0$, provided that an estimate of the form \eqref{intro_model_iled} holds for the (suitably weighted) right hand side of \eqref{intro_model_transport_eqn} and that the initial energy $\mathbb{E}_0$ is finite and controls the initial energy flux 
\begin{equation*}
\int_{\underline{C}_{v_0}} \sin\theta\, du\,d\theta \, d\phi\,\Omega^2\,|\overset{\text{\scalebox{.6}{$(1)$}}}{\Gamma}|^2 \, .
\end{equation*}
We remark that if the right hand side of \eqref{intro_model_transport_eqn} satisfies a \emph{degenerate} integrated decay estimate, meaning an estimate of the form \eqref{intro_model_iled} with an additional degenerating factor inside the integral, then the estimate \eqref{intro_model_iled} inherits such a degeneracy.~Since top-order estimates for the gauge invariant linearised quantities \eqref{intro_alphas_gauge_inv_quantities} degenerate at trapping $r=3M$, degenerate estimates for linearised connection coefficients and curvature components do inevitably appear in our scheme.~We also point out that, since suitably weighted angular operators commute trivially with the equation \eqref{intro_model_transport_eqn}, an estimate like \eqref{intro_model_iled} holds for suitably weighted angular derivatives of the linearised connection coefficient.\footnote{Again, provided appropriate control over angular derivatives of the right hand side of \eqref{intro_model_transport_eqn} and over the initial energy flux of angular derivatives of the linearised connection coefficient.}

\medskip

The proof of the estimate \eqref{intro_model_iled} follows two steps.~One first proves an integrated energy decay estimate in a bounded (in $r$) spacetime region which extends from the event horizon.~As a second step, one upgrades the local integrated energy decay estimate obtained in the first step to a global (in $r$) estimate. 

\medskip

The red-shift term in the equation \eqref{intro_model_transport_eqn} is crucial to prove the first local estimate, for this reason known as \emph{red-shift estimate}.~Indeed, from equation \eqref{intro_model_transport_eqn}, one can derive the inequality
\begin{equation} \label{intro_ineq_redshift}
\nablasl_4|\overset{\text{\scalebox{.6}{$(1)$}}}{\Gamma}|^2+ \omegah|\overset{\text{\scalebox{.6}{$(1)$}}}{\Gamma}|^2\lesssim |\overset{\text{\scalebox{.6}{$(1)$}}}{F}|^2
\end{equation}
over a spacetime region $[u(r_1,v),\infty]\times [v_0,\infty]\times \mathbb{S}^2_{u,v}$, with $2M<r_1<\infty$ (see shaded region in Figure \ref{fig:redshift_estimate}).~By integrating \eqref{intro_ineq_redshift} over $[u(r_1,v),\infty]\times [v_0,v_*]\times \mathbb{S}^2_{u,v}$, the first term on the left hand side yields the difference of positive energy fluxes 
\begin{equation*}
\int_{\underline{C}_{v_*}\cap\left\lbrace u\geq u(r_1,v_*) \right\rbrace} \sin\theta\, du\,d\theta \, d\phi\,\Omega^2\,|\overset{\text{\scalebox{.6}{$(1)$}}}{\Gamma}|^2 - \int_{\underline{C}_{v_0}\cap\left\lbrace u\geq u(r_1,v_0) \right\rbrace} \sin\theta\, du\,d\theta \, d\phi\,\Omega^2\,|\overset{\text{\scalebox{.6}{$(1)$}}}{\Gamma}|^2 \, ,
\end{equation*}
where the first energy flux can be dropped and the second is taken to the right hand side and controlled (together with the spacetime integral of the right hand side of \eqref{intro_ineq_redshift}) by the initial energy $\mathbb{E}_0$.~The positivity of the red-shift factor $k\omegah$ is crucial to guarantee that one estimates a positive quantity.~The fact that the equation \eqref{intro_model_transport_eqn} is a transport equation in the outgoing direction suffices for the extension of the local integrated decay estimate to a global estimate.~For this second step, the red-shift term does not play a role, whereas the second term on the left hand side of the equation \eqref{intro_model_transport_eqn} dictates the precise weighted linearised quantity that one can estimate globally in $r$ (note, in particular, the role of $p$). 

\medskip

\begin{figure}
\centering

\begin{tikzpicture}

\draw (-2,0)--(0,2);
\draw[dashed] (0,2)--(2,0);
\draw[dashed] (-2,0)--(0,-2);
\draw[dashed] (0,-2)--(2,0);
\draw (0,-0.4)--(1.2,0.8);
\draw (0,-0.4)--(-1.2,0.8);

\node at (-1,1.4) {$\mathcal{H}^+$};
\node at (1.1,1.4) {$\mathcal{I}^+$};
\node at (0.15,1) {${}_{r=r_1}$};
\node at (1.1,0.2) {$C_{u_0}$};
\node at (-1,0.2) {$\underline{C}_{v_0}$};

\draw (0,2) circle (1.5pt);
\draw (2,0) circle (1.5pt);
\draw (0,-2) circle (1.5pt);

\filldraw[gray, draw=black] (-0.27,-0.13) -- (-1.2,0.8) -- (0,2) -- (-0.25,1) -- (-0.3,0.6) -- (-0.3,0.3) -- (-0.3,0.05) -- cycle;

\end{tikzpicture}

\caption{}
\label{fig:redshift_estimate}

\end{figure}

\medskip

In addition to the red-shifted equations, our scheme also exploits the \emph{no-shifted} and \emph{blue-shifted} outgoing transport equations (i.e.~$k=0$ and $k\in\mathbb{Z}_-$ respectively).~These equations are either commuted with transversal derivatives or multiplied by a growing (towards the event horizon) factor to obtain \emph{red-shifted} equations of the form \eqref{intro_model_transport_eqn} for higher order or rescaled versions of the original linearised quantity (see some examples below).~The commutation procedures generate lower order terms, whose control requires the estimates for the horizon energy fluxes discussed in Section \ref{sec_overview_horizon_fluxes}.~We point out that, in our scheme, none of the gauge dependent linearised quantities is controlled by integrating an ingoing transport equation in the ingoing direction.

\medskip

We now briefly comment on the integrated energy decay estimates for some specific linearised quantities: 
\begin{itemize}
\item\textbf{Integrated decay for the linearised outgoing shear.}~Our scheme starts from proving integrated energy decay for the linearised outgoing shear in Section \ref{sec_iled_chih}.~As already noted in Section \ref{sec_overview_system}, the transport equation \eqref{intro_4_chih} for the linearised outgoing shear coincides with the corresponding linearised equation in \cite{DHR}, and can thus be analogously exploited to derive integrated energy decay estimates.~We briefly recall the argument from Section 13.3 of \cite{DHR}.

The transport equation \eqref{intro_4_chih} is \emph{blue-shifted}.~By commuting twice the equation with a $\nablasl_3$-derivative, we obtain a red-shifted transport equation of the form \eqref{intro_model_transport_eqn} for two $\nablasl_3$-derivatives of the linearised outgoing shear, where lower order terms also appear on the left hand side.~The presence of the lower order terms implies that a red-shift estimate can only be proven in a region sufficiently close to the event horizon and has to exploit the horizon energy fluxes \eqref{intro_fluxes_outgoing_shear}.~For any $\delta>0$, $v\geq v_0$ and $u\geq u_0$, one can then extend the local estimate to the global integrated energy decay estimate
\begin{equation} \label{intro_iled_chih}
\int_{v_0}^v d\bar{v} \int_{u_0}^u d\bar{u} \int_{\mathbb{S}^2_{\bar{u},\bar{v}}} \sin\theta\, d\theta \, d\phi \, \frac{\Omega^2}{r^{1+\delta}} \, |r^2\chiho|^2 \lesssim \mathbb{E}_0 \, .
\end{equation}
The argument crucially relies on previous control over the gauge invariant linearised quantity appearing on the right hand side of \eqref{intro_4_chih}.

The commutation procedure to estimate the linearised outgoing shear is typically referred to as \emph{red-shift commutation}, in that it exploits the presence of the red-shift effect at the event horizon.~The technique is closely related to the vector field commutation developed in \cite{Dafermos_Rodnianski_decay_waves_schwarzschild} for the wave equation on Schwarzschild exterior spacetimes.~See Section 2.3 in \cite{DHR}.

\item\textbf{Integrated decay for the linearised ingoing shear.}~In Section \ref{sec_iled_chibh}, we obtain integrated energy decay for the linearised ingoing shear
\begin{equation*}
\int_{v_0}^v d\bar{v} \int_{u_0}^u d\bar{u} \int_{\mathbb{S}^2_{\bar{u},\bar{v}}} \sin\theta\, d\theta \, d\phi \, \frac{\Omega^2}{r^{1+\delta}} \, |r\chibho|^2 \lesssim \mathbb{E}_0 
\end{equation*}
by exploiting the \emph{red-shifted} transport equation \eqref{intro_4_chibh}.\footnote{We remark that, since the equation \eqref{intro_4_chibh} is already red-shifted, no preliminary red-shift commutation is needed.}~As previously noted in Section \ref{sec_overview_system}, the equation only couples with the linearised outgoing shear, which we already control at this stage of the scheme.

This step of the scheme is where our proof departs from the approach of \cite{DHR} to the analysis of initial-data normalised solutions.~After proving integrated energy decay for the linearised outgoing shear, the proof of \cite{DHR} can only establish uniform \emph{boundedness} for the linearised ingoing shear.~To obtain integrated energy \emph{decay}, the proof of \cite{DHR} has to ultimately resort to a future gauge re-normalisation.\footnote{As already observed in Section \ref{sec_overview_system}, the second variational formula for the linearised ingoing shear in \cite{DHR} only couples with a gauge invariant linearised quantity (cf.~equation \eqref{intro_3_chib}), and can thus be immediately exploited in the analysis of the gauge dependent part of the system.~The equation does not, however, allow to obtain \emph{decay} for the linearised ingoing shear of an initial-data normalised solution.}

\item\textbf{Integrated decay for the remaining linearised quantities.}~Once integrated energy decay for both the linearised outgoing and ingoing shears is established, one can obtain integrated energy decay for \emph{all} the remaining gauge dependent linearised quantities.~See Section \ref{sec_iled_remaining_quantities}.

We comment on some of the remaining steps in the scheme.~The fully decoupled \emph{blue-shifted} equation \eqref{intro_lin_4_trchi} allows to prove integrated energy decay for the rescaled linearised quantity\footnote{In fact, the linearised quantity \eqref{intro_rescaled_trchi} can be estimated at any stage of our scheme.}
\begin{equation} \label{intro_rescaled_trchi}
\mathcal{A}^{[3]}r^2\slashed{\mathcal{D}}^{\star}_2\nablasl(\Omega^{-4}(\trchio)) \, .
\end{equation}
The linearised quantity \eqref{intro_rescaled_trchi} satisfies the red-shifted transport equation
\begin{equation*}
\nablasl_4(\mathcal{A}^{[3]}r^2\slashed{\mathcal{D}}^{\star}_2\nablasl(\Omega^{-4}(\trchio)))+(\text{tr}\chi)\,(\mathcal{A}^{[3]}r^2\slashed{\mathcal{D}}^{\star}_2\nablasl(\Omega^{-4}(\trchio)))+\omegah\,(\mathcal{A}^{[3]}r^2\slashed{\mathcal{D}}^{\star}_2\nablasl(\Omega^{-4}(\trchio)))=0 \, ,
\end{equation*}
which can be used to achieve the integrated energy decay estimate
\begin{equation} \label{intro_iled_trchi}
\int_{v_0}^v d\bar{v} \int_{u_0}^{u} d\bar{u}\int_{\mathbb{S}^2_{\bar{u},\bar{v}}} \sin\theta \,d\theta \, d\phi \,\frac{\Omega^2}{r^{1+\delta}} \, | \mathcal{A}^{[3]}r^2\slashed{\mathcal{D}}^{\star}_2\nablasl(r^2\Omega^{-4}(\trchio) ) | ^2 \lesssim \mathbb{E}_0 \, .
\end{equation}
We remark that the normalisation identities \eqref{intro_normalisation_trchi} guarantee that the linearised quantity \eqref{intro_rescaled_trchi} remains uniformly bounded on the initial ingoing null cone $\underline{C}_{v_0}$, including on the initial horizon sphere $\mathbb{S}^2_{\infty,v_0}$.

The estimate \eqref{intro_iled_trchi} allows to employ the red-shifted transport equation
\begin{equation*}
\nablasl_4(\mathcal{A}^{[3]}r^2\slashed{\mathcal{D}}^{\star}_2\nablasl(\Omega^{-2}(\textup{tr}\overset{\text{\scalebox{.6}{$(1)$}}}{\slashed{g}})))+\omegah\,(\mathcal{A}^{[3]}r^2\slashed{\mathcal{D}}^{\star}_2\nablasl(\Omega^{-2}(\textup{tr}\overset{\text{\scalebox{.6}{$(1)$}}}{\slashed{g}})))=2\,r^{-2}\Omega^2\mathcal{A}^{[3]}r^2\slashed{\mathcal{D}}^{\star}_2\nablasl(r^2\Omega^{-4}(\trchio))
\end{equation*}
to prove the integrated energy decay estimate
\begin{equation*}
\int_{v_0}^v d\bar{v} \int_{u_0}^{u} d\bar{u}\int_{\mathbb{S}^2_{\bar{u},\bar{v}}} \sin\theta \,d\theta \, d\phi \,\frac{\Omega^2}{r^{1+\delta}} \, |  \mathcal{A}^{[3]}r^2\slashed{\mathcal{D}}^{\star}_2\nablasl(\Omega^{-2}(\textup{tr}\overset{\text{\scalebox{.6}{$(1)$}}}{\slashed{g}})) | ^2 \lesssim \mathbb{E}_0 \, .
\end{equation*}
We note that the linearised first variational formula in its original form \eqref{4_trg} is \emph{no-shifted}, and thus it would not yield decay for the non-rescaled linearised quantity.~The first of the normalisation identities \eqref{intro_normalisation_metric} ensures that the rescaled linearised quantity that we estimate remains uniformly bounded on the initial ingoing null cone $\underline{C}_{v_0}$, including on the initial horizon sphere $\mathbb{S}^2_{\infty,v_0}$.

The linearised first variational formulae in the outgoing direction also allow to prove the integrated energy decay estimate
\begin{equation} \label{intro_iled_ghat}
\int_{v_0}^v d\bar{v} \int_{u_0}^{u} d\bar{u}\int_{\mathbb{S}^2_{\bar{u},\bar{v}}} \sin\theta \,d\theta \, d\phi \,\frac{\Omega^2}{r^{1+\delta}} \, | \mathcal{A}^{[3]}\overset{\text{\scalebox{.6}{$(1)$}}}{\widehat{\slashed{g}}} | ^2  \lesssim \mathbb{E}_0 \, .
\end{equation}
As for the equation \eqref{4_trg}, the equation \eqref{4_ghat} is \emph{no-shifted}.~To obtain the estimate \eqref{intro_iled_ghat}, one has to commute once with a $\nablasl_3$-derivative the equation \eqref{4_ghat} to obtain the red-shifted transport equation 
\begin{equation*}
\nablasl_4(\nablasl_3(\mathcal{A}^{[3]}\overset{\text{\scalebox{.6}{$(1)$}}}{\widehat{\slashed{g}}}))+\omegah\,(\nablasl_3(\mathcal{A}^{[3]}\overset{\text{\scalebox{.6}{$(1)$}}}{\widehat{\slashed{g}}}))= 2r^{-2} \nablasl_3(\mathcal{A}^{[3]}(r^2\chiho))+4r^{-3} (\mathcal{A}^{[3]}(r^2\chiho)) \, .
\end{equation*}
Integrated energy decay is first established for the transversal derivative and then upgraded to the estimate \eqref{intro_iled_ghat}.~This step of the scheme makes essential use of the control over the horizon energy flux \eqref{intro_fluxes_metric}, which in turn relies on the normalisation identities \eqref{intro_normalisation_metric}.

\item\textbf{Top-order estimates for the linearised curvature components.}~The integrated energy decay estimates for the linearised ingoing and outgoing shears described above only yield a (non-degenerate) version of the estimates \eqref{intro_iled_beta} and \eqref{intro_iled_betab} for \emph{four} angular derivatives.~The estimates \eqref{intro_iled_beta} and \eqref{intro_iled_betab} are achieved in Section \ref{sec_top_order_estimates_curvature} and require a refinement of the estimates for both the linearised shears.~The refined estimates exploit the linearised elliptic equations in the system (e.g.~the Codazzi equation \eqref{intro_lin_codazzi}) and the lower order version of the estimates \eqref{intro_iled_beta} and \eqref{intro_iled_betab} already obtained.~This concludes the analysis of the gauge dependent linearised quantities.

\item\textbf{Early estimates for the linearised induced metric and frame coefficients.}~Our scheme to prove integrated energy decay for the gauge dependent linearised quantities is not the only one allowed by the system of linearised gravity.~For reasons discussed in Appendix \ref{sec_early_estimates}, it is relevant to note that most of the integrated energy decay estimates for the linearised induced metric and frame coefficients can be obtained earlier than in our scheme and, in particular, \emph{without} exploiting prior integrated energy decay estimates for the linearised \emph{ingoing} shear.
\end{itemize}

In Section \ref{sec_proof_main_th}, we combine the energy estimates obtained as part of the gauge invariant and gauge dependent analysis into a proof of Theorem \ref{intro_main_theorem}.

\section{The Schwarzschild exterior manifold} \label{sec_The_Schwarzschild_exterior_manifold}

In this section, we introduce the Schwarzschild exterior manifold and define some geometric quantities associated to it.~We assume the formalism from Section 4 of \cite{benomio_kerr_system}.

\medskip

Let $(\mathcal{M},g_{M})$ be the Schwarzschild exterior manifold with Eddington--Finkelstein (double null) differentiable structure $(u,v,\theta^A)$ on $\mathcal{M}\setminus\mathcal{H}^+$ such that
\begin{equation} \label{schw_m}
g_M=-4\left( 1-\frac{2M}{r(u,v)} \right)du\, dv + r^2(u,v) \slashed{\gamma}_{\theta^A\theta^B}d\theta^A\, d\theta^B \, ,
\end{equation}
with $\mathcal{H}^+:=\partial\mathcal{M}$ denoting the future event horizon, $r(u,v)$ the standard Schwarzschild radius and $\slashed{\gamma}$ the round metric on the unit two-sphere.~The definition of both $(\mathcal{M},g_{M})$ and the differentiable structure $(u,v,\theta^A)$ can be found in Sections 4.1 and 4.2 of \cite{DHR}.~The Schwarzschild exterior manifold can also be obtained by setting $|a|=0$ for the Kerr exterior manifold defined in Section 5 of \cite{benomio_kerr_system}.~We define the two-spheres
\begin{equation*}
\mathbb{S}^2_{u,v}:= \left\lbrace u , v \right\rbrace \times \mathbb{S}^2 \, .
\end{equation*}

\medskip

We define the scalar function $$\Omega_M^2(u,v):=1-\frac{2M}{r(u,v)}$$ on $\mathcal{M}\setminus\mathcal{H}^+$ and the local null frame $$\mathcal{N}_{\text{as}}=(e^{\text{as}}_1,e^{\text{as}}_2,e^{\text{as}}_3,e^{\text{as}}_4)$$ on $\mathcal{M}\setminus\mathcal{H}^+$, with
\begin{align} \label{def_e4_e3}
e^{\text{as}}_4&:=\partial_v \, , &  e^{\text{as}}_3&:=\Omega^{-2}_M\,\partial_u 
\end{align}
and $(e^{\text{as}}_1,e^{\text{as}}_2)$ \emph{orthonormal} vector fields.~We note that $\Omega_M^2$ is a smooth function on $\mathcal{M}\setminus\mathcal{H}^+$ and can be smoothly extended to $\mathcal{H}^+$.~We also note that $\mathcal{N}_{\text{as}}$ is a regular frame on $\mathcal{M}\setminus\mathcal{H}^+$ and extends \emph{regularly} to a non-degenerate frame on $\mathcal{H}^+$.

\medskip

\begin{remark}
Our choice of null frame vector fields \eqref{def_e4_e3} differs from the one in \cite{DHR}, where the null frame vector fields do not extend regularly to $\mathcal{H}^+$.~The connection coefficients and curvature components of $g_M$ relative to our frame $\mathcal{N}_{\textup{as}}$ are regular quantities, with no need for the regularising weights employed in \cite{DHR}.
\end{remark}

\medskip

We assume the definitions for the induced metric, connection coefficients and curvature components and the notion of $\mathfrak{D}_{\mathcal{N}_{\text{as}}}$ tensors from Section 4 of \cite{benomio_kerr_system}.~In the present work, we shall refer to $\mathfrak{D}_{\mathcal{N}_{\text{as}}}$ tensors as $\mathbb{S}^2_{u,v}$ \emph{tensors}.~The non-vanishing connection coefficients and curvature components of $g_M$ relative to $\mathcal{N}_{\text{as}}$ are the smooth scalar functions
\begin{align*}
\omegah_M&=\frac{2M}{r^2} \, ,  &
(\text{tr}\chi)_M &=\frac{2\,\Omega_M^2}{r}  \, ,  &  (\text{tr}\chib)_M &=-\frac{2}{r} \, ,  & \rho_M &=-\frac{2M}{r^3} \, .
\end{align*}
In particular, we note that $\omegabh_M=0$.~The connection coefficients and curvature components are regular quantities on $\mathcal{M}\setminus \mathcal{H}^+$ which extend regularly to $\mathcal{H}^+$.

\medskip

For any finite $u_0,v_0\in\mathbb{R}$, we define the ingoing and outgoing null cones
\begin{align*}
\underline{C}_{v_0}&:= \left\lbrace u\geq u_0, v= v_0 \right\rbrace \, , &  C_{u_0}&:= \left\lbrace u=u_0 ,v\geq v_0\right\rbrace
\end{align*}
emanating (to the future) from the two-sphere $\mathbb{S}^2_{u_0,v_0}$ and the union of two null cones $$\mathcal{S}_{u_0,v_0}:=\underline{C}_{v_0}\cup C_{u_0} \, .$$~We also define the spacetime region 
$$\mathcal{R}_{u_0,v_0}:= \bigcup_{u\geq u_0 , v\geq v_0}\mathbb{S}^2_{u,v} \, .$$

\subsection{Angular operators on spheres and commutation formulae}

In this section, we recall some useful identities.~We assume the definition of the angular operators and the formulae from Sections 4.4 and 4.6 of \cite{benomio_kerr_system}.

\medskip

Consider a smooth scalar function $f$ and a $\mathbb{S}^2_{u,v}$ one-tensor $\varsigma$.~We have the identities
\begin{align*}
\slashed{\text{div}}\nablasl f &=\slashed{\Delta}f \, , & \slashed{\text{div}}{}^{\star}\nablasl f &=0  \\
\slashed{\text{curl}}{}^{\star}\nablasl f &=-\slashed{\Delta}f\, , & \slashed{\text{curl}}\nablasl f &=0 
\end{align*}
and
\begin{equation*}
\slashed{\text{div}}\,\slashed{\mathcal{D}}^{\star}_2\,\varsigma =-\frac{1}{2}\,(\slashed{\Delta}+\slashed{K})\,\varsigma \, , 
\end{equation*}
with $\slashed{K}$ the Gauss curvature of the $\mathbb{S}^2_{u,v}$ spheres.~We recall that $\slashed{\Delta}$ is the Laplacian associated to $\slashed{g}_M$.~We define the Laplacian on the unit sphere $\Delta_{\mathbb{S}^2}:=r^2\,\slashed{\Delta}$.

\medskip

We have the commutation formulae
\begin{align*}
[\nablasl_4,\nablasl]\,\varsigma &= -\frac{1}{2}\,(\text{tr}\chi)\nablasl\varsigma \, , &
[\nablasl_3,\nablasl]\,\varsigma &= -\frac{1}{2}\,(\text{tr}\chib)\nablasl\varsigma \, , &
[\nablasl_3,\nablasl_4]\,\varsigma &=\omegah\nablasl_3\varsigma 
\end{align*}
and, in particular,
\begin{align*}
[\nablasl_4,r\nablasl]\,\varsigma &= 0 \, , &
[\nablasl_3,r\nablasl]\,\varsigma &= 0 \, , &
[\Omega^2_M\nablasl_3,\nablasl_4]\,\varsigma &=0 
\end{align*}
and
\begin{align*}
[\nablasl_4,r\,\slashed{\text{div}}]\,\varsigma &= 0 \, , &
[\nablasl_4,r\,\slashed{\text{curl}}]\,\varsigma &= 0 \, , &
[\nablasl_4,r\,\slashed{\mathcal{D}}^{\star}_2]\,\varsigma &=0 \, , & [\nablasl_4,r^2\,\slashed{\Delta}]\,\varsigma &=0 \, ,\\
[\nablasl_3,r\,\slashed{\text{div}}]\,\varsigma &= 0 \, , &
[\nablasl_3,r\,\slashed{\text{curl}}]\,\varsigma &= 0 \, , &
[\nablasl_3,r\,\slashed{\mathcal{D}}^{\star}_2]\,\varsigma &=0 \, , & [\nablasl_3,r^2\,\slashed{\Delta}]\,\varsigma &=0 \, .
\end{align*}
We define the angular operators $\mathcal{A}^{[i]}$ acting on symmetric traceless $\mathbb{S}^2_{u,v}$ two-tensors, with $i\in\mathbb{N}_0$, such that $\mathcal{A}^{[i]}= \text{Id}$ and satisfying the inductive relations
\begin{align*}
\mathcal{A}^{[2i+1]}&=r\,\slashed{\text{div}}\,\mathcal{A}^{[2i]} \, , & \mathcal{A}^{[2i+2]}&=r^2\,\slashed{\mathcal{D}}^{\star}_2\,\slashed{\text{div}}\,\mathcal{A}^{[2i]} \, .
\end{align*}
We have the commutation formulae
\begin{align*}
[\nablasl_4,\mathcal{A}^{[i]}]\,\varsigma &= 0 \, , &
[\nablasl_3,\mathcal{A}^{[i]}]\,\varsigma &= 0 \, .
\end{align*}

\subsection{Norms, spherical harmonics and elliptic estimates on spheres}

In this section, we define some relevant norms and recall some useful facts about spherical harmonics and elliptic estimates on spheres.~We assume the definitions and propositions from Sections 4.4.2 and 4.4.3 of \cite{DHR} for the $\ell$-spherical harmonics $Y^{\ell}_m(\theta,\phi)$ associated to the $\mathbb{S}^2_{u,v}$-spheres, the $\ell$-spherical projections $f|_{\ell=0}$, $f|_{\ell=1}$, $f|_{\ell\geq 2}$, $\varsigma|_{\ell=1}$ and $\varsigma|_{\ell\geq 2}$ of smooth scalar functions $f$ and $\mathbb{S}^2_{u,v}$ one and two-tensors $\varsigma$ and the elliptic estimates for angular operators acting on $S^2_{u,v}$ tensors.

\medskip

Consider a $\mathbb{S}^2_{u,v}$ $k$-tensor $\varsigma$.~We define the pointwise norm
\begin{equation*}
|\varsigma|^2:=\slashed{g}_M^{A_1B_1}\cdots\slashed{g}_M^{A_kB_k}\varsigma_{A_1\cdots A_k} \varsigma_{B_1\cdots B_k}
\end{equation*}
and, for any finite $u_0,v_0\in\mathbb{R}$, the $L^2$-norms on null cones
\begin{align}
\lVert \varsigma \rVert^2_{L^2(\underline{C}_{v_0})} &:= \int_{u_0}^{\infty}d\bar{u}\int_{\mathbb{S}^2_{u,v_0}}\sin\theta\,d\theta\,d\phi\,\Omega^2 \, |\varsigma|^2  \, , \label{def_norm_ingoing_cone}\\
\lVert \varsigma \rVert^2_{L^2(C_{u_0})} &:= \int_{v_0}^{\infty}d\bar{v}\int_{\mathbb{S}^2_{u_0,v}}\sin\theta\,d\theta\,d\phi\, |\varsigma|^2 \, , \label{def_norm_outgoing_cone}
\end{align}
with $(\theta,\phi)$ standard spherical coordinates on the $\mathbb{S}^2_{u,v}$-spheres.~We note that \ul{the norms} \eqref{def_norm_ingoing_cone} \ul{and} \eqref{def_norm_outgoing_cone} \ul{do not contain any weight in $r$}.

\medskip

We recall the identities
\begin{align*}
\slashed{\Delta} Y^{\ell}_m &=-\frac{1}{r^2}\,\ell(\ell+1)\,Y^{\ell}_m \, , &  \Delta_{\mathbb{S}^2} Y^{\ell}_m&= -\ell(\ell+1)\,Y^{\ell}_m 
\end{align*}
and the fact that the spherical harmonics $Y^{\ell=0}_{m=0}$, $Y^{\ell=1}_{m=-1}(\theta,\phi)$, $Y^{\ell=1}_{m=0}(\theta,\phi)$ and $Y^{\ell=1}_{m=1}(\theta,\phi)$ form a basis for all pairs of scalar functions in the kernel of the angular operator $r^2\slashed{\mathcal{D}}^{\star}_2\slashed{\mathcal{D}}^{\star}_1$ on the unit sphere.~In particular, the symmetric traceless $S^2_{u,v}$ two-tensors
\begin{align*}
&r^2\slashed{\mathcal{D}}^{\star}_2\nablasl f \, , & &r\slashed{\mathcal{D}}^{\star}_2\varsigma
\end{align*} 
identically vanish for any scalar function $f=f|_{\ell=0,1}$ and $\mathbb{S}^2_{u,v}$ one-tensor $\varsigma=\varsigma|_{\ell=1}$.~Furthermore, we recall that, for any symmetric traceless $\mathbb{S}^2_{u,v}$ two-tensor $\varsigma$, its $\ell=1$-spherical projection and the $\ell=0,1$-spherical projection of the scalar functions $\slashed{\text{div}}\,\slashed{\text{div}}\,\varsigma$ and $\slashed{\text{curl}}\,\slashed{\text{div}}\,\varsigma$ identically vanish.

\section{The system of linearised gravity} \label{sec_system_linearised_gravity}

The system of linearised gravity on the Schwarzschild exterior manifold $(\mathcal{M},g_M)$ is obtained by setting $|a|=0$ in the system of linearised gravity derived in \cite{benomio_kerr_system}.~A complete list of the unknowns in the system is presented in Table \ref{table:linearised_unknowns}.~All the unknowns are regular quantities on $\mathcal{M}\setminus\mathcal{H}^+$ which extend regularly to $\mathcal{H}^+$.

\medskip

\begin{table}[H]
\centering
\begin{tabular}{ |c||c|c|c| } 
 \hline
 {} & Scalar function & $\mathbb{S}^2_{u,v}$ one-tensor & $\mathbb{S}^2_{u,v}$ two-tensor \\
 \hline\hline
Metric, Frame & $\overset{\text{\scalebox{.6}{$(1)$}}}{\mathfrak{\underline{\mathfrak{f}}}}_4,\overset{\text{\scalebox{.6}{$(1)$}}}{\mathfrak{\underline{\mathfrak{f}}}}_3,(\text{tr}\overset{\text{\scalebox{.6}{$(1)$}}}{\slashed{g}})$ &  $\overset{\text{\scalebox{.6}{$(1)$}}}{\mathfrak{\underline{\mathfrak{f}}}},\overset{\text{\scalebox{.6}{$(1)$}}}{\mathfrak{\slashed{\mathfrak{f}}}}_{3},\overset{\text{\scalebox{.6}{$(1)$}}}{\mathfrak{\slashed{\mathfrak{f}}}}_{4}$ & $\overset{\text{\scalebox{.6}{$(1)$}}}{\widehat{\slashed{g}}}$ \\
\hline
Connection & $(\overset{\text{\scalebox{.6}{$(1)$}}}{\text{tr}\chi}),(\overset{\text{\scalebox{.6}{$(1)$}}}{\slashed{\varepsilon}\cdot\chi}),(\overset{\text{\scalebox{.6}{$(1)$}}}{\text{tr}\chib}),(\overset{\text{\scalebox{.6}{$(1)$}}}{\slashed{\varepsilon}\cdot\chib}),\overset{\text{\scalebox{.6}{$(1)$}}}{\omegab}$ & $\overset{\text{\scalebox{.6}{$(1)$}}}{\yb},\overset{\text{\scalebox{.6}{$(1)$}}}{\eta},\overset{\text{\scalebox{.6}{$(1)$}}}{\zeta}$ & $\overset{\text{\scalebox{.6}{$(1)$}}}{\chih},\overset{\text{\scalebox{.6}{$(1)$}}}{\chibh}$  \\
\hline
Curvature & $\overset{\text{\scalebox{.6}{$(1)$}}}{\widetilde{\slashed{K}}},\overset{\text{\scalebox{.6}{$(1)$}}}{\rho},\overset{\text{\scalebox{.6}{$(1)$}}}{\sigma}$ & $\overset{\text{\scalebox{.6}{$(1)$}}}{\beta},\overset{\text{\scalebox{.6}{$(1)$}}}{\betab}$  & $\overset{\text{\scalebox{.6}{$(1)$}}}{\alpha},\overset{\text{\scalebox{.6}{$(1)$}}}{\alphab}$ \\
\hline
\end{tabular}
\caption{Unknowns of the system of linearised gravity.}
\label{table:linearised_unknowns}
\end{table}

\medskip

A \emph{solution} $\mathfrak{S}$ to the system of linearised gravity is a collection of all linearised quantities listed in Table \ref{table:linearised_unknowns} solving the system of equations.~The sum of any two solutions $\mathfrak{S}_1$ and $\mathfrak{S}_2$, obtained by summing the corresponding linearised quantities relative to the two solutions, remains, by linearity, a solution to the system of linearised gravity.

\medskip

The full system of linearised gravity reads as follows.~We have the transport equations
\begin{align}
\nablasl_4\overset{\text{\scalebox{.6}{$(1)$}}}{\mathfrak{\underline{\mathfrak{f}}}}_4 &= 0 \, , \label{4_fb4}\\
\nablasl_4 \overset{\text{\scalebox{.6}{$(1)$}}}{\mathfrak{\underline{\mathfrak{f}}}}_3+\omegah \overset{\text{\scalebox{.6}{$(1)$}}}{\mathfrak{\underline{\mathfrak{f}}}}_3 &=  \omegabo \, , \label{4_f3}\\
\nablasl_4\overset{\text{\scalebox{.6}{$(1)$}}}{\mathfrak{\underline{\mathfrak{f}}}}-\frac{1}{2}\,(\text{tr}\chi)\overset{\text{\scalebox{.6}{$(1)$}}}{\mathfrak{\underline{\mathfrak{f}}}}+\omegah \overset{\text{\scalebox{.6}{$(1)$}}}{\mathfrak{\underline{\mathfrak{f}}}} 
 &= -2 \etao \label{4_fA}
\end{align}
\begin{align}
\nablasl_4 \overset{\text{\scalebox{.6}{$(1)$}}}{\mathfrak{\slashed{\mathfrak{f}}}}_{4}+\frac{1}{2}\,(\text{tr}\chi) \overset{\text{\scalebox{.6}{$(1)$}}}{\mathfrak{\slashed{\mathfrak{f}}}}_{4}-\omegah \overset{\text{\scalebox{.6}{$(1)$}}}{\mathfrak{\slashed{\mathfrak{f}}}}_{4}&=0  \, , \label{4_fA4}\\
\nablasl_3 \overset{\text{\scalebox{.6}{$(1)$}}}{\mathfrak{\slashed{\mathfrak{f}}}}_{3}+\frac{1}{2}\,(\text{tr}\chib)\overset{\text{\scalebox{.6}{$(1)$}}}{\mathfrak{\slashed{\mathfrak{f}}}}_{3} &= \nablasl \overset{\text{\scalebox{.6}{$(1)$}}}{\mathfrak{\underline{\mathfrak{f}}}}_3+\ybo \label{3_fA3}
\end{align}
and
\begin{align}
\nablasl_4 \overset{\text{\scalebox{.6}{$(1)$}}}{\mathfrak{\slashed{\mathfrak{f}}}}_{3}+\frac{1}{2}\,(\text{tr}\chi) \overset{\text{\scalebox{.6}{$(1)$}}}{\mathfrak{\slashed{\mathfrak{f}}}}_{3}&= \zetao \, , \label{4_fA3} \\
\nablasl_3 \overset{\text{\scalebox{.6}{$(1)$}}}{\mathfrak{\slashed{\mathfrak{f}}}}_{4}+\frac{1}{2}\,(\text{tr}\chib)\overset{\text{\scalebox{.6}{$(1)$}}}{\mathfrak{\slashed{\mathfrak{f}}}}_{4} &= \nablasl \overset{\text{\scalebox{.6}{$(1)$}}}{\mathfrak{\underline{\mathfrak{f}}}}_4+\etao-\zetao -\omegah \overset{\text{\scalebox{.6}{$(1)$}}}{\mathfrak{\slashed{\mathfrak{f}}}}_{3}  \label{3_fA4}
\end{align}
and the elliptic equations
\begin{align}
2\,\slashed{\text{curl}}\overset{\text{\scalebox{.6}{$(1)$}}}{\mathfrak{\slashed{\mathfrak{f}}}}_{4} &= (\overset{\text{\scalebox{.6}{$(1)$}}}{\slashed{\varepsilon}\cdot\chi})  \, ,  \label{A_fA4}\\
2\,\slashed{\text{curl}}\overset{\text{\scalebox{.6}{$(1)$}}}{\mathfrak{\slashed{\mathfrak{f}}}}_{3} &=  (\overset{\text{\scalebox{.6}{$(1)$}}}{\slashed{\varepsilon}\cdot\chib}) \, . \label{A_fA3}
\end{align}
We have the linearised first variational formulae
\begin{align}
\nablasl_4\,\overset{\text{\scalebox{.6}{$(1)$}}}{\widehat{\slashed{g}}} &=  2 \chiho  \, ,\label{4_ghat} \\ 
\nablasl_3\, \overset{\text{\scalebox{.6}{$(1)$}}}{\widehat{\slashed{g}}}  &=  2 \chibho 
  +2\,\slashed{\mathcal{D}}_2^{\star}\overset{\text{\scalebox{.6}{$(1)$}}}{\mathfrak{\underline{\mathfrak{f}}}} \, , \label{3_ghat} 
\end{align}
\begin{align}
\nablasl_4(\text{tr}\overset{\text{\scalebox{.6}{$(1)$}}}{\slashed{g}}) &= 2\,(\trchio) \, , \label{4_trg} \\ 
\nablasl_3 (\text{tr}\overset{\text{\scalebox{.6}{$(1)$}}}{\slashed{g}})  &=  2\,(\trchibo)-2\,\slashed{\text{div}}\overset{\text{\scalebox{.6}{$(1)$}}}{\mathfrak{\underline{\mathfrak{f}}}}  -2\,(\text{tr}\chi)\overset{\text{\scalebox{.6}{$(1)$}}}{\mathfrak{\underline{\mathfrak{f}}}}_{3}-2\,(\text{tr}\chib)\overset{\text{\scalebox{.6}{$(1)$}}}{\mathfrak{\underline{\mathfrak{f}}}}_{4}  \, , \label{3_trgo}
\end{align}
the linearised second variational formulae
\begin{align}
\nablasl_4 \chiho +(\text{tr}\chi)\chiho-\omegah \chiho =& \, -\alphao \, , \label{4_chiho}\\
\nablasl_3 \chibho+(\text{tr}\chib)\chibho =& \, -2\,\slashed{\mathcal{D}}_2^{\star} \ybo    -\alphabo  \label{3_chibho}
\end{align}
and the linearised Raychaudhuri equations
\begin{align}
\nablasl_4 (\trchio)+ (\text{tr}\chi)(\trchio)-\omegah \, (\trchio) =& \,  0 \, , \label{4_trchio}\\
\nablasl_3 (\trchibo)+ (\text{tr}\chib)(\trchibo) =& \, 2\, \slashed{\text{div}} \ybo +(\text{tr}\chib)\omegabo    \label{3_trchibo}\\ & -\nablasl_3(\text{tr}\chib)\overset{\text{\scalebox{.6}{$(1)$}}}{\mathfrak{\underline{\mathfrak{f}}}}_4 -\nablasl_4(\text{tr}\chib)\overset{\text{\scalebox{.6}{$(1)$}}}{\mathfrak{\underline{\mathfrak{f}}}}_3   \, . \nonumber
\end{align}
We have the linearised mixed transport equations 
\begin{align}
\nablasl_4 \chibho+\frac{1}{2}\,(\text{tr}\chi)\chibho+\omegah\chibho =& \, -\frac{1}{2}\,(\text{tr}\chib)\chiho \, , \label{4_chibho} \\
\nablasl_3 \chiho+\frac{1}{2}\,(\text{tr}\chib)\chiho =& \, -2\,\slashed{\mathcal{D}}_2^{\star} \etao  -\frac{1}{2}\,(\text{tr}\chi)\chibho  \, , \label{3_chiho}
\end{align}
\begin{align}
\nablasl_4( \trchibo)+ \frac{1}{2}\,(\text{tr}\chi)(\trchibo)+\omegah \, (\trchibo) =& \, - \frac{1}{2}\,(\text{tr}\chib)(\trchio)+2\rhoo \, ,  \label{4_trchibo}\\
\nablasl_3 (\trchio)+\frac{1}{2}\,(\text{tr}\chib)(\trchio) =& \, - \frac{1}{2}(\text{tr}\chi)(\trchibo) -(\text{tr}\chi)\omegabo +2\,\slashed{\text{div}}\etao+2\rhoo \label{3_trchio} \\
& -\nablasl_4(\text{tr}\chi)\overset{\text{\scalebox{.6}{$(1)$}}}{\mathfrak{\underline{\mathfrak{f}}}}_3  -\nablasl_3(\text{tr}\chi)\overset{\text{\scalebox{.6}{$(1)$}}}{\mathfrak{\underline{\mathfrak{f}}}}_4    \nonumber
\end{align}
and the transport equations
\begin{align}
\nablasl_4 (\overset{\text{\scalebox{.6}{$(1)$}}}{\slashed{\varepsilon}\cdot\chi})+(\text{tr}\chi)(\overset{\text{\scalebox{.6}{$(1)$}}}{\slashed{\varepsilon}\cdot\chi})-\omegah \, (\overset{\text{\scalebox{.6}{$(1)$}}}{\slashed{\varepsilon}\cdot\chi})   =& \, 0 \, , \label{4_atrchio}\\
\nablasl_3 (\overset{\text{\scalebox{.6}{$(1)$}}}{\slashed{\varepsilon}\cdot\chib})+(\text{tr}\chib)(\overset{\text{\scalebox{.6}{$(1)$}}}{\slashed{\varepsilon}\cdot\chib})  =& \,   2\, \slashed{\text{curl}} \, \ybo \, , \label{3_atrchibo}
\end{align}
\begin{align}
\nablasl_4 (\overset{\text{\scalebox{.6}{$(1)$}}}{\slashed{\varepsilon}\cdot\chib}) +\frac{1}{2}\,(\text{tr}\chi)(\overset{\text{\scalebox{.6}{$(1)$}}}{\slashed{\varepsilon}\cdot\chib})+\omegah \, (\overset{\text{\scalebox{.6}{$(1)$}}}{\slashed{\varepsilon}\cdot\chib})  =& \, -\frac{1}{2}\,(\text{tr}\chib)(\overset{\text{\scalebox{.6}{$(1)$}}}{\slashed{\varepsilon}\cdot \chi})    +2\sigmao \, , \label{4_atrchibo} \\
\nablasl_3 (\overset{\text{\scalebox{.6}{$(1)$}}}{\slashed{\varepsilon}\cdot\chi}) +\frac{1}{2}\,(\text{tr}\chib)(\overset{\text{\scalebox{.6}{$(1)$}}}{\slashed{\varepsilon}\cdot\chi}) =& \, -\frac{1}{2}\,(\text{tr}\chi)(\overset{\text{\scalebox{.6}{$(1)$}}}{\slashed{\varepsilon}\cdot \chib})+2\,\slashed{\text{curl}}\etao -2\sigmao  \, . \label{3_atrchio}
\end{align}
We have the equations
\begin{align}
\nablasl_4 \etao +\frac{1}{2}\,(\text{tr}\chi)\etao     &=   - \betao  \, ,  \label{4_etao}\\
 \nablasl_4\ybo +2\omegah \ybo   &=       -\frac{1}{2}\,(\text{tr}\chib) \etao - \betabo  \, , \label{4_ybo}\\
\nablasl_4\omegabo+2\omegah\omegabo  &= -2\rhoo      -(\nablasl_4\omegah)\overset{\text{\scalebox{.6}{$(1)$}}}{\mathfrak{\underline{\mathfrak{f}}}}_3  -(\nablasl_3\omegah)\overset{\text{\scalebox{.6}{$(1)$}}}{\mathfrak{\underline{\mathfrak{f}}}}_4  \label{4_omegabo}
\end{align}
and the equations for the torsion
\begin{align}
\nablasl_4 \zetao +\frac{1}{2}\,(\text{tr}\chi)\zetao +\omegah \zetao   &= -\betao -(\nablasl_3\omegah)\overset{\text{\scalebox{.6}{$(1)$}}}{\mathfrak{\slashed{\mathfrak{f}}}}_{4}-(\nablasl_4\omegah)\overset{\text{\scalebox{.6}{$(1)$}}}{\mathfrak{\slashed{\mathfrak{f}}}}_{3}     \, ,  \label{4_zetao} \\
\nablasl_3 \zetao +\frac{1}{2}\,(\text{tr}\chib) \zetao   &= \nablasl \omegabo -\frac{1}{2}\,(\text{tr}\chib) \etao  +\frac{1}{2}\,(\text{tr}\chi) \ybo-\omegah \ybo    -\betabo \, , \label{3_zetao}\\
\slashed{\text{curl}}\zetao &=  \frac{1}{4}\,((\text{tr}\chi)-2\,\omegah)(\overset{\text{\scalebox{.6}{$(1)$}}}{\slashed{\varepsilon}\cdot\chib})  -\frac{1}{4}\,(\text{tr}\chib)(\overset{\text{\scalebox{.6}{$(1)$}}}{\slashed{\varepsilon}\cdot\chi})  +\sigmao \, . \label{curl_zetao}
\end{align}
We have the linearised Codazzi equations
\begin{align}
\slashed{\text{div}}\chiho =& \,\frac{1}{2}\,\nablasl (\trchio) -\frac{1}{2}\,{}^{\star}(\nablasl (\overset{\text{\scalebox{.6}{$(1)$}}}{\slashed{\varepsilon}\cdot\chi}))  +\frac{1}{2}\,(\text{tr}\chi)\zetao  -\betao \label{codazzi_chiho}\\
& +\frac{1}{2}\,(\nablasl_3(\text{tr}\chi))\overset{\text{\scalebox{.6}{$(1)$}}}{\mathfrak{\slashed{\mathfrak{f}}}}_{4}+\frac{1}{2}\,(\nablasl_4(\text{tr}\chi))\overset{\text{\scalebox{.6}{$(1)$}}}{\mathfrak{\slashed{\mathfrak{f}}}}_{3} \, ,  \nonumber \\
\slashed{\text{div}}\chibho =& \, \frac{1}{2}\,\nablasl (\trchibo)-\frac{1}{2}\,{}^{\star} (\nablasl (\overset{\text{\scalebox{.6}{$(1)$}}}{\slashed{\varepsilon}\cdot\chib}))-\frac{1}{2}\,(\text{tr}\chib)\zetao +\betabo  \label{codazzi_chibho}\\
& +\frac{1}{2}\,(\nablasl_3(\text{tr}\chib))\overset{\text{\scalebox{.6}{$(1)$}}}{\mathfrak{\slashed{\mathfrak{f}}}}_{4}+\frac{1}{2}\,(\nablasl_4(\text{tr}\chib))\overset{\text{\scalebox{.6}{$(1)$}}}{\mathfrak{\slashed{\mathfrak{f}}}}_{3}  \nonumber
\end{align}
and the linearised Gauss equation
\begin{align}
\overset{\text{\scalebox{.6}{$(1)$}}}{\widetilde{\slashed{K}}} =& \, -\frac{1}{4}\,(\text{tr}\chib)(\trchio)-\frac{1}{4}\,(\text{tr}\chi)(\trchibo)  -\rhoo \, . \label{gauss_eqn} 
\end{align}
The linearised Bianchi equations read
\begin{gather}
\nablasl_3\alphao+\frac{1}{2}\,(\text{tr}\chib)\alphao  = \, -2\,\slashed{\mathcal{D}}_2^{\star}\betao-3\,\rho\,\chiho   \, ,\label{3_alphao}\\
\nablasl_4\betao+2 \, (\text{tr}\chi)\betao-\omegah\betao =\, \slashed{\text{div}}\alphao \, , \label{4_betao}
\end{gather}
\begin{align}
\nablasl_3\betao+(\text{tr} \chib)\betao =& \,\slashed{\mathcal{D}}_1^{\star}(-\rhoo,\sigmao)+3\, \rho\etao +(\nablasl_3\rho)\overset{\text{\scalebox{.6}{$(1)$}}}{\mathfrak{\slashed{\mathfrak{f}}}}_{4}+(\nablasl_4\rho)\overset{\text{\scalebox{.6}{$(1)$}}}{\mathfrak{\slashed{\mathfrak{f}}}}_{3} \, , \label{3_betao}
\end{align}
\begin{gather}
\nablasl_4 \rhoo+\frac{3}{2}\,(\text{tr}\chi)\rhoo  =\,\slashed{\text{div}}\betao-\frac{3}{2}\,\rho\,(\trchio) \, , \label{4_rhoo}\\
\nablasl_4 \sigmao+\frac{3}{2}\,(\text{tr}\chi)\sigmao =\, -\slashed{\text{curl}}\betao +\frac{3}{2}\,\rho\,(\overset{\text{\scalebox{.6}{$(1)$}}}{\slashed{\varepsilon}\cdot\chi}) \, ,  \label{4_sigmao}
\end{gather}
\begin{gather}
\nablasl_3 \rhoo+\frac{3}{2}\,(\text{tr}\chib)\rhoo  = \, -\slashed{\text{div}}\betabo-\frac{3}{2}\,\rho\,(\trchibo)   -(\nablasl_4\rho)\overset{\text{\scalebox{.6}{$(1)$}}}{\mathfrak{\underline{\mathfrak{f}}}}_3  -(\nablasl_3\rho)\overset{\text{\scalebox{.6}{$(1)$}}}{\mathfrak{\underline{\mathfrak{f}}}}_4 \, , \label{3_rhoo} \\
\nablasl_3 \sigmao+\frac{3}{2}\,(\text{tr}\chib)\sigmao  = \, -\slashed{\text{curl}}\betabo-\frac{3}{2}\,\rho\,(\overset{\text{\scalebox{.6}{$(1)$}}}{\slashed{\varepsilon}\cdot\chib}) \, , \label{3_sigmao}
\end{gather}
\begin{align}
\nablasl_4\betabo+(\text{tr}\chi)\betabo+\omegah\betabo  =& \,\slashed{\mathcal{D}}_1^{\star}(\rhoo,\sigmao)-(\nablasl_3\rho)\overset{\text{\scalebox{.6}{$(1)$}}}{\mathfrak{\slashed{\mathfrak{f}}}}_{4}-(\nablasl_4\rho)\overset{\text{\scalebox{.6}{$(1)$}}}{\mathfrak{\slashed{\mathfrak{f}}}}_{3} \, , \label{4_betabo}  
\end{align}
\begin{gather}
\nablasl_3\betabo+2\,(\text{tr}\chib)\betabo = \, -\slashed{\text{div}}\alphabo -3\,\rho\ybo \, , \label{3_betabo}\\ 
\nablasl_4\alphabo+\frac{1}{2}\,(\text{tr}\chi)\alphabo+2\,\omegah\alphabo =  \,  2\,\slashed{\mathcal{D}}_2^{\star}\betabo-3\,\rho\chibho  \, . \label{4_alphabo} 
\end{gather}

\section{Initial data and well-posedness of the linearised system} \label{sec_initial_data_well_posedness}

Let us recall the definitions of $\mathcal{S}_{u_0,v_0}$ and $\mathcal{R}_{u_0,v_0}$ from Section \ref{sec_The_Schwarzschild_exterior_manifold}.~We give the following definition.

\medskip

\begin{definition}[Smooth seed initial data]
A \emph{smooth seed initial data set} on $\mathcal{S}_{u_0,v_0}$ for the linearised system of equations consists of the following geometric quantities:
\begin{align*}
\overset{\text{\scalebox{.6}{$(1)$}}}{\widehat{\slashed{g}}}& & &\text{on $\mathcal{S}_{u_0,v_0}$,}\\
\overset{\text{\scalebox{.6}{$(1)$}}}{\mathfrak{\underline{\mathfrak{f}}}}_4 \, , \, \overset{\text{\scalebox{.6}{$(1)$}}}{\mathfrak{\underline{\mathfrak{f}}}}_3 \, , \, \overset{\text{\scalebox{.6}{$(1)$}}}{\mathfrak{\underline{\mathfrak{f}}}} \, , \,  \overset{\text{\scalebox{.6}{$(1)$}}}{\mathfrak{\slashed{\mathfrak{f}}}}_{4} \, , \, \overset{\text{\scalebox{.6}{$(1)$}}}{\mathfrak{\slashed{\mathfrak{f}}}}_{3} \, , \, \omegabo & & &\text{on $\underline{C}_{v_0}$,} \\
(\textup{tr}\overset{\text{\scalebox{.6}{$(1)$}}}{\slashed{g}}) \, , \, (\overset{\text{\scalebox{.6}{$(1)$}}}{\textup{tr}\chi}) \, , \, (\overset{\text{\scalebox{.6}{$(1)$}}}{\textup{tr}\chib}) \, , \, \zetao  & & &\text{on $\mathbb{S}^2_{\infty,v_0}$.} 
\end{align*}
\end{definition}

\medskip

We have the following well-posedness statement for the linearised system of equations.~The proof is presented in Appendix \ref{appendix_well_posedness}.

\medskip

\begin{prop}[Well-posedness] \label{prop_well_posedness}
For any smooth seed initial data set on $\mathcal{S}_{u_0,v_0}$, there exists a unique smooth solution $\mathfrak{S}$ to the linearised system of equations on $\mathcal{R}_{u_0,v_0}$ which agrees with the given initial data set on $\mathcal{S}_{u_0,v_0}$.
\end{prop}

\medskip

Let
\begin{equation*}
\slashed{D}_{n_1,n_2}\varsigma := (r\nablasl)^{n_1}(r\nablasl_4)^{n_2}\varsigma
\end{equation*}
for any $\mathbb{S}^2_{u,v}$ tensor $\varsigma$ and any $n_1\geq 0$ and $n_2\geq 0$.~We now define the notion of (pointwise) \emph{asymptotic flatness} for smooth seed initial data.

\medskip

\begin{definition}[Pointwise asymptotic flatness] \label{def_asymptotic_flat_data}
A smooth seed initial data set is \emph{(pointwise) asymptotically flat} with weight $s$ to order $n$ if there exists a constant $C_{n_1,n_2}>0$ such that
\begin{equation} \label{ineq_asymptotic_flat}
|\overset{\text{\scalebox{.6}{$(1)$}}}{\widehat{\slashed{g}}}|+|\slashed{D}_{n_1,n_2}(r^2\nablasl_4\overset{\text{\scalebox{.6}{$(1)$}}}{\widehat{\slashed{g}}})|+|\slashed{D}_{n_1,n_2}(r^{3+s}r^{-2}\Omega^2(\nablasl_4(r^2\Omega^{-2}\nablasl_4\overset{\text{\scalebox{.6}{$(1)$}}}{\widehat{\slashed{g}}}))|\leq C_{n_1,n_2}
\end{equation}
on the initial outgoing null cone $C_{u_0}$ for some $0< s \leq 1$ and any $n_1\geq 0$ and $n_2\geq 0$, with $n_1+n_2\leq n$.
\end{definition}

\medskip

\begin{remark}
The well-posed statement of Proposition \ref{prop_well_posedness} does not assume that the smooth seed initial data are asymptotically flat, whereas the main result of Theorem \ref{th_main_theorem} does require asymptotic flatness as defined in Definition \ref{def_asymptotic_flat_data}.
\end{remark}

\medskip

\begin{remark}
Asymptotic flatness is propagated by the equations.~For smooth seed initial data which are asymptotically flat to sufficiently high order $n$, the inequality \eqref{ineq_asymptotic_flat} holds (at order $k$ sufficiently lower than $n$) on any outgoing null cone $C_{u}$ with $u\geq u_0$.~A precise statement and proof for the propagation of asymptotic flatness are not given in the present paper.~The corresponding statement in \cite{DHR} is Theorem A.1.
\end{remark}

\medskip

\begin{remark}
Our Definition \ref{def_asymptotic_flat_data} may be compared to Definition 8.2 of asymptotic flatness in \cite{DHR}.~We note that our inequality \eqref{ineq_asymptotic_flat} coincides, mutatis mutandis, with the inequality (188) in \cite{DHR}.~We also note that, for our seed initial data, the linearised quantity controlled by the inequality \eqref{ineq_asymptotic_flat} is the only one prescribed on the initial outgoing null cone $C_{u_0}$.
\end{remark}

\section{Initial-data normalised solutions} \label{sec_normalised_solutions}

We define \emph{initial-data normalised solutions} to the linearised system of equations.

\medskip

\begin{definition}[Initial-data normalised solutions] \label{def_initial_gauge_norm}
An \emph{initial-data normalised solution} $\mathfrak{S}_N$ to the linearised system of equations is a solution satisfying the following normalisation identities:
\begin{itemize}
\item On $\underline{C}_{v_0}$ (including $\mathbb{S}^2_{\infty,v_0}$):~The \emph{ingoing-cone normalisation identities}
\begin{align*}
\overset{\text{\scalebox{.6}{$(1)$}}}{\mathfrak{\underline{\mathfrak{f}}}}_4|_{\ell\geq 2} &= 0 \, , & \overset{\text{\scalebox{.6}{$(1)$}}}{\mathfrak{\slashed{\mathfrak{f}}}}_{4}|_{\ell\geq 2} &=0   
\end{align*}
and the \emph{auxiliary ingoing-cone normalisation identities}
\begin{align*}
\overset{\text{\scalebox{.6}{$(1)$}}}{\mathfrak{\underline{\mathfrak{f}}}}_4|_{\ell=0}&=-4M^2\rhoo|_{\ell=0}(\infty,v_0)   \, , &
\overset{\text{\scalebox{.6}{$(1)$}}}{\mathfrak{\underline{\mathfrak{f}}}}_4|_{\ell=1}&=0 \, , \\
\overset{\text{\scalebox{.6}{$(1)$}}}{\mathfrak{\underline{\mathfrak{f}}}}_3|_{\ell = 0,1}&=0 \, , & (\slashed{\textup{curl}}\overset{\text{\scalebox{.6}{$(1)$}}}{\mathfrak{\underline{\mathfrak{f}}}})|_{\ell=1}&= -\frac{8}{3}\,M\, \sigmao|_{\ell=1}(\infty,v_0) \, , \\
\overset{\text{\scalebox{.6}{$(1)$}}}{\mathfrak{\slashed{\mathfrak{f}}}}_{4}|_{\ell=1}&=\frac{4}{3}\,M^3\,\Omega^2_M\,{}^{\star}\nablasl \sigmao|_{\ell=1}(\infty,v_0) \, , &
\overset{\text{\scalebox{.6}{$(1)$}}}{\mathfrak{\slashed{\mathfrak{f}}}}_{3}|_{\ell=1}&=\frac{4}{3}\,M^2\left(r+M\right){}^{\star}\nablasl\sigmao|_{\ell=1}(\infty,v_0) \, , \\
\omegabo|_{\ell=0,1}&=0 \, .
\end{align*}

\item On $\mathbb{S}^2_{\infty,v_0}$:~The \emph{horizon normalisation identities}
\begin{align*}
(\textup{tr}\overset{\text{\scalebox{.6}{$(1)$}}}{\slashed{g}})|_{\ell\geq  2}&=0 \, , & \slashed{\textup{curl}}\,\slashed{\textup{div}}\overset{\text{\scalebox{.6}{$(1)$}}}{\widehat{\slashed{g}}}&= 2 \sigmao|_{\ell\geq 2} \, , \\
(\overset{\text{\scalebox{.6}{$(1)$}}}{\textup{tr}\chi})|_{\ell\geq 2}&=0 \, , & (\slashed{\textup{div}}\etao+\rhoo)|_{\ell\geq 2}&=0 
\end{align*}
and the \emph{auxiliary horizon normalisation identities}
\begin{align*}
(\textup{tr}\overset{\text{\scalebox{.6}{$(1)$}}}{\slashed{g}})|_{\ell= 1}&=0 \, ,  &
(\overset{\text{\scalebox{.6}{$(1)$}}}{\textup{tr}\chi})|_{\ell= 0,1}&=0 \, , \\
(\overset{\text{\scalebox{.6}{$(1)$}}}{\textup{tr}\chib})|_{\ell= 0}&=4M\rhoo|_{\ell= 0}  \, , &
(\overset{\text{\scalebox{.6}{$(1)$}}}{\textup{tr}\chib})|_{\ell= 1}&=0  \, , \\
(\slashed{\textup{div}}\etao+\rhoo)|_{\ell=1}&=0 \, .
\end{align*}
\end{itemize}
\end{definition}

\medskip

\begin{remark}
In the main result of Theorem \ref{th_main_theorem}, we consider initial-data normalised solutions $\mathfrak{S}_N$.~As we shall prove in Proposition \ref{prop_initial_data_normalisation}, this can be done without loss of generality.
\end{remark}

\medskip

\begin{remark}
The auxiliary normalisation identities in Definition \ref{def_initial_gauge_norm} are needed to ensure that the projection to $\ell=0,1$-spherical modes of any initial-data normalised solution $\mathfrak{S}_N$ coincides with a suitable \emph{reference linearised Kerr solution}.~See Definition \ref{def_ref_kerr} and Proposition \ref{prop_proj_kernel_kerr}.
\end{remark}

\medskip

\begin{remark}
Our Definition \ref{def_initial_gauge_norm} may be compared to Definition 9.1 in \cite{DHR}.~Our definition does not include normalisation identities on the initial outgoing null cone $C_{u_0}$ or round-sphere conditions at infinity. 
\end{remark}

\subsection{Pure gauge solutions and initial-data gauge normalisation} \label{sec_pure_gauge_solutions}

In this section, we define the classes of \emph{pure gauge solutions} to the linearised system of equations (Sections \ref{sec_coord_pure_gauge_solns} and \ref{sec_frame_pure_gauge_solns}) and show how they can be employed to initial-data normalise any given solution to the system (Section \ref{sec_achieve_renorm}).~See Section \ref{sec_overview_pure_gauge_solutions} of the overview for a discussion about the meaning and derivation of pure gauge solutions.

\subsubsection{Coordinate pure gauge solutions} \label{sec_coord_pure_gauge_solns}

Consider the smooth scalar functions $(h_1,\Omega^2_Mh_2)$ and $\mathbb{S}^2_{u,v}$ vector field $h$ on $\mathcal{M}\setminus\mathcal{H}^+$ which extend regularly to $\mathcal{H}^+$ and satisfy the system of ODEs\footnote{The notation $h^{\theta^A}$ denotes the $\theta^A$-(coordinate) component of $h$.~The reader should note the difference with the frame index notation $h^A=h(\slashed{g}^{AB}e_B^{\text{as}})$.}
\begin{align}
\nablasl_4(\Omega^2_Mh_2)-\frac{2M}{r^2} (\Omega^2_Mh_2) &=0 \, ,  \label{ODE_pure_gauge_1}\\
\nablasl_4^2 h_1 +\frac{2M}{r^2}(\nablasl_4 h_1)+\frac{4M}{r^3} (\Omega^2_Mh_2-\Omega^2_Mh_1)&=0 \, , \label{ODE_pure_gauge_2}\\
\nablasl_4^2 h^{\theta^A}+2\left(\frac{r-3M}{r^2}\right) \nablasl_4 h^{\theta^A} &=0 \, . \label{ODE_pure_gauge_3}
\end{align}

\medskip

\begin{definition}[Coordinate pure gauge solutions] \label{def_coordinate_pure_gauge_soln}
We define \emph{coordinate pure gauge solutions} as the family of solutions $\mathsf{CG}(h_1,h_2,h)$ to the linearised system of equations with linearised frame coefficients and induced metric
\begin{align*}
\overset{\text{\scalebox{.6}{$(1)$}}}{\mathfrak{\slashed{\mathfrak{f}}}}_{4}&= -\nablasl(\Omega^2_M h_2)+ \frac{1}{2}\,(\nablasl_4 h)_{\flat}-\frac{\Omega^2_M}{2\,r}\, h_{\flat} \, ,  & \overset{\text{\scalebox{.6}{$(1)$}}}{\mathfrak{\slashed{\mathfrak{f}}}}_{3}&= -\nablasl h_1 \, , \\
\overset{\text{\scalebox{.6}{$(1)$}}}{\mathfrak{\underline{\mathfrak{f}}}}_4 &=  -\nablasl_3(\Omega^2_M h_2)-\nablasl_4 h_1 -\frac{2M}{r^2}\,h_1 \, ,   &
\overset{\text{\scalebox{.6}{$(1)$}}}{\mathfrak{\underline{\mathfrak{f}}}}_3 &=-\nablasl_3 h_1 \, ,  \\
\overset{\text{\scalebox{.6}{$(1)$}}}{\mathfrak{\underline{\mathfrak{f}}}} &=-(\nablasl_3 h)_{\flat}-\frac{1}{r}\, h_{\flat}  \, ,   \\
(\textup{tr}\overset{\text{\scalebox{.6}{$(1)$}}}{\slashed{g}}) &= \frac{4\Omega_M^2}{r}(h_1-h_2)+2\,(\slashed{\textup{div}}\,h) \, ,  &
\overset{\text{\scalebox{.6}{$(1)$}}}{\widehat{\slashed{g}}}&=-4\,(\slashed{\mathcal{D}}_2^{\star}\,h)_{\flat} \, .
\end{align*}
\end{definition}

\medskip

The linearised connection coefficients and curvature components of coordinate pure gauge solutions can be determined through the linearised system of equations.~We note that
\begin{align*}
(\trchio)&=-2\,\frac{\Omega_M^2(r-4M)}{r^3}(h_1-h_2)+\frac{2\,\Omega_M^2}{r}\nablasl_4 h_1+\slashed{\textup{div}}(\nablasl_4\,h)_{\flat}-\frac{\Omega_M^2}{r}\,(\slashed{\textup{div}}\,h) \, , 
\end{align*}
\begin{align*}
(\trchibo)&= \frac{2\,\Omega_M^2}{r^2}\,(h_1-h_2)  +\frac{2}{r}\,(\nablasl_4 h_1) \, , & \omegabo&=-\nablasl_3\nablasl_4 h_1 \, , \\
(\overset{\text{\scalebox{.6}{$(1)$}}}{\slashed{\varepsilon}\cdot\chi}) &=\nablasl_4 (\slashed{\textup{curl}}\,h) \, , & (\overset{\text{\scalebox{.6}{$(1)$}}}{\slashed{\varepsilon}\cdot\chib})&=0 \, , \\
\chiho&=-2\,\slashed{\mathcal{D}}_2^{\star}(\nablasl_4h)_{\flat}+\frac{2\,\Omega^2_M}{r}(\slashed{\mathcal{D}}_2^{\star}\,h)_{\flat}  \, ,  &
\chibho &=  -\nablasl_3(\slashed{\mathcal{D}}_2^{\star}\,h)_{\flat}  
\end{align*}
and
\begin{align*}
\alphao&=0 \, , &  \alphabo &=0 \, .
\end{align*}
On the event horizon $\mathcal{H}^+$, we have
\begin{align*}
(\slashed{\textup{div}} \etao+\rhoo)(\infty,v) =& \,  \frac{1}{2}\nablasl_3(\slashed{\textup{div}}(\nablasl_4\,h)_{\flat})(\infty,v) -\frac{1}{4M}\,\slashed{\textup{div}}(\nablasl_4\,h)_{\flat}(\infty,v)  +\frac{1}{8M^2}\,(\slashed{\textup{div}}\,h)(\infty,v)    \\
&-\frac{3}{2M^3}\,(\Omega_M^2h_2)(\infty,v) \, , \\
\rhoo(\infty,v)=& \, -\frac{1}{4\,M}\,\slashed{\textup{div}}(\nablasl_4 h)_{\flat}(\infty,v)-\frac{3}{8\,M^3}\,(\Omega_M^2h_2)(\infty,v)  \, , \\
     \sigmao(\infty,v) =& \, -\frac{1}{4\,M}\,\slashed{\textup{curl}}( \nablasl_4 h)_{\flat} (\infty,v) \, .
\end{align*}

\medskip

\begin{remark}
The fact that coordinate pure gauge solutions, as defined in Definition \ref{def_coordinate_pure_gauge_soln}, are indeed solutions to the linearised system of equations is a check left to the reader. 
\end{remark}

\subsubsection{Frame pure gauge solutions} \label{sec_frame_pure_gauge_solns}

Consider the $\mathbb{S}^2_{u,v}$ vector field $k$ on $\mathcal{M}\setminus\mathcal{H}^+$ which extends regularly to $\mathcal{H}^+$ and satisfies the system of ODEs
\begin{equation} \label{ODE_k}
\nablasl_4 k^A+\frac{2M}{r^2}\,k^A=0 \, .
\end{equation}

\medskip

\begin{definition}[Frame pure gauge solutions] \label{def_frame_pure_gauge_soln}
We define \emph{frame pure gauge solutions} as the family of solutions $\mathsf{FG}(k)$ to the linearised system of equations such that
\begin{align*}
\overset{\text{\scalebox{.6}{$(1)$}}}{\mathfrak{\underline{\mathfrak{f}}}} &=k_{\flat} \, ,  &
\overset{\text{\scalebox{.6}{$(1)$}}}{\mathfrak{\slashed{\mathfrak{f}}}}_{3} &= \frac{1}{2}\, k_{\flat} \, , \\
\etao &= \frac{\Omega^2_M}{2\,r}\,k_{\flat} \, , &
\zetao &= \frac{r-4M}{2\,r^2}\,k_{\flat} \, , \\
 \ybo &= \frac{1}{2}\, (\nablasl_3 k)_{\flat}-\frac{1}{2\,r}\,k_{\flat}  \, , & \chibho &= -(\slashed{\mathcal{D}}_2^{\star}\,k)_{\flat} \, ,\\
(\overset{\text{\scalebox{.6}{$(1)$}}}{\textup{tr}\chib}) &= \slashed{\textup{div}}\,k \, , & (\overset{\text{\scalebox{.6}{$(1)$}}}{\slashed{\varepsilon}\cdot\chib})&=\slashed{\textup{curl}}\, k \, , \\
\overset{\text{\scalebox{.6}{$(1)$}}}{\widetilde{\slashed{K}}} &= -\frac{\Omega_M^2}{2\,r}\,\slashed{\textup{div}}\,k  \, ,  \\
\betabo &= -\slashed{\textup{div}}\,(\slashed{\mathcal{D}}_2^{\star}\,k)_{\flat}-\frac{1}{2}\nablasl \slashed{\textup{div}}\,k+\frac{1}{2}{}^{\star}\nablasl\slashed{\textup{curl}}\, k-\frac{r-3M}{r^3}\,k_{\flat} 
\end{align*}
and all the remaining linearised frame coefficients, connection coefficients and curvature components identically vanishing.
\end{definition}

\medskip

\begin{remark}
For any frame pure gauge solution, we have
\begin{align*}
\alphao&=0  \, , & \alphabo&=0 \, , & (\slashed{\textup{div}}\etao+\rhoo)(\infty,v) &=0 \, .
\end{align*}
\end{remark}

\medskip

\begin{remark}
The fact that frame pure gauge solutions, as defined in Definition \ref{def_frame_pure_gauge_soln}, are indeed solutions to the linearised system of equations is a check left to the reader. 
\end{remark}

\subsubsection{Achieving the initial-data gauge normalisation} \label{sec_achieve_renorm}

We prove the following key proposition.

\medskip

\begin{prop} \label{prop_initial_data_normalisation}
For any solution $\mathfrak{S}$ to the linearised system of equations arising from smooth, asymptotically flat (to sufficiently high order) seed initial data, there exist a coordinate pure gauge solution $\mathsf{CG}(h_1,h_2,h)$ and a frame pure gauge solution $\mathsf{FG}(k)$ such that the solution
\begin{equation}
\mathfrak{S}_N=\mathfrak{S}+\mathsf{CG}(h_1,h_2,h)+\mathsf{FG}(k) \label{sum_initial_data_normalisation}
\end{equation}
is an initial-data normalised solution to the linearised system of equations.~The solution $\mathfrak{S}_N$ arises from smooth, asymptotically flat seed initial data. 
\end{prop}

\medskip

\begin{proof}
We consider a solution $\mathfrak{S}$ to the linearised system of equations.~To achieve the initial-data normalisation, we need to suitably construct the quantities $h_1$, $h_2$, $h$ and $k$ on $\mathcal{M}$.~We note that the ODEs \eqref{ODE_pure_gauge_1}, \eqref{ODE_pure_gauge_2}, \eqref{ODE_pure_gauge_3} and \eqref{ODE_k} only allow to choose
\begin{equation}
h_1 \, , \, \Omega^2_M h_2 \, , \, h^{\theta^A} \, , \, \nablasl_4 h_1 \, , \, \nablasl_4h^{\theta^A} \, , \, k^A \label{pure_gauge_quantities}
\end{equation} 
freely on $\underline{C}_{v_0}$.~Our proof constructs the quantities \eqref{pure_gauge_quantities} on $\underline{C}_{v_0}$.~The fact that the quantities constructed achieve \eqref{sum_initial_data_normalisation} is a check left to the reader.

\medskip

We divide the construction into three parts, each corresponding to the construction of the $\ell=0$, $\ell=1$ and $\ell\geq 2$-spherical modes of the quantities \eqref{pure_gauge_quantities} respectively.~The quantities $h$, $\nablasl_4h$ and $k$ will be treated as $\mathbb{S}^2_{u,v}$ one-tensors (denoted $h_{\flat}$, $\nablasl_4h_{\flat}$ and $k_{\flat}$), and thus are supported on $\ell\geq 1$-spherical modes.

\medskip

\underline{$\ell=0$-spherical modes}:~We choose $(h_1)|_{\ell=0}$, $(\Omega^2_M h_2)|_{\ell=0}$ and $(\nablasl_4 h_1)|_{\ell=0}$ on $\underline{C}_{v_0}$ as solutions to the system of ODEs in the $e_3^{\text{as}}$-direction
\begin{align*}
\nablasl_3 (h_1)|_{\ell=0} =& \, (\overset{\text{\scalebox{.6}{$(1)$}}}{\mathfrak{\underline{\mathfrak{f}}}}_3)_{\mathfrak{S}}|_{\ell=0} \, ,  \\
  \nablasl_3(\Omega^2_M h_2)|_{\ell=0} =& \, -(\nablasl_4 h_1)|_{\ell=0} -\frac{2M}{r^2}\,(h_1)|_{\ell=0}+(\overset{\text{\scalebox{.6}{$(1)$}}}{\mathfrak{\underline{\mathfrak{f}}}}_4)_{\mathfrak{S}}|_{\ell=0} \\ &+4M^2(\rhoo_{\mathfrak{S}}|_{\ell=0}-\frac{3}{8\,M^3}\,(\Omega_M^2h_2)|_{\ell=0})(\infty,v_0) \, , \\
 \nablasl_3(\nablasl_4 h_1)|_{\ell=0} =& \, (\omegabo)_{\mathfrak{S}}|_{\ell=0}
\end{align*}
with initial conditions on $\mathbb{S}^2_{\infty,v_0}$ such that
\begin{align*}
(\Omega_M^2h_2)|_{\ell= 0}(\infty,v_0) &= 2\,M^2 (\trchio)_{\mathfrak{S}}|_{\ell= 0}(\infty,v_0) \, , \\
 (\nablasl_4 h_1)|_{\ell= 0}(\infty,v_0) &=4M^2\rhoo_{\mathfrak{S}}|_{\ell=0}(\infty,v_0)-2\,M \,(\trchio)_{\mathfrak{S}}|_{\ell= 0}(\infty,v_0)-M\,(\trchibo)_{\mathfrak{S}}|_{\ell= 0}(\infty,v_0) \, .
\end{align*}
One can choose $(h_1)|_{\ell=0}(\infty,v_0)$ arbitrarily.

\medskip

\underline{$\ell=1$-spherical modes}:~We choose $(h_1)|_{\ell=1}$, $(\Omega^2_M h_2)|_{\ell=1}$ and $(\nablasl_4 h_1)|_{\ell=1}$ on $\underline{C}_{v_0}$ as solutions to the system of ODEs in the $e_3^{\text{as}}$-direction
\begin{align*}
\nablasl_3 (h_1)|_{\ell= 1} &= (\overset{\text{\scalebox{.6}{$(1)$}}}{\mathfrak{\underline{\mathfrak{f}}}}_3)_{\mathfrak{S}}|_{\ell= 1}  \, , \\
   \nablasl_3(\Omega^2_M h_2)|_{\ell= 1}&= -(\nablasl_4 h_1)|_{\ell= 1} -\frac{2M}{r^2}\,(h_1)|_{\ell= 1} +(\overset{\text{\scalebox{.6}{$(1)$}}}{\mathfrak{\underline{\mathfrak{f}}}}_4)_{\mathfrak{S}}|_{\ell= 1} \, ,   \\
   \nablasl_3(\nablasl_4 h_1)|_{\ell= 1} &= (\omegabo)_{\mathfrak{S}}|_{\ell= 1}   
\end{align*}
with initial conditions on $\mathbb{S}^2_{\infty,v_0}$ solving the elliptic system
\begin{align*}
2\,\slashed{\Delta}(\Omega^2_M h_2)|_{\ell=1}(\infty,v_0)-\frac{1}{2\,M^2}(\Omega_M^2h_2)|_{\ell=1}(\infty,v_0)  =& \, 2\,(\slashed{\textup{div}}\overset{\text{\scalebox{.6}{$(1)$}}}{\mathfrak{\slashed{\mathfrak{f}}}}_{4})_{\mathfrak{S}}|_{\ell= 1}(\infty,v_0) \\ &-(\trchio)_{\mathfrak{S}}|_{\ell=1}(\infty,v_0) \, , \\
2\,\slashed{\Delta}\, h_1|_{\ell= 1}(\infty,v_0)  - \frac{1}{2\,M^2}\,(\Omega_M^2h_2)|_{\ell=1}(\infty,v_0)  +\frac{1}{M}\,(\nablasl_4 h_1)|_{\ell=1}(\infty,v_0)  =& \, 2\,(\slashed{\textup{div}}\overset{\text{\scalebox{.6}{$(1)$}}}{\mathfrak{\slashed{\mathfrak{f}}}}_{3})_{\mathfrak{S}}|_{\ell= 1}(\infty,v_0)\\ & -(\trchibo)_{\mathfrak{S}}|_{\ell=1}(\infty,v_0) \, , 
\end{align*}
\begin{align*}
 \slashed{\Delta}(\nablasl_4 h_1)|_{\ell= 1}(\infty,v_0) +\frac{1}{2M}\slashed{\Delta}\,h_1|_{\ell= 1}&(\infty,v_0)    +\frac{3}{2M^3}\,(\Omega_M^2h_2)|_{\ell=1}(\infty,v_0) \\ 
 =&\,  -\nablasl_3(\slashed{\textup{div}}\overset{\text{\scalebox{.6}{$(1)$}}}{\mathfrak{\slashed{\mathfrak{f}}}}_{4})_{\mathfrak{S}}|_{\ell= 1} (\infty,v_0)+\frac{1}{2M}(\slashed{\textup{div}}\overset{\text{\scalebox{.6}{$(1)$}}}{\mathfrak{\slashed{\mathfrak{f}}}}_{4})_{\mathfrak{S}}|_{\ell= 1}(\infty,v_0)    \\ & +\slashed{\Delta}(\overset{\text{\scalebox{.6}{$(1)$}}}{\mathfrak{\underline{\mathfrak{f}}}}_4)_{\mathfrak{S}}|_{\ell= 1}(\infty,v_0) +(\slashed{\textup{div}} \etao+\rhoo)_{\mathfrak{S}}|_{\ell=1}(\infty,v_0) \, .
\end{align*}
We then choose $k_{\flat}|_{\ell= 1}$ on $\underline{C}_{v_0}$ such that
\begin{align*}
( k_{\flat})|_{\ell= 1} =& \,  2\,\nablasl (h_1)|_{\ell= 1} \\ & -2\,(\overset{\text{\scalebox{.6}{$(1)$}}}{\mathfrak{\slashed{\mathfrak{f}}}}_{3})_{\mathfrak{S}}|_{\ell= 1}  +\frac{8}{3}\,M^2\left(r+M\right){}^{\star}\nablasl(\sigmao_{\mathfrak{S}}|_{\ell=1}+\frac{1}{2\,M}\,(\slashed{\textup{curl}}\,\overset{\text{\scalebox{.6}{$(1)$}}}{\mathfrak{\slashed{\mathfrak{f}}}}_{4})_{\mathfrak{S}}|_{\ell=1})(\infty,v_0) \, .
\end{align*}
We choose $(\slashed{\textup{div}}\, h)|_{\ell=1}$ and $(\slashed{\textup{curl}}\, h)|_{\ell=1}$ on $\underline{C}_{v_0}$ such that 
\begin{align*}
 \nablasl_3(\slashed{\textup{curl}}\, h)|_{\ell=1}  =& \, (\slashed{\textup{curl}}\, k)|_{\ell=1} \\ & + (\slashed{\textup{curl}}\,\overset{\text{\scalebox{.6}{$(1)$}}}{\mathfrak{\underline{\mathfrak{f}}}})_{\mathfrak{S}}|_{\ell=1}  +\frac{8}{3}\,M (\sigmao_{\mathfrak{S}}|_{\ell=1}+\frac{1}{2\,M}\,(\slashed{\textup{curl}}\,\overset{\text{\scalebox{.6}{$(1)$}}}{\mathfrak{\slashed{\mathfrak{f}}}}_{4})_{\mathfrak{S}}|_{\ell=1})(\infty,v_0)  
 \end{align*}
and
\begin{equation*}
(\slashed{\textup{div}}\,h)|_{\ell=1}(\infty,v_0) =   \frac{1}{M}(\Omega_M^2h_2)|_{\ell=1}(\infty,v_0)  -\frac{1}{2}\,(\textup{tr}\overset{\text{\scalebox{.6}{$(1)$}}}{\slashed{g}})_{\mathfrak{S}}|_{\ell=1}(\infty,v_0) \, . 
\end{equation*}
One can arbitrarily choose $(\slashed{\textup{curl}}\,h)|_{\ell=1}(\infty,v_0)$ and $(\slashed{\textup{div}}\,h)|_{\ell=1}$ away from $\mathbb{S}^2_{\infty,v_0}$.~We conclude by choosing $(\nablasl_4 h)_{\flat}|_{\ell=1}$ such that
 \begin{align*}
  (\nablasl_4 h)_{\flat}|_{\ell=1} =& \,  \frac{\Omega^2_M}{r}\, (h_{\flat})|_{\ell=1}+2\,\nablasl(\Omega^2_M h_2)|_{\ell=1} \\ &-2\,(\overset{\text{\scalebox{.6}{$(1)$}}}{\mathfrak{\slashed{\mathfrak{f}}}}_{4})_{\mathfrak{S}}|_{\ell= 1}   +\frac{8}{3}\,\Omega^2_M\,M^3\,{}^{\star}\nablasl (\sigmao_{\mathfrak{S}}|_{\ell=1}+\frac{1}{2\,M}\,(\slashed{\textup{curl}}\,\overset{\text{\scalebox{.6}{$(1)$}}}{\mathfrak{\slashed{\mathfrak{f}}}}_{4})_{\mathfrak{S}}|_{\ell=1})(\infty,v_0) \, . 
\end{align*}

\medskip

\underline{$\ell\geq 2$-spherical modes}:~We set $h_1|_{\ell\geq 2}\equiv 0$ on $\underline{C}_{v_0}$.~We choose $(\Omega^2_M h_2)|_{\ell\geq 2}$ on $\underline{C}_{v_0}$ as the solution to the ODE in the $e_3^{\text{as}}$-direction 
\begin{equation*}
\nablasl_3(\Omega^2_M h_2)|_{\ell\geq 2}+(\nablasl_4h_1)|_{\ell\geq 2}= (\overset{\text{\scalebox{.6}{$(1)$}}}{\mathfrak{\underline{\mathfrak{f}}}}_4)_{\mathfrak{S}}|_{\ell\geq 2}
\end{equation*}
with initial condition on $\mathbb{S}^2_{\infty,v_0}$ such that
\begin{align*}
2\,\slashed{\Delta}(\Omega^2_M h_2)|_{\ell\geq 2}(\infty,v_0)-\frac{1}{2\,M^2}(\Omega_M^2h_2)|_{\ell\geq 2}(\infty,v_0)  =& \, 2\,(\slashed{\text{div}}\,\overset{\text{\scalebox{.6}{$(1)$}}}{\mathfrak{\slashed{\mathfrak{f}}}}_{4})_{\mathfrak{S}}|_{\ell\geq 2}(\infty,v_0) \\ & - (\trchio)_{\mathfrak{S}}|_{\ell\geq 2}(\infty,v_0) 
\end{align*}
and $(\nablasl_4h_1)|_{\ell\geq 2}=(\nablasl_4h_1)|_{\ell\geq 2}(\theta^A)$ on $\underline{C}_{v_0}$ such that
\begin{align*}
 \slashed{\Delta}(\nablasl_4 h_1)|_{\ell\geq 2}   +\frac{3}{2M^3}\,(\Omega_M^2h_2)|_{\ell\geq 2}(\infty,v_0)  
 =&\,  -\nablasl_3(\slashed{\textup{div}}\overset{\text{\scalebox{.6}{$(1)$}}}{\mathfrak{\slashed{\mathfrak{f}}}}_{4})_{\mathfrak{S}}|_{\ell\geq 2} (\infty,v_0)+\frac{1}{2M}(\slashed{\textup{div}}\overset{\text{\scalebox{.6}{$(1)$}}}{\mathfrak{\slashed{\mathfrak{f}}}}_{4})_{\mathfrak{S}}|_{\ell\geq 2}(\infty,v_0)    \\ & +\slashed{\Delta}(\overset{\text{\scalebox{.6}{$(1)$}}}{\mathfrak{\underline{\mathfrak{f}}}}_4)_{\mathfrak{S}}|_{\ell\geq 2}(\infty,v_0) +(\slashed{\textup{div}} \etao+\rhoo)_{\mathfrak{S}}|_{\ell\geq 2}(\infty,v_0)   \, .
\end{align*}
We choose $h_{\flat}|_{\ell\geq 2}=h_{\flat}|_{\ell\geq 2}(\theta^A)$ on $\underline{C}_{v_0}$ such that\footnote{By considering the unique representation $h=\slashed{\mathcal{D}}_1^{\star}(q_1,q_2)$ for scalar functions $q_i=q_i(\theta^A)$, one can check that the equations \eqref{aux_gauge_achieve_l_geq_2_1} and \eqref{aux_gauge_achieve_l_geq_2_2} form an elliptic system on spheres for $q_1$ and $q_2$ which can indeed be solved.}
\begin{equation}
(\slashed{\textup{div}}\,h)|_{\ell\geq 2} = \frac{1}{M}\,(\Omega_M^2h_2)|_{\ell\geq 2}(\infty,v_0) -\frac{1}{2}\,(\textup{tr}\overset{\text{\scalebox{.6}{$(1)$}}}{\slashed{g}})_{\mathfrak{S}}|_{\ell\geq 2}(\infty,v_0)  \label{aux_gauge_achieve_l_geq_2_1}
\end{equation}
and
\begin{align}
\slashed{\text{curl}}\,\slashed{\text{div}}\,\slashed{\mathcal{D}}_2^{\star}\,h_{\flat}=& \, \frac{1}{4}\,(\slashed{\text{curl}}\,\slashed{\text{div}}\overset{\text{\scalebox{.6}{$(1)$}}}{\widehat{\slashed{g}}})_{\mathfrak{S}}(\infty,v_0)  -\frac{1}{2}\,\sigmao_{\mathfrak{S}}|_{\ell\geq 2}(\infty,v_0)  \label{aux_gauge_achieve_l_geq_2_2} \\
&-\frac{1}{4M}\,(\slashed{\text{curl}}\,\overset{\text{\scalebox{.6}{$(1)$}}}{\mathfrak{\slashed{\mathfrak{f}}}}_{4})_{\mathfrak{S}}|_{\ell\geq 2} (\infty,v) \, .  \nonumber
\end{align}
We then chose $(\slashed{\text{div}}\,(\nablasl_4 h))|_{\ell\geq 2}$ and $(\slashed{\text{curl}}\,(\nablasl_4 h))|_{\ell\geq 2}$ on $\underline{C}_{v_0}$ such that
\begin{equation*}
(\slashed{\text{div}}\,(\nablasl_4 h))|_{\ell\geq 2}  = 2\,\slashed{\Delta}(\Omega^2_M h_2)|_{\ell\geq 2}+\frac{\Omega^2_M}{r}\,(\slashed{\text{div}} \,h )|_{\ell\geq 2}-2\,(\slashed{\text{div}}\,\overset{\text{\scalebox{.6}{$(1)$}}}{\mathfrak{\slashed{\mathfrak{f}}}}_{4})_{\mathfrak{S}}|_{\ell\geq 2}
\end{equation*}
and
\begin{equation*}
(\slashed{\text{curl}}\,(\nablasl_4 h))|_{\ell\geq 2} = \frac{\Omega^2_M}{r}\,(\slashed{\text{curl}} \,h )|_{\ell\geq 2}-2\,(\slashed{\text{curl}}\,\overset{\text{\scalebox{.6}{$(1)$}}}{\mathfrak{\slashed{\mathfrak{f}}}}_{4})_{\mathfrak{S}}|_{\ell\geq 2} \, .
\end{equation*}
One can arbitrarily choose $k_{\flat}|_{\ell\geq 2}$ on $\underline{C}_{v_0}$.
\end{proof}

\medskip

\begin{remark} \label{rmk_non_uniqueness_pure_gauge_solns}
As noted in the proof of Proposition \ref{prop_initial_data_normalisation}, given a solution $\mathfrak{S}$ to the linearised system of equations, one has some freedom in constructing the geometric quantities $(h_1,h_2,h,k)$ which achieve \eqref{sum_initial_data_normalisation}.~In other words, given a solution $\mathfrak{S}$, the linear combination of coordinate and frame pure gauge solutions which achieves \eqref{sum_initial_data_normalisation} is not unique.~This is a manifestation of the fact that our Definition \ref{def_initial_gauge_norm} of initial-data normalised solutions does not fix the residual gauge freedom completely.~In fact, one can enforce the uniqueness of the pure gauge solutions in Proposition \ref{prop_initial_data_normalisation} by imposing extra normalisation identities in Definition \ref{def_initial_gauge_norm}, which would not however play any other relevant role in the problem.~This aspect may be compared with Theorem 9.1 in \cite{DHR}, where the pure gauge solution employed to achieve the initial-data normalisation is unique. 
\end{remark}

\medskip

\begin{remark}
One can check that if the smooth seed initial data relative to the given solution $\mathfrak{S}$ are asymptotically flat with weight $s$ and to order $n$, then the smooth seed initial data relative to the initial-data normalised solution $\mathfrak{S}_N$ are asymptotically flat with weight $s$ and to order $n-2$.
\end{remark}

\subsection{Reference linearised solutions} \label{sec_reference_linearised_solutions}

In this section, we define the \emph{reference linearised solutions} to the linearised system of equations.~See Section \ref{sec_overview_reference_linearised_solutions} of the overview for a discussion about the meaning and derivation of these solutions. 

\medskip

\begin{definition}[Reference linearised Schwarzschild solutions] \label{def_ref_schw}
Let $\mathfrak{m}\in\mathbb{R}$.~We define the \emph{reference linearised Schwarzschild solutions} as the one-parameter family of solutions $\mathsf{S}(\mathfrak{m})$ to the linearised system of equations such that
\begin{align*}
\overset{\text{\scalebox{.6}{$(1)$}}}{\mathfrak{\underline{\mathfrak{f}}}}_4&=\mathfrak{m} \, , & (\textup{tr}\overset{\text{\scalebox{.6}{$(1)$}}}{\slashed{g}})&=-2\,\mathfrak{m} \, , &
(\overset{\text{\scalebox{.6}{$(1)$}}}{\textup{tr}\chib})&=(\textup{tr}\chib)_M\,\mathfrak{m} \, ,  &
\rhoo&=\rho_M\,\mathfrak{m} \, , & \overset{\text{\scalebox{.6}{$(1)$}}}{\widetilde{\slashed{K}}}&= \slashed{K}_M\,\mathfrak{m} 
\end{align*}
on $\mathcal{R}_{u_0,v_0}$ and all the remaining linearised frame coefficients, connection coefficients and curvature components identically vanishing.
\end{definition}

\medskip

\begin{definition}[$\ell=1$-reference linearised Kerr solutions] \label{def_ref_kerr}
Let $\mathfrak{a}\in\mathbb{R}$ and $m=-1,0,1$.~We define the \emph{$\ell=1$-reference linearised Kerr solutions} as the two-parameter family of solutions $\mathsf{K}(\mathfrak{a},m)$ to the linearised system of equations such that
\begin{align*}
\overset{\text{\scalebox{.6}{$(1)$}}}{\mathfrak{\slashed{\mathfrak{f}}}}_{4,m} &=\frac{\mathfrak{a} \Omega^2_M}{2 r}\,r\,{}^{\star}\nablasl Y_m^{\ell=1} \, , &
\overset{\text{\scalebox{.6}{$(1)$}}}{\mathfrak{\slashed{\mathfrak{f}}}}_{3,m} &=\frac{\mathfrak{a}}{2 M r}\left(r+M\right)r\,{}^{\star}\nablasl Y_m^{\ell=1} \, , \\ \overset{\text{\scalebox{.6}{$(1)$}}}{\mathfrak{\underline{\mathfrak{f}}}}_{m} &=-\frac{\mathfrak{a} r}{2 M^2}\,r\,{}^{\star}\nablasl Y_m^{\ell=1} \, , &
(\overset{\text{\scalebox{.6}{$(1)$}}}{\slashed{\varepsilon}\cdot\chi})_m&=\frac{2 \mathfrak{a} \Omega^2_M}{r^2} \, Y_m^{\ell=1} \, , \\
(\overset{\text{\scalebox{.6}{$(1)$}}}{\slashed{\varepsilon}\cdot\chib})_m&=\frac{2 \mathfrak{a}}{M r^2}\left(r+M\right) Y_m^{\ell=1} \, ,  &
\etao_{m}&=\frac{\mathfrak{a}}{2  M r}\,r\,{}^{\star}\nablasl Y_m^{\ell=1} \, ,  \\ \ybo_{m} &= -\frac{\mathfrak{a}}{2  M   r}\,r\,{}^{\star}\nablasl Y_m^{\ell=1} \, , & 
\zetao_{m} &=\frac{\mathfrak{a} \Omega^2_M}{2  M  r}\,r\,{}^{\star}\nablasl Y_m^{\ell=1} \, , \\
\betabo_{m} &=\frac{3 \mathfrak{a}}{r^3}\,r\,{}^{\star}\nablasl Y_m^{\ell=1} \, , & \sigmao_m &=  \frac{6 \mathfrak{a}  M}{r^4} \, Y_m^{\ell=1}
\end{align*}
on $\mathcal{R}_{u_0,v_0}$ and all the remaining linearised frame coefficients, connection coefficients and curvature components identically vanishing.
\end{definition}

\medskip

\begin{remark}
We note that, in particular, both $\alphao$ and $\alphabo$ identically vanish for any reference linearised Schwarzschild and $\ell=1$-reference linearised Kerr solution.
\end{remark}

\medskip

\begin{remark}
The fact that the reference linearised Schwarzschild and $\ell=1$-reference linearised Kerr solutions, as defined in Definitions \ref{def_ref_schw} and \ref{def_ref_kerr}, are indeed solutions to the system of equations is a check left to the reader. 
\end{remark}

\medskip

\begin{remark}
The reference linearised Schwarzschild and $\ell=1$-reference linearised Kerr solutions are initial-data normalised.~See Definition \ref{def_initial_gauge_norm}.
\end{remark}

\medskip

We conclude with the following definition and remark.

\medskip

\begin{definition}[Reference linearised Kerr solutions]
Let $\mathfrak{m},\mathfrak{s}_{-1},\mathfrak{s}_{0},\mathfrak{s}_{-1}\in\mathbb{R}$.~We define the \emph{reference linearised Kerr solution} $\mathcal{K}_{\mathfrak{m},\mathfrak{s}_{m}}$ with parameters $(\mathfrak{m},\mathfrak{s}_{-1},\mathfrak{s}_{0},\mathfrak{s}_{-1})$ as the sum of the reference linearised Schwarzschild solution $\mathsf{S}(\mathfrak{m})$ with a linear combination of $\ell=1$-reference linearised Kerr solutions $\mathsf{K}(\mathfrak{a},m)$ such that
\begin{equation*}
\sigmao[\mathcal{K}_{\mathfrak{m},\mathfrak{s}_{m}}]=\sum_m \mathfrak{s}_{m}\cdot  \sigmao_{m}[\mathsf{K}(\mathfrak{a},m)] \, .
\end{equation*}
\end{definition}

\medskip

\begin{remark}
A reference linearised Kerr solution $\mathcal{K}_{\mathfrak{m},\mathfrak{s}_{m}}$ is only supported on $\ell=0,1$-spherical modes.
\end{remark}

\subsection{Properties of initial-data normalised solutions}

In this section, we state the main global properties of initial-data normalised solutions.~Propositions \ref{glob_prop_l_01} and \ref{glob_prop_l_2} can be easily checked using the system of linearised equations.

\medskip

\begin{prop}[Global properties $\ell=0,1$-spherical modes] \label{glob_prop_l_01}
Consider an initial-data normalised solution $\mathfrak{S}_N$.~Then, $\mathfrak{S}_N$ satisfies the following identities:
\begin{itemize}
\item On $\mathcal{R}_{u_0,v_0}$ (including $\mathcal{H}^+_{v\geq v_0}$):~The identities
\begin{align*}
\overset{\text{\scalebox{.6}{$(1)$}}}{\mathfrak{\underline{\mathfrak{f}}}}_4|_{\ell= 1}&=0 \, , & (\overset{\text{\scalebox{.6}{$(1)$}}}{\slashed{\varepsilon}\cdot\chi})|_{\ell=0}&=0 \, , & (\overset{\text{\scalebox{.6}{$(1)$}}}{\slashed{\varepsilon}\cdot\chib})|_{\ell=0}&=0 \, .
\end{align*}
\item On $\mathcal{H}_{v\geq v_0}^+$:~The identities
\begin{align*}
(\textup{tr}\overset{\text{\scalebox{.6}{$(1)$}}}{\slashed{g}})|_{\ell= 1}&=0 \, , &
(\overset{\text{\scalebox{.6}{$(1)$}}}{\textup{tr}\chi})|_{\ell=0,1}&=0 \, , \\
(\overset{\text{\scalebox{.6}{$(1)$}}}{\slashed{\varepsilon}\cdot\chi})|_{\ell=1}&=0 \, , &
(\slashed{\textup{div}}\etao+\rhoo)|_{\ell = 1}&=0 \, .
\end{align*}
\end{itemize}
\end{prop}

\medskip

\begin{prop}[Global properties $\ell\geq 2$-spherical modes] \label{glob_prop_l_2}
Consider an initial-data normalised solution $\mathfrak{S}_N$.~Then, $\mathfrak{S}_N$ satisfies the following identities:
\begin{itemize}
\item On $\mathcal{R}_{u_0,v_0}$ (including $\mathcal{H}^+_{v\geq v_0}$):~The identities
\begin{align*}
\overset{\text{\scalebox{.6}{$(1)$}}}{\mathfrak{\underline{\mathfrak{f}}}}_4|_{\ell\geq 2}&=0 \, , & \overset{\text{\scalebox{.6}{$(1)$}}}{\mathfrak{\slashed{\mathfrak{f}}}}_{4}|_{\ell\geq 2}&=0 \, ,  \\
 \overset{\text{\scalebox{.6}{$(1)$}}}{\mathfrak{\slashed{\mathfrak{f}}}}_{3}|_{\ell\geq 2}&=\frac{r^2}{2M}\,(\etao-\zetao)|_{\ell\geq 2} \, , & 
  (\overset{\text{\scalebox{.6}{$(1)$}}}{\slashed{\varepsilon}\cdot\chi})|_{\ell\geq 2}&=0 \, .
\end{align*}
\item On $\mathcal{H}^+_{v\geq v_0}$:~The identities
\begin{align*}
(\textup{tr}\overset{\text{\scalebox{.6}{$(1)$}}}{\slashed{g}})|_{\ell\geq  2}&=0 \, , & \slashed{\textup{curl}}\,\slashed{\textup{div}}\overset{\text{\scalebox{.6}{$(1)$}}}{\widehat{\slashed{g}}}&= 2 \sigmao|_{\ell\geq 2} \, , \\
(\overset{\text{\scalebox{.6}{$(1)$}}}{\textup{tr}\chi})|_{\ell\geq 2}&=0 \, , & (\slashed{\textup{div}}\etao+\rhoo)|_{\ell \geq 2}&=0 \, , &
(\slashed{\textup{curl}}\etao)|_{\ell\geq 2} &=\sigmao|_{\ell\geq 2} \, .
\end{align*}
\end{itemize}
\end{prop}

\medskip

\begin{prop}  \label{prop_vanish_Omega4_trchio}
Consider an initial-data normalised solution $\mathfrak{S}_N$.~Then, the linearised quantities 
\begin{align*}
&\Omega^{-4}_M\,(\overset{\text{\scalebox{.6}{$(1)$}}}{\textup{tr}\chi})|_{\ell\geq 1} \, , & &\Omega^{-4}_M\,\mathcal{A}^{[n]}r^2\slashed{\mathcal{D}}_2^{\star}\nablasl(\overset{\text{\scalebox{.6}{$(1)$}}}{\textup{tr}\chi})
\end{align*}
are uniformly bounded on $\underline{C}_{v_0}$.
\end{prop}

\medskip

\begin{proof}
This fact can be deduced from the vanishing of both $(\trchio)|_{\ell\geq 1}$ and $\nablasl_3(\trchio)|_{\ell\geq 1}$ on $\mathbb{S}^2_{\infty,v_0}$, the latter following from equation \eqref{3_trchio} and the properties $(\slashed{\textup{div}}\etao+\rhoo)|_{\ell \geq 1}=0$ and $\overset{\text{\scalebox{.6}{$(1)$}}}{\mathfrak{\underline{\mathfrak{f}}}}_4|_{\ell= 1}=0$ on $\mathbb{S}^2_{\infty,v_0}$.
\end{proof}

\medskip

We conclude the section with a key proposition.

\medskip

\begin{prop} \label{prop_proj_kernel_kerr}
Consider an initial-data normalised solution $\mathfrak{S}_N$.~Then, the $\ell=0,1$-spherical projection of $\mathfrak{S}_N$ is a reference linearised Kerr solution $\mathcal{K}_{\mathfrak{m},\mathfrak{s}_{m}}$ with parameters $(\mathfrak{m},\mathfrak{s}_{m})$ such that
\begin{align*}
\rhoo[\mathfrak{S}_N]|_{\ell=0}(\infty,v_0) &=-\frac{\mathfrak{m}}{4M^2} \, , & \sigmao[\mathfrak{S}_N]|_{\ell=1}(\infty,v_0)&=\sum_m \mathfrak{s}_{m} Y_m^{\ell=1} \, .
\end{align*}
\end{prop}

\medskip

\begin{proof}
The symmetric traceless $\mathbb{S}^2_{u,v}$ two-tensors
\begin{equation*}
\overset{\text{\scalebox{.6}{$(1)$}}}{\widehat{\slashed{g}}} \, , \, \chiho \, , \, \chibho \, , \, \alphao \, , \, \alphabo \, ,
\end{equation*}
as well as the scalar functions obtained by taking their $\slashed{\text{div}}\,\slashed{\text{div}}$ and $\slashed{\text{curl}}\,\slashed{\text{div}}$, are supported on $\ell\geq 2$-spherical modes. 

\medskip

We treat the $\ell=0$ and $\ell=1$-spherical modes for the remaining linearised quantities separately.~Before we start, it is useful to recall the identity
\begin{equation} \label{aux_expression_K}
2\,\overset{\text{\scalebox{.6}{$(1)$}}}{\widetilde{\slashed{K}}}=-\frac{1}{2}\,\slashed{\Delta}(\textup{tr}\overset{\text{\scalebox{.6}{$(1)$}}}{\slashed{g}})+\slashed{\text{div}}\,\slashed{\text{div}}\overset{\text{\scalebox{.6}{$(1)$}}}{\widehat{\slashed{g}}}-\frac{1}{r^2}\,(\textup{tr}\overset{\text{\scalebox{.6}{$(1)$}}}{\slashed{g}}) \, .
\end{equation}

\medskip

\underline{$\ell=0$-spherical modes}:~It suffices to prove
\begin{gather*}
\overset{\text{\scalebox{.6}{$(1)$}}}{\mathfrak{\underline{\mathfrak{f}}}}_4|_{\ell=0}=\overset{\text{\scalebox{.6}{$(1)$}}}{\mathfrak{\underline{\mathfrak{f}}}}_3|_{\ell=0}=(\textup{tr}\overset{\text{\scalebox{.6}{$(1)$}}}{\slashed{g}})|_{\ell=0}=0 \, , \\
(\trchio)|_{\ell=0}=(\overset{\text{\scalebox{.6}{$(1)$}}}{\slashed{\varepsilon}\cdot\chi})|_{\ell=0}=(\trchibo)|_{\ell=0}=(\overset{\text{\scalebox{.6}{$(1)$}}}{\slashed{\varepsilon}\cdot\chib})|_{\ell=0}=\omegabo|_{\ell=0}=0 \, , \\
\overset{\text{\scalebox{.6}{$(1)$}}}{\widetilde{\slashed{K}}}|_{\ell=0}=\rhoo|_{\ell=0}=\sigmao|_{\ell=0}=0
\end{gather*}
on $\mathcal{R}_{u_0,v_0}$ for any initial-data normalised solution $\mathfrak{S}_N$ such that 
\begin{equation} \label{aux_zero_rho_horizon}
\rhoo[\mathfrak{S}_N]|_{\ell=0}(\infty,v_0)=0 \, .
\end{equation}
We note that one can immediately conclude $(\overset{\text{\scalebox{.6}{$(1)$}}}{\slashed{\varepsilon}\cdot\chi})|_{\ell=0}=(\overset{\text{\scalebox{.6}{$(1)$}}}{\slashed{\varepsilon}\cdot\chib})|_{\ell=0}=\sigmao|_{\ell=0}=0$ on $\mathcal{R}_{u_0,v_0}$ from the equations \eqref{A_fA4}, \eqref{A_fA3} and \eqref{curl_zetao}.

\medskip

Using Definition \ref{def_initial_gauge_norm}, the identity \eqref{aux_zero_rho_horizon} implies $\overset{\text{\scalebox{.6}{$(1)$}}}{\mathfrak{\underline{\mathfrak{f}}}}_4|_{\ell=0}=0$ on $\underline{C}_{v_0}$ and $(\trchibo)|_{\ell=0}=0$ on $\mathbb{S}^2_{\infty,v_0}$.~We also recall that, by Definition \ref{def_initial_gauge_norm}, we have $\overset{\text{\scalebox{.6}{$(1)$}}}{\mathfrak{\underline{\mathfrak{f}}}}_3|_{\ell=0}=\omegabo|_{\ell=0}=0$ on $\underline{C}_{v_0}$.~Using equation \eqref{3_trchibo}, one has $(\trchibo)|_{\ell=0}=0$ on $\underline{C}_{v_0}$.~Then, $\rhoo|_{\ell=0}=0$ on $\underline{C}_{v_0}$ by equation \eqref{3_rhoo}.~By equation \eqref{3_trchio}, $(\trchio)|_{\ell=0}$ on $\underline{C}_{v_0}$.~We also have $\overset{\text{\scalebox{.6}{$(1)$}}}{\widetilde{\slashed{K}}}|_{\ell=0}=0$ on $\underline{C}_{v_0}$ by equation \eqref{gauss_eqn}, which implies $(\textup{tr}\overset{\text{\scalebox{.6}{$(1)$}}}{\slashed{g}})|_{\ell=0}=0$ on $\underline{C}_{v_0}$ by the identity \eqref{aux_expression_K}.~This concludes the vanishing of all quantities on $\underline{C}_{v_0}$. 

\medskip

Using equation \eqref{4_fb4} and \eqref{4_trchio}, one has $\overset{\text{\scalebox{.6}{$(1)$}}}{\mathfrak{\underline{\mathfrak{f}}}}_4|_{\ell=0}=(\trchio)|_{\ell=0}=0$ on $\mathcal{R}_{u_0,v_0}$.~Equation \eqref{4_rhoo} then gives $\rhoo|_{\ell=0}=0$ on $\mathcal{R}_{u_0,v_0}$ and equation \eqref{4_trg} gives $(\textup{tr}\overset{\text{\scalebox{.6}{$(1)$}}}{\slashed{g}})|_{\ell=0}=0$ on $\mathcal{R}_{u_0,v_0}$.~Equation \eqref{4_trchibo} implies $(\trchibo)|_{\ell=0}=0$ on $\mathcal{R}_{u_0,v_0}$.~Equation \eqref{gauss_eqn} gives $\overset{\text{\scalebox{.6}{$(1)$}}}{\widetilde{\slashed{K}}}|_{\ell=0}=0$ on $\mathcal{R}_{u_0,v_0}$.~Equations \eqref{4_f3} and \eqref{4_omegabo} now form an homogeneous system of coupled ODEs in the $e_4^{\text{as}}$-direction for $\overset{\text{\scalebox{.6}{$(1)$}}}{\mathfrak{\underline{\mathfrak{f}}}}_3|_{\ell=0}$ and $\omegabo|_{\ell=0}$.~Since the quantities identically vanish on $\underline{C}_{v_0}$, one can conclude $\overset{\text{\scalebox{.6}{$(1)$}}}{\mathfrak{\underline{\mathfrak{f}}}}_3|_{\ell=0}=\omegabo|_{\ell=0}=0$ on $\mathcal{R}_{u_0,v_0}$.

\medskip

\underline{$\ell=1$-spherical modes}:~It suffices to prove
\begin{gather*}
\overset{\text{\scalebox{.6}{$(1)$}}}{\mathfrak{\underline{\mathfrak{f}}}}_4|_{\ell=1}=\overset{\text{\scalebox{.6}{$(1)$}}}{\mathfrak{\underline{\mathfrak{f}}}}_3|_{\ell=1}=(\textup{tr}\overset{\text{\scalebox{.6}{$(1)$}}}{\slashed{g}})|_{\ell=1}=0  \, ,\\
(\trchio)|_{\ell=1}=(\overset{\text{\scalebox{.6}{$(1)$}}}{\slashed{\varepsilon}\cdot\chi})|_{\ell=1}=(\trchibo)|_{\ell=1}=(\overset{\text{\scalebox{.6}{$(1)$}}}{\slashed{\varepsilon}\cdot\chib})|_{\ell=1}=\omegabo|_{\ell=1}=0 \, , \\
\overset{\text{\scalebox{.6}{$(1)$}}}{\widetilde{\slashed{K}}}|_{\ell=1}=\rhoo|_{\ell=1}=\sigmao|_{\ell=1}=0
\end{gather*}
and
\begin{gather*}
\overset{\text{\scalebox{.6}{$(1)$}}}{\mathfrak{\underline{\mathfrak{f}}}}|_{\ell=1}=\overset{\text{\scalebox{.6}{$(1)$}}}{\mathfrak{\slashed{\mathfrak{f}}}}_{4}|_{\ell=1}=\overset{\text{\scalebox{.6}{$(1)$}}}{\mathfrak{\slashed{\mathfrak{f}}}}_{3}|_{\ell=1}=0 \, , \\
\ybo|_{\ell=1}=\etao|_{\ell=1}=\zetao|_{\ell=1}=0 \, , \\
\betao|_{\ell=1}=\betabo|_{\ell=1}=0
\end{gather*}
on $\mathcal{R}_{u_0,v_0}$ for any initial-data normalised solution $\mathfrak{S}_N$ such that 
\begin{equation} \label{aux_zero_sigma_horizon}
\sigmao[\mathfrak{S}_N]|_{\ell=1}(\infty,v_0)=0 \, .
\end{equation}

\medskip

Using Definition \ref{def_initial_gauge_norm}, the identity \eqref{aux_zero_sigma_horizon} implies $\overset{\text{\scalebox{.6}{$(1)$}}}{\mathfrak{\slashed{\mathfrak{f}}}}_{4}|_{\ell=1}=\overset{\text{\scalebox{.6}{$(1)$}}}{\mathfrak{\slashed{\mathfrak{f}}}}_{3}|_{\ell=1}=(\slashed{\textup{curl}}\overset{\text{\scalebox{.6}{$(1)$}}}{\mathfrak{\underline{\mathfrak{f}}}})|_{\ell=1}=0$ on $\underline{C}_{v_0}$.~We also recall that, by Definition \ref{def_initial_gauge_norm}, we have $\overset{\text{\scalebox{.6}{$(1)$}}}{\mathfrak{\underline{\mathfrak{f}}}}_4|_{\ell=1}=\overset{\text{\scalebox{.6}{$(1)$}}}{\mathfrak{\underline{\mathfrak{f}}}}_3|_{\ell=1}=\omegabo|_{\ell=1}=0$ on $\underline{C}_{v_0}$.~Equations \eqref{A_fA4} and \eqref{A_fA3} imply $(\overset{\text{\scalebox{.6}{$(1)$}}}{\slashed{\varepsilon}\cdot\chi})|_{\ell=1}=(\overset{\text{\scalebox{.6}{$(1)$}}}{\slashed{\varepsilon}\cdot\chib})|_{\ell=1}=0$ on $\underline{C}_{v_0}$, while equation \eqref{3_fA4} implies 
\begin{equation} \label{aux_etao_zetao}
\etao|_{\ell=1}=\zetao|_{\ell=1}
\end{equation}
on $\underline{C}_{v_0}$.~Equation \eqref{3_fA3} yields $\ybo|_{\ell=1}=0$ on $\underline{C}_{v_0}$.~Using equations \eqref{3_atrchio} and \eqref{curl_zetao}, we have $\slashed{\text{curl}}\etao|_{\ell=1}=\slashed{\text{curl}}\zetao|_{\ell=1}=0$ on $\mathbb{S}^2_{\infty,v_0}$.~Using the identity \eqref{aux_expression_K} for $\overset{\text{\scalebox{.6}{$(1)$}}}{\widetilde{\slashed{K}}}|_{\ell=1}$ and the normalisation identity $(\textup{tr}\overset{\text{\scalebox{.6}{$(1)$}}}{\slashed{g}})|_{\ell=1}(\infty,v_0)=0$, we have $\overset{\text{\scalebox{.6}{$(1)$}}}{\widetilde{\slashed{K}}}|_{\ell=1}=0$ on $\mathbb{S}^2_{\infty,v_0}$.~By applying equation \eqref{gauss_eqn}, we get $\rhoo|_{\ell=1}=0$ on $\mathbb{S}^2_{\infty,v_0}$.~Using the normalisation identity $(\slashed{\text{div}}\etao+\rhoo)|_{\ell=1}(\infty,v_0)=0$, we get $\slashed{\text{div}}\etao|_{\ell=1}=\slashed{\text{div}}\zetao|_{\ell=1}=0$ on $\mathbb{S}^2_{\infty,v_0}$.~Using the equation \eqref{codazzi_chiho} and \eqref{codazzi_chibho}, we have $\slashed{\text{div}}\betao|_{\ell=1}=\slashed{\text{div}}\betabo|_{\ell=1}=\slashed{\text{curl}}\betao|_{\ell=1}=\slashed{\text{curl}}\betabo|_{\ell=1}=0$ on $\mathbb{S}^2_{\infty,v_0}$.~This concludes the vanishing of all quantities on $\mathbb{S}^2_{\infty,v_0}$.

\medskip

Equation \eqref{3_betabo} gives $\slashed{\text{div}}\betabo|_{\ell=1}=\slashed{\text{curl}}\betabo|_{\ell=1}=0$ on $\underline{C}_{v_0}$.~Equation \eqref{3_zetao}, together with the identity \eqref{aux_etao_zetao}, give $\slashed{\text{div}}\zetao|_{\ell=1}=\slashed{\text{div}}\etao|_{\ell=1}=\slashed{\text{curl}}\zetao|_{\ell=1}=\slashed{\text{curl}}\etao|_{\ell=1}=0$ on $\underline{C}_{v_0}$.~Equation \eqref{3_trchibo} gives $(\trchibo)|_{\ell=1}=0$ on $\underline{C}_{v_0}$.~Equations \eqref{3_rhoo} and \eqref{3_sigmao} give $\rhoo|_{\ell=1}=\sigmao|_{\ell=1}=0$ on $\underline{C}_{v_0}$.~Equation \eqref{3_trchio} then gives $(\trchio)|_{\ell=1}=0$ on $\underline{C}_{v_0}$, which, through equation \eqref{codazzi_chiho}, yields $\slashed{\text{div}}\betao|_{\ell=1}=\slashed{\text{curl}}\betao|_{\ell=1}=0$ on $\underline{C}_{v_0}$.~Equation \eqref{gauss_eqn} gives $\overset{\text{\scalebox{.6}{$(1)$}}}{\widetilde{\slashed{K}}}|_{\ell=1}=0$ on $\underline{C}_{v_0}$, which then implies $(\textup{tr}\overset{\text{\scalebox{.6}{$(1)$}}}{\slashed{g}})|_{\ell=1}=0$ on $\underline{C}_{v_0}$ by the identity \eqref{aux_expression_K}.~Equation \eqref{3_trgo} gives $(\slashed{\textup{div}}\overset{\text{\scalebox{.6}{$(1)$}}}{\mathfrak{\underline{\mathfrak{f}}}})|_{\ell=1}=0$ on $\underline{C}_{v_0}$.~This concludes the vanishing of all quantities on $\underline{C}_{v_0}$.

\medskip

Equation \eqref{4_fb4} gives $\overset{\text{\scalebox{.6}{$(1)$}}}{\mathfrak{\underline{\mathfrak{f}}}}_4|_{\ell=1}=0$ on $\mathcal{R}_{u_0,v_0}$.~Equations \eqref{4_fA4}, \eqref{4_trchio} and \eqref{4_atrchio} give $\overset{\text{\scalebox{.6}{$(1)$}}}{\mathfrak{\slashed{\mathfrak{f}}}}_{4}|_{\ell=1}=0$ and $(\trchio)|_{\ell=1}=(\overset{\text{\scalebox{.6}{$(1)$}}}{\slashed{\varepsilon}\cdot\chi})|_{\ell=1}=0$ on $\mathcal{R}_{u_0,v_0}$.~Equation \eqref{4_trg} then gives $(\textup{tr}\overset{\text{\scalebox{.6}{$(1)$}}}{\slashed{g}})|_{\ell=1}=0$ on $\mathcal{R}_{u_0,v_0}$.~Equation \eqref{4_betao} gives $\slashed{\text{div}}\betao|_{\ell=1}=\slashed{\text{curl}}\betao|_{\ell=1}=0$ on $\mathcal{R}_{u_0,v_0}$.~Equations \eqref{4_rhoo} and \eqref{4_sigmao} then give $\rhoo|_{\ell=1}=\sigmao|_{\ell=1}=0$ on $\mathcal{R}_{u_0,v_0}$.~Equations \eqref{4_trchibo} and \eqref{4_atrchibo} give $(\trchibo)|_{\ell=1}=(\overset{\text{\scalebox{.6}{$(1)$}}}{\slashed{\varepsilon}\cdot\chib})|_{\ell=1}=0$ on $\mathcal{R}_{u_0,v_0}$.~Equation \eqref{4_etao} yields $\etao|_{\ell=1}=0$.~Then, equation \eqref{4_fA} gives $\overset{\text{\scalebox{.6}{$(1)$}}}{\mathfrak{\underline{\mathfrak{f}}}}|_{\ell=1}=0$ on $\mathcal{R}_{u_0,v_0}$.~We have $\overset{\text{\scalebox{.6}{$(1)$}}}{\widetilde{\slashed{K}}}|_{\ell=1}=0$ on $\mathcal{R}_{u_0,v_0}$ from equation \eqref{gauss_eqn}.~Equations \eqref{4_f3}, \eqref{4_fA3}, \eqref{4_ybo}, \eqref{4_omegabo}, \eqref{4_zetao} and \eqref{4_betabo} now form an homogeneous system of coupled ODEs in the $e_4^{\text{as}}$-direction for $$\overset{\text{\scalebox{.6}{$(1)$}}}{\mathfrak{\underline{\mathfrak{f}}}}_3|_{\ell=1} \, , \, \overset{\text{\scalebox{.6}{$(1)$}}}{\mathfrak{\slashed{\mathfrak{f}}}}_{3}|_{\ell=1} \, , \, \ybo|_{\ell=1} \, , \, \omegabo|_{\ell=1} \, , \, \zetao|_{\ell=1} \, , \, \betabo|_{\ell=1} \, .$$ Since all the quantities identically vanish on $\underline{C}_{v_0}$, one can conclude 
\begin{align*}
\overset{\text{\scalebox{.6}{$(1)$}}}{\mathfrak{\underline{\mathfrak{f}}}}_3|_{\ell=1}&=\omegabo|_{\ell=1}=0 \, , & \overset{\text{\scalebox{.6}{$(1)$}}}{\mathfrak{\slashed{\mathfrak{f}}}}_{3}|_{\ell=1}&=\ybo|_{\ell=1}=\zetao|_{\ell=1}=\betabo|_{\ell=1}=0
\end{align*}
on $\mathcal{R}_{u_0,v_0}$.
\end{proof}

\section{The main theorem}  \label{sec_main_theorem}

In this section, we state the main theorem of the present work.

\subsection{Energies and norms}

We define the initial energy fluxes for the (derived) gauge invariant linearised quantities\footnote{The definition of the initial energy fluxes for the (derived) gauge invariant linearised quantities is analogous to the one presented in Sections 10.1.1 and 10.2.1 of \cite{DHR}, but with the regularising weights for the linearised quantities removed.}
\begin{align}
\mathbb{F}^{n,T,\nablasl}_{0}[\overset{\text{\scalebox{.6}{$(1)$}}}{\Psi}] := & \, \sum_{i+j\leq n}\lVert \nablasl_3 (T^i(r\nablasl_A)^j\overset{\text{\scalebox{.6}{$(1)$}}}{\Psi}) \rVert_{L^2(\underline{C}_{v_0})}^2+\lVert r\nablasl (T^i(r\nablasl_A)^j\overset{\text{\scalebox{.6}{$(1)$}}}{\Psi}) \rVert_{L^2(\underline{C}_{v_0})}^2+\lVert T^i(r\nablasl_A)^j\overset{\text{\scalebox{.6}{$(1)$}}}{\Psi} \rVert_{L^2(\underline{C}_{v_0})}^2 \label{flux_1} \\
&+\lVert r\nablasl_4 (T^i(r\nablasl_A)^j\overset{\text{\scalebox{.6}{$(1)$}}}{\Psi}) \rVert_{L^2(C_{u_0})}^2+\lVert \nablasl (T^i(r\nablasl_A)^j\overset{\text{\scalebox{.6}{$(1)$}}}{\Psi}) \rVert_{L^2(C_{u_0})}^2+\lVert  r^{-1}(T^i(r\nablasl_A)^j\overset{\text{\scalebox{.6}{$(1)$}}}{\Psi}) \rVert_{L^2(C_{u_0})}^2 \, , \nonumber\\[5pt]
\mathbb{F}^{n,T,\nablasl}_{0}[\overset{\text{\scalebox{.6}{$(1)$}}}{\underline{\Psi}}] := & \, \sum_{i+j\leq n}\lVert \nablasl_3 (T^i(r\nablasl_A)^j\overset{\text{\scalebox{.6}{$(1)$}}}{\underline{\Psi}}) \rVert_{L^2(\underline{C}_{v_0})}^2+\lVert r\nablasl (T^i(r\nablasl_A)^j\overset{\text{\scalebox{.6}{$(1)$}}}{\underline{\Psi}}) \rVert_{L^2(\underline{C}_{v_0})}^2+\lVert T^i(r\nablasl_A)^j\overset{\text{\scalebox{.6}{$(1)$}}}{\underline{\Psi}} \rVert_{L^2(\underline{C}_{v_0})}^2 \label{flux_1b}\\
&+\lVert r\nablasl_4 (T^i(r\nablasl_A)^j\overset{\text{\scalebox{.6}{$(1)$}}}{\underline{\Psi}}) \rVert_{L^2(C_{u_0})}^2+\lVert \nablasl (T^i(r\nablasl_A)^j\overset{\text{\scalebox{.6}{$(1)$}}}{\underline{\Psi}}) \rVert_{L^2(C_{u_0})}^2+\lVert  r^{-1}(T^i(r\nablasl_A)^j\overset{\text{\scalebox{.6}{$(1)$}}}{\underline{\Psi}}) \rVert_{L^2(C_{u_0})}^2 \, , \nonumber
\end{align}
\begin{align}
\mathbb{F}^{n,T,\nablasl}_{0}[\overset{\text{\scalebox{.6}{$(1)$}}}{\Psi},\overset{\text{\scalebox{.6}{$(1)$}}}{\psi}] := & \, \mathbb{F}^{n,T,\nablasl}_{0}[\overset{\text{\scalebox{.6}{$(1)$}}}{\Psi}]+\sum_{i+j\leq n}\lVert  r^{4-\delta}(T^i(r\nablasl_A)^j\overset{\text{\scalebox{.6}{$(1)$}}}{\psi}) \rVert_{L^2(C_{u_0})}^2 \, ,  \label{flux_2}\\[5pt]
\mathbb{F}^{n,T,\nablasl}_{0}[\overset{\text{\scalebox{.6}{$(1)$}}}{\underline{\Psi}},\overset{\text{\scalebox{.6}{$(1)$}}}{\underline{\psi}}] := & \, \mathbb{F}^{n,T,\nablasl}_{0}[\overset{\text{\scalebox{.6}{$(1)$}}}{\underline{\Psi}}]+\sum_{i+j\leq n}\lVert  r^{3}(T^i(r\nablasl_A)^j\overset{\text{\scalebox{.6}{$(1)$}}}{\underline{\psi}}) \rVert_{L^2(\underline{C}_{v_0})}^2 \, , \label{flux_2b}
\end{align}
\begin{align}
\mathbb{F}^{n,T,\nablasl}_{0}[\overset{\text{\scalebox{.6}{$(1)$}}}{\Psi},\overset{\text{\scalebox{.6}{$(1)$}}}{\psi},\alphao] := & \, \mathbb{F}^{n,T,\nablasl}_{0}[\overset{\text{\scalebox{.6}{$(1)$}}}{\Psi},\overset{\text{\scalebox{.6}{$(1)$}}}{\psi}]+\sum_{i+j\leq n}\lVert  r^{3-\delta}(T^i(r\nablasl_A)^j\overset{\text{\scalebox{.6}{$(1)$}}}{\alpha}) \rVert_{L^2(C_{u_0})}^2 \, , \label{flux_3}\\[5pt]
\mathbb{F}^{n,T,\nablasl}_{0}[\overset{\text{\scalebox{.6}{$(1)$}}}{\underline{\Psi}},\overset{\text{\scalebox{.6}{$(1)$}}}{\underline{\psi}},\alphabo] := & \, \mathbb{F}^{n,T,\nablasl}_{0}[\overset{\text{\scalebox{.6}{$(1)$}}}{\underline{\Psi}},\overset{\text{\scalebox{.6}{$(1)$}}}{\underline{\psi}}]+\sum_{i+j\leq n}\lVert  T^i(r\nablasl_A)^j\overset{\text{\scalebox{.6}{$(1)$}}}{\underline{\alpha}} \rVert_{L^2(\underline{C}_{v_0})}^2 \, , \label{flux_3b}
\end{align}
\begin{align}
\mathbb{F}^{n,T,\nablasl}_{0}[\overset{\text{\scalebox{.6}{$(1)$}}}{\Psi},\mathfrak{D}\overset{\text{\scalebox{.6}{$(1)$}}}{\psi},\mathfrak{D}\alphao] := & \, \mathbb{F}^{n,T,\nablasl}_{0}[\overset{\text{\scalebox{.6}{$(1)$}}}{\Psi},\overset{\text{\scalebox{.6}{$(1)$}}}{\psi},\alphao] \label{flux_4}\\
&+\sum_{i+j\leq n} \lVert r^{4-\delta}\cdot r\nablasl_4 (T^i(r\nablasl_A)^j\overset{\text{\scalebox{.6}{$(1)$}}}{\psi}) \rVert_{L^2(C_{u_0})}^2+\lVert r^{4-\delta} \nablasl_3 (T^i(r\nablasl_A)^j \overset{\text{\scalebox{.6}{$(1)$}}}{\psi}) \rVert_{L^2(C_{u_0})}^2 \nonumber \\ &+\lVert r^{4-\delta}\cdot r\nablasl_A(T^i(r\nablasl_A)^j \overset{\text{\scalebox{.6}{$(1)$}}}{\psi}) \rVert_{L^2(C_{u_0})}^2 +\lVert r^{3-\delta}\cdot r\nablasl_4 (T^i(r\nablasl_A)^j\overset{\text{\scalebox{.6}{$(1)$}}}{\alpha}) \rVert_{L^2(C_{u_0})}^2 \nonumber\\
&+\lVert r^{3-\delta} \nablasl_3 (T^i(r\nablasl_A)^j\overset{\text{\scalebox{.6}{$(1)$}}}{\alpha} )\rVert_{L^2(C_{u_0})}^2  +\lVert r^{3-\delta}\cdot r\nablasl_A (T^i(r\nablasl_A)^j\overset{\text{\scalebox{.6}{$(1)$}}}{\alpha}) \rVert_{L^2(C_{u_0})}^2 \, , \nonumber\\[5pt]
\mathbb{F}^{n,T,\nablasl}_{0}[\overset{\text{\scalebox{.6}{$(1)$}}}{\underline{\Psi}},\mathfrak{D}\overset{\text{\scalebox{.6}{$(1)$}}}{\underline{\psi}},\mathfrak{D}\alphabo] := & \, \mathbb{F}^{n,T,\nablasl}_{0}[\overset{\text{\scalebox{.6}{$(1)$}}}{\underline{\Psi}},\overset{\text{\scalebox{.6}{$(1)$}}}{\underline{\psi}},\alphabo] \label{flux_4b}\\
&+\sum_{i+j\leq n} \lVert r^{3}\cdot r\nablasl_4 (T^i(r\nablasl_A)^j\overset{\text{\scalebox{.6}{$(1)$}}}{\underline{\psi}}) \rVert_{L^2(\underline{C}_{v_0})}^2+\lVert r^{3} \nablasl_3 (T^i(r\nablasl_A)^j \overset{\text{\scalebox{.6}{$(1)$}}}{\underline{\psi}}) \rVert_{L^2(\underline{C}_{v_0})}^2 \nonumber\\ &+\lVert r^{3}\cdot r\nablasl_A(T^i(r\nablasl_A)^j \overset{\text{\scalebox{.6}{$(1)$}}}{\underline{\psi}}) \rVert_{L^2(\underline{C}_{v_0})}^2 +\lVert  r\nablasl_4 (T^i(r\nablasl_A)^j\overset{\text{\scalebox{.6}{$(1)$}}}{\underline{\alpha}}) \rVert_{L^2(\underline{C}_{v_0})}^2 \nonumber\\
&+\lVert  \nablasl_3 (T^i(r\nablasl_A)^j\overset{\text{\scalebox{.6}{$(1)$}}}{\underline{\alpha}} )\rVert_{L^2(\underline{C}_{v_0})}^2  +\lVert  r\nablasl_A (T^i(r\nablasl_A)^j\overset{\text{\scalebox{.6}{$(1)$}}}{\underline{\alpha}}) \rVert_{L^2(\underline{C}_{v_0})}^2  \, , \nonumber
\end{align}
with $T$ the stationary Killing vector field, $n\in\mathbb{N}_0$, $\delta\geq 0$ and the linearised quantities $\overset{\text{\scalebox{.6}{$(1)$}}}{\psi}$, $\overset{\text{\scalebox{.6}{$(1)$}}}{\Psi}$, $\overset{\text{\scalebox{.6}{$(1)$}}}{\underline{\psi}}$ and $\overset{\text{\scalebox{.6}{$(1)$}}}{\underline{\Psi}}$ defined in Section \ref{sec_analysis_gauge_invariant}.~We define the initial energy flux for the gauge dependent linearised quantities
\begin{align}
\mathbb{D}_0 := & \, \lVert \nablasl_3^2(\mathcal{A}^{[3]} \chiho) \rVert_{L^2(\underline{C}_{v_0})}^2 +\lVert \mathcal{A}^{[3]} \chibho \rVert_{L^2(\underline{C}_{v_0})}^2 +\lVert \mathcal{A}^{[3]}\slashed{\mathcal{D}}_2^{\star} \overset{\text{\scalebox{.6}{$(1)$}}}{\mathfrak{\slashed{\mathfrak{f}}}}_{3}\rVert_{L^2(\underline{C}_{v_0})}^2 +\lVert \mathcal{A}^{[3]}\slashed{\mathcal{D}}_2^{\star}\zetao \rVert_{L^2(\underline{C}_{v_0})}^2  \label{def_initial_energy_norm_D} \\
& +\lVert \mathcal{A}^{[3]} \slashed{\mathcal{D}}_2^{\star}\nablasl(\Omega^{-4}(\trchio))\rVert_{L^2(\underline{C}_{v_0})}^2 +\lVert \mathcal{A}^{[3]} \slashed{\mathcal{D}}_2^{\star}\ybo\rVert_{L^2(\underline{C}_{v_0})}^2 + \lVert \mathcal{A}^{[3]} \slashed{\mathcal{D}}_2^{\star}\nablasl(\trchibo)\rVert_{L^2(\underline{C}_{v_0})}^2 \nonumber \\
&+ \lVert \mathcal{A}^{[3]} \slashed{\mathcal{D}}_2^{\star}\nablasl(\overset{\text{\scalebox{.6}{$(1)$}}}{\slashed{\varepsilon}\cdot\chib})\rVert_{L^2(\underline{C}_{v_0})}^2 +\lVert \mathcal{A}^{[3]} \slashed{\mathcal{D}}_2^{\star} \overset{\text{\scalebox{.6}{$(1)$}}}{\mathfrak{\underline{\mathfrak{f}}}}\rVert_{L^2(\underline{C}_{v_0})}^2 +\lVert \slashed{\textup{div}} \,\slashed{\mathcal{D}}_2^{\star}\nablasl\omegabo \rVert_{L^2(\underline{C}_{v_0})}^2  \nonumber \\ 
&+ \lVert \slashed{\textup{div}}\, \slashed{\mathcal{D}}_2^{\star}\nablasl  \overset{\text{\scalebox{.6}{$(1)$}}}{\mathfrak{\underline{\mathfrak{f}}}}_3 \rVert_{L^2(\underline{C}_{v_0})}^2 + \lVert \mathcal{A}^{[3]}\slashed{\mathcal{D}}_2^{\star}\nablasl (\Omega^{-2}(\textup{tr}\overset{\text{\scalebox{.6}{$(1)$}}}{\slashed{g}})) \rVert_{L^2(\underline{C}_{v_0})}^2  + \lVert \nablasl_3(\mathcal{A}^{[3]}\overset{\text{\scalebox{.6}{$(1)$}}}{\widehat{\slashed{g}}}) \rVert_{L^2(\underline{C}_{v_0})}^2 \, . \nonumber
\end{align}
We define the initial energy
\begin{equation}
\mathbb{E}_0 := \mathbb{F}^{2,T,\nablasl}_{0}[\overset{\text{\scalebox{.6}{$(1)$}}}{\Psi},\mathfrak{D}\overset{\text{\scalebox{.6}{$(1)$}}}{\psi},\mathfrak{D}\alphao]+\mathbb{F}^{2,T,\nablasl}_{0}[\overset{\text{\scalebox{.6}{$(1)$}}}{\underline{\Psi}},\mathfrak{D}\overset{\text{\scalebox{.6}{$(1)$}}}{\underline{\psi}},\mathfrak{D}\alphabo]+ \mathbb{D}_0 \, . \label{def_initial_energy_norm}
\end{equation}
We define the spacetime energies
\begin{align*}
&\mathbb{I}^{n,T,\nablasl}_{\mathcal{I},\delta}[\overset{\text{\scalebox{.6}{$(1)$}}}{\Psi}] \, , \, \mathbb{I}^{n,T,\nablasl}_{\textup{deg}}[\overset{\text{\scalebox{.6}{$(1)$}}}{\Psi}] \, , \, \mathbb{I}^{n,T,\nablasl}_{\textup{master}}[\overset{\text{\scalebox{.6}{$(1)$}}}{\Psi},\mathfrak{D}\overset{\text{\scalebox{.6}{$(1)$}}}{\psi},\mathfrak{D}\alphao] \, , & &\mathbb{I}^{n,T,\nablasl}_{\mathcal{I},\delta}[\overset{\text{\scalebox{.6}{$(1)$}}}{\underline{\Psi}}] \, , \, \mathbb{I}^{n,T,\nablasl}_{\textup{deg}}[\overset{\text{\scalebox{.6}{$(1)$}}}{\underline{\Psi}}] \, , \, \mathbb{I}^{n,T,\nablasl}_{\textup{master}}[\overset{\text{\scalebox{.6}{$(1)$}}}{\underline{\Psi}},\mathfrak{D}\overset{\text{\scalebox{.6}{$(1)$}}}{\underline{\psi}},\mathfrak{D}\alphabo]
\end{align*}
as defined in Sections 10.1.1 and 10.2.1 of \cite{DHR}, but with the regularising weights for the linearised quantities removed.

\subsection{The linear stability statement}

The following is the main result of the paper.

\medskip

\begin{theorem}[Linear stability of the Schwarzschild solution] \label{th_main_theorem}
Consider an initial-data normalised solution $\mathfrak{S}_N$ to the system of linearised equations arising from smooth, asymptotically flat seed initial data on $\mathcal{S}_{u_0,v_0}$.~Assume that the initial energy $\mathbb{E}_0$ is finite.~Then, 
\begin{itemize}
\item The uniform boundedness estimates for the linearised curvature components
\begin{align}
\sup_{u\geq u_0}\int_{v_0}^v d\bar{v} \int_{\mathbb{S}^2_{\bar{u},\bar{v}}} \sin\theta \,d\theta \, d\phi \,\frac{1}{r^{2}} \,  |\mathcal{A}^{[5]}(r^3\alphao)| ^2 &\lesssim \mathbb{E}_0 \, , \label{th_est_AAA}\\
\sup_{v\geq v_0}  \int_{u_0}^{u} d\bar{u}\int_{\mathbb{S}^2_{\bar{u},\bar{v}}} \sin\theta \,d\theta \, d\phi \, \Omega^2 \,   |\mathcal{A}^{[5]}(r \alphabo)| ^2  &\lesssim \mathbb{E}_0 \, , \\
\sup_{u\geq u_0}\int_{v_0}^v d\bar{v} \int_{\mathbb{S}^2_{\bar{u},\bar{v}}} \sin\theta \,d\theta \, d\phi \,\frac{1}{r^{2}} \, |\mathcal{A}^{[4]}r\slashed{\mathcal{D}}_2^{\star}(r^3\betao)| ^2  &\lesssim \mathbb{E}_0 \, , \\
\sup_{v\geq v_0} \int_{u_0}^{u} d\bar{u}\int_{\mathbb{S}^2_{\bar{u},\bar{v}}} \sin\theta \,d\theta \, d\phi \, \Omega^2 \,  |\mathcal{A}^{[4]}r\slashed{\mathcal{D}}_2^{\star}(r^2\betabo)| ^2  &\lesssim \mathbb{E}_0 
\end{align}
and
\begin{align}
\sup_{u\geq u_0} \int_{v_0}^{v} d\bar{v}\int_{\mathbb{S}^2_{\bar{u},\bar{v}}} \sin\theta \,d\theta \, d\phi \,\frac{1}{r^{2}}\,  |\mathcal{A}^{[3]}r^2\slashed{\mathcal{D}}_2^{\star}\slashed{\mathcal{D}}_1^{\star}(r^3\rhoo,r^3\sigmao)| ^2 &\lesssim \mathbb{E}_0 \, , \\
\sup_{v\geq v_0} \int_{u_0}^{u} d\bar{u}\int_{\mathbb{S}^2_{\bar{u},\bar{v}}} \sin\theta \,d\theta \, d\phi \,\Omega^2 \,  |\mathcal{A}^{[3]}r^2\slashed{\mathcal{D}}_2^{\star}\slashed{\mathcal{D}}_1^{\star}(r^3\rhoo,r^3\sigmao)| ^2 &\lesssim \mathbb{E}_0  \label{th_est_BBB}
\end{align}
hold.
\item Uniform boundedness estimates (in terms of $\mathbb{E}_0$) for weighted $L^2$-fluxes on null cones of up to five angular derivatives of all linearised connection coefficients, frame coefficients and induced metric hold.
\item For any $\delta>0$, $v\geq v_0$ and $u\geq u_0$, the integrated energy decay estimates for the linearised curvature components
\begin{align}
\int_{v_0}^v d\bar{v} \int_{u_0}^{u} d\bar{u}\int_{\mathbb{S}^2_{\bar{u},\bar{v}}} \sin\theta \,d\theta \, d\phi \,\frac{\Omega^2}{r^{1+\delta}} \left( 1-\frac{3M}{r}\right)^2  |\mathcal{A}^{[5]}(r^3\alphao)| ^2 &\lesssim \mathbb{E}_0 \, , \label{th_est_AA}\\
\int_{v_0}^v d\bar{v} \int_{u_0}^{u} d\bar{u}\int_{\mathbb{S}^2_{\bar{u},\bar{v}}} \sin\theta \,d\theta \, d\phi \,\frac{\Omega^2}{r^{1+\delta}} \left( 1-\frac{3M}{r}\right)^2  |\mathcal{A}^{[5]}(r \alphabo)| ^2  &\lesssim \mathbb{E}_0 \, , \\
\int_{v_0}^v d\bar{v} \int_{u_0}^{u} d\bar{u}\int_{\mathbb{S}^2_{\bar{u},\bar{v}}} \sin\theta \,d\theta \, d\phi \,\frac{\Omega^2}{r^{1+\delta}}\left( 1-\frac{3M}{r}\right)^2  |\mathcal{A}^{[4]}r\slashed{\mathcal{D}}_2^{\star}(r^3\betao)| ^2  &\lesssim \mathbb{E}_0 \, , \\
\int_{v_0}^v d\bar{v} \int_{u_0}^{u} d\bar{u}\int_{\mathbb{S}^2_{\bar{u},\bar{v}}} \sin\theta \,d\theta \, d\phi \,\frac{\Omega^2}{r^{1+\delta}}\left( 1-\frac{3M}{r}\right)^2  |\mathcal{A}^{[4]}r\slashed{\mathcal{D}}_2^{\star}(r^2\betabo)| ^2  &\lesssim \mathbb{E}_0 \, , \\
\int_{v_0}^v d\bar{v} \int_{u_0}^{u} d\bar{u}\int_{\mathbb{S}^2_{\bar{u},\bar{v}}} \sin\theta \,d\theta \, d\phi \,\frac{\Omega^2}{r^{1+\delta}} \left( 1-\frac{3M}{r}\right)^2  |\mathcal{A}^{[3]}r^2\slashed{\mathcal{D}}_2^{\star}\slashed{\mathcal{D}}_1^{\star}(r^3\rhoo,r^3\sigmao)| ^2 &\lesssim \mathbb{E}_0 \label{th_est_BB}
\end{align}
hold.~In particular, analogous integrated energy decay estimates, but without degenerating factor, hold for one less angular derivative of all linearised curvature components.
\item For any $\delta>0$, $v\geq v_0$ and $u\geq u_0$, the integrated energy decay estimates for the linearised connection coefficients
\begin{align}
\int_{v_0}^v d\bar{v} \int_{u_0}^u d\bar{u} \int_{\mathbb{S}^2_{\bar{u},\bar{v}}}\sin\theta\, d\theta \, d\phi \, \frac{\Omega^2}{r^{1+\delta}} |\mathcal{A}^{[3]}(r^2\chiho)|^2  &\lesssim \mathbb{E}_0 \, , \label{th_est_A}\\
\int_{v_0}^v d\bar{v} \int_{u_0}^{u} d\bar{u}\int_{\mathbb{S}^2_{\bar{u},\bar{v}}} \sin\theta \,d\theta \, d\phi \,\frac{\Omega^2}{r^{1+\delta}} \, |\mathcal{A}^{[3]}(r\chibho)| ^2 &\lesssim \mathbb{E}_0 \, , \\
\int_{v_0}^v d\bar{v} \int_{u_0}^{u} d\bar{u}\int_{\mathbb{S}^2_{\bar{u},\bar{v}}} \sin\theta \,d\theta \, d\phi \,\frac{\Omega^2}{r^{1+\delta}} \, | \mathcal{A}^{[3]}r^2\slashed{\mathcal{D}}_2^{\star}\nablasl (r^2\Omega^{-4}(\overset{\text{\scalebox{.6}{$(1)$}}}{\textup{tr}\chi}) ) | ^2 &\lesssim \mathbb{E}_0 \, , \\
\int_{v_0}^v d\bar{v} \int_{u_0}^{u} d\bar{u}\int_{\mathbb{S}^2_{\bar{u},\bar{v}}} \sin\theta \,d\theta \, d\phi \,\frac{\Omega^2}{r^{1+\delta}} \left( 1-\frac{3M}{r}\right)^2  |\mathcal{A}^{[3]}r^2\slashed{\mathcal{D}}_2^{\star}\nablasl(r(\overset{\text{\scalebox{.6}{$(1)$}}}{\textup{tr}\chib}))| ^2   &\lesssim \mathbb{E}_0 \, , \\
\int_{v_0}^v d\bar{v} \int_{u_0}^{u} d\bar{u}\int_{\mathbb{S}^2_{\bar{u},\bar{v}}} \sin\theta \,d\theta \, d\phi \,\frac{\Omega^2}{r^{1+\delta}} \left( 1-\frac{3M}{r}\right)^2  |\mathcal{A}^{[3]}r^2\slashed{\mathcal{D}}_2^{\star}\nablasl(r(\overset{\text{\scalebox{.6}{$(1)$}}}{\slashed{\varepsilon}\cdot\chib}))| ^2  &\lesssim \mathbb{E}_0 \, , \\
\int_{v_0}^v d\bar{v} \int_{u_0}^{u} d\bar{u}\int_{\mathbb{S}^2_{\bar{u},\bar{v}}} \sin\theta \,d\theta \, d\phi \,\frac{\Omega^2}{r^{1+\delta}}\, |r\,\slashed{\textup{div}}\,r^2\slashed{\mathcal{D}}_2^{\star}\nablasl\omegabo| ^2   &\lesssim \mathbb{E}_0 \, , \\
\int_{v_0}^v d\bar{v} \int_{u_0}^{u} d\bar{u}\int_{\mathbb{S}^2_{\bar{u},\bar{v}}} \sin\theta \,d\theta \, d\phi \,\frac{\Omega^2}{r^{1+\delta}} \, |\mathcal{A}^{[3]}r\slashed{\mathcal{D}}_2^{\star}\ybo| ^2  &\lesssim \mathbb{E}_0 \, , \\
\int_{v_0}^v d\bar{v} \int_{u_0}^{u} d\bar{u}\int_{\mathbb{S}^2_{\bar{u},\bar{v}}} \sin\theta \,d\theta \, d\phi \,\frac{\Omega^2}{r^{1+\delta}}\, |\mathcal{A}^{[3]}r\slashed{\mathcal{D}}_2^{\star}(r\etao)| ^2 &\lesssim \mathbb{E}_0 \, , \\
\int_{v_0}^v d\bar{v} \int_{u_0}^{u} d\bar{u}\int_{\mathbb{S}^2_{\bar{u},\bar{v}}} \sin\theta \,d\theta \, d\phi \,\frac{\Omega^2}{r^{1+\delta}} \,  |\mathcal{A}^{[3]}r\slashed{\mathcal{D}}_2^{\star}(r\zetao)| ^2 &\lesssim \mathbb{E}_0 \, ,
\end{align}
for the linearised frame coefficients 
\begin{align}
\int_{v_0}^v d\bar{v} \int_{u_0}^{u} d\bar{u}\int_{\mathbb{S}^2_{\bar{u},\bar{v}}} \sin\theta \,d\theta \, d\phi \,\frac{\Omega^2}{r^{1+\delta}}\, |\mathcal{A}^{[3]}r \slashed{\mathcal{D}}_2^{\star} \overset{\text{\scalebox{.6}{$(1)$}}}{\mathfrak{\slashed{\mathfrak{f}}}}_{3} | ^2 &\lesssim \mathbb{E}_0 \, , \\
\int_{v_0}^v d\bar{v} \int_{u_0}^{u} d\bar{u}\int_{\mathbb{S}^2_{\bar{u},\bar{v}}} \sin\theta \,d\theta \, d\phi \,\frac{\Omega^2}{r^{1+\delta}}   | \mathcal{A}^{[3]}r\slashed{\mathcal{D}}_2^{\star}  (r^{-1} \overset{\text{\scalebox{.6}{$(1)$}}}{\mathfrak{\underline{\mathfrak{f}}}} ) | ^2 &\lesssim \mathbb{E}_0 \, , \\
\int_{v_0}^v d\bar{v} \int_{u_0}^{u} d\bar{u}\int_{\mathbb{S}^2_{\bar{u},\bar{v}}} \sin\theta \,d\theta \, d\phi \,\frac{\Omega^2}{r^{1+\delta}} \, | r\,\slashed{\textup{div}}\,r^2\slashed{\mathcal{D}}_2^{\star}\nablasl  (r^{-1} \overset{\text{\scalebox{.6}{$(1)$}}}{\mathfrak{\underline{\mathfrak{f}}}}_3 ) | ^2 &\lesssim \mathbb{E}_0
\end{align}
and for the linearised induced metric
\begin{align}
\int_{v_0}^v d\bar{v} \int_{u_0}^{u} d\bar{u}\int_{\mathbb{S}^2_{\bar{u},\bar{v}}} \sin\theta \,d\theta \, d\phi \,\frac{\Omega^2}{r^{1+\delta}} \, | \mathcal{A}^{[3]}r^2\slashed{\mathcal{D}}_2^{\star}\nablasl (\Omega^{-2}(\textup{tr}\overset{\text{\scalebox{.6}{$(1)$}}}{\slashed{g}})) | ^2 &\lesssim \mathbb{E}_0 \, , \\
\int_{v_0}^v d\bar{v} \int_{u_0}^{u} d\bar{u}\int_{\mathbb{S}^2_{\bar{u},\bar{v}}} \sin\theta \,d\theta \, d\phi \,\frac{\Omega^2}{r^{1+\delta}} \, | \mathcal{A}^{[3]}\overset{\text{\scalebox{.6}{$(1)$}}}{\widehat{\slashed{g}}} | ^2 &\lesssim \mathbb{E}_0  \label{th_est_B}
\end{align}
hold.~Moreover, the estimates \eqref{th_est_A}-\eqref{th_est_B} can be upgraded to (possibly non-degenerate) integrated energy decay estimates (in terms of $\mathbb{E}_0$) for up to five angular derivatives of all linearised connection coefficients, frame coefficients and induced metric.
\end{itemize}
\end{theorem}

\medskip

\begin{remark}
In the present work, we prove the estimates \eqref{th_est_AA}-\eqref{th_est_BB} and \eqref{th_est_A}-\eqref{th_est_B}.~The proof of the estimates \eqref{th_est_AAA}-\eqref{th_est_BBB} can be easily reconstructed by the reader from our proof of the decay estimates \eqref{th_est_AA}-\eqref{th_est_BB}.~The procedure to upgrade the estimates \eqref{th_est_A}-\eqref{th_est_B} to estimates for higher angular derivatives of the linearised quantities is standard (see, for instance, Propositions \ref{iled_chiho_bis} and \ref{iled_chibho_refined}).~We also remark that some of the estimates stated in our main theorem are not expected to be optimal near infinity, meaning that analogous estimates with stronger weights in $r$ can be proven.
\end{remark}

\medskip

\begin{remark}
Theorem \ref{th_main_theorem} controls the $\ell\geq 2$-spherical projection of all linearised quantities in the linearised system of equations.~In view of Proposition \ref{prop_proj_kernel_kerr}, the theorem can be interpreted as the full linear stability result.
\end{remark}

\medskip

\begin{remark}
For smooth seed initial data which are asymptotically flat with suitable weight $s$ and order $n$ (see Definition \ref{def_asymptotic_flat_data}), the initial energy $\mathbb{E}_0$ is, in fact, finite.~We note, in particular, that Proposition \ref{prop_vanish_Omega4_trchio} and the horizon normalisation identities of Proposition \ref{def_initial_gauge_norm} guarantee the finiteness of the initial energy fluxes in \eqref{def_initial_energy_norm_D} containing a growing factor towards the event horizon.
\end{remark}

\medskip

\begin{remark}
The top-order term in the initial energy $\mathbb{E}_0$ is at the level of five derivatives of linearised curvature components, and thus the (degenerate) integrated decay estimates for the linearised curvature components in Theorem \ref{th_main_theorem} do \emph{not} lose derivatives.
\end{remark}

\medskip

\begin{remark}
In the initial energy flux for the gauge dependent linearised quantities, some of the terms can be dropped by using the linearised system of equations.~The form \eqref{def_initial_energy_norm_D} is however more convenient to read off the initial fluxes needed for the proof of Theorem \ref{th_main_theorem}.~We also note that the initial energy $\mathbb{E}_0$ vanishes for smooth seed initial data corresponding to a reference linearised Kerr solution.
\end{remark}

\section{Analysis of the gauge invariant linearised quantities}  \label{sec_analysis_gauge_invariant}

The analysis of the gauge invariant linearised quantities in the system relies on Sections 7, 11 and 12 of \cite{DHR}.~In this section, we explain how the results of \cite{DHR} can be applied to our setting. 

\medskip

We define the (regular) symmetric traceless $\mathbb{S}^2_{u,v}$ two-tensors 
\begin{align*}
\overset{\text{\scalebox{.6}{$(1)$}}}{\psi} &:= -\frac{1}{2\,r}\nablasl_3(r\alphao) \, , &  \overset{\text{\scalebox{.6}{$(1)$}}}{P}&:=\frac{1}{r^3}\nablasl_3(\overset{\text{\scalebox{.6}{$(1)$}}}{\psi}r^3) \, , & \overset{\text{\scalebox{.6}{$(1)$}}}{\Psi}&:=r^5\overset{\text{\scalebox{.6}{$(1)$}}}{P} \, ,  \\
\overset{\text{\scalebox{.6}{$(1)$}}}{\underline{\psi}} &:= \frac{1}{2\,r\,\Omega^4_M}\nablasl_4(r\,\Omega^4_M\alphabo) \, , &  \overset{\text{\scalebox{.6}{$(1)$}}}{\underline{P}}&:=-\frac{1}{r^3\Omega^2}\nablasl_4(\overset{\text{\scalebox{.6}{$(1)$}}}{\underline{\psi}}r^3\Omega) \, , & \overset{\text{\scalebox{.6}{$(1)$}}}{\underline{\Psi}}&:=r^5\overset{\text{\scalebox{.6}{$(1)$}}}{\underline{P}} \, .
\end{align*}

\medskip

The \emph{Teukolsky equation of spin $+2$} for a symmetric traceless $\mathbb{S}^2_{u,v}$ two-tensor is the equation (173) in \cite{DHR}.~The \emph{Teukolsky equation of spin $-2$} for a symmetric traceless $\mathbb{S}^2_{u,v}$ two-tensor is the equation (174) in \cite{DHR}.~The \emph{Regge--Wheeler equation} for a symmetric traceless $\mathbb{S}^2_{u,v}$ two-tensor is the equation (175) in \cite{DHR}.

\medskip

The following proposition is the analogue of Proposition 7.4.1 of \cite{DHR} in our setting.~The proof is a check left to the reader.

\medskip

\begin{prop}
Consider a solution $\mathfrak{S}$ to the linearised system of equations.~Then, the linearised curvature components $\alphao$ and $\alphabo$ satisfy the Teukolsky equations of spin $+2$ and $-2$ respectively, the quantities $\overset{\text{\scalebox{.6}{$(1)$}}}{P}$ and $\overset{\text{\scalebox{.6}{$(1)$}}}{\underline{P}}$ satisfy the Regge--Wheeler equation and the identities
\begin{align*}
\overset{\text{\scalebox{.6}{$(1)$}}}{\psi} &= \slashed{\mathcal{D}}_2^{\star}\betao+\frac{3}{2}\,\rho\,\chiho \, , &  \overset{\text{\scalebox{.6}{$(1)$}}}{\underline{\psi}} &= \slashed{\mathcal{D}}_2^{\star}\betabo-\frac{3}{2}\,\rho\,\chibho   
\end{align*}
and
\begin{align*}
\overset{\text{\scalebox{.6}{$(1)$}}}{P}&= \slashed{\mathcal{D}}_2^{\star}\slashed{\mathcal{D}}_1^{\star}(-\rhoo,\sigmao) -\frac{3}{4}\,\rho\,(\textup{tr}\chib)\chiho-\frac{3}{4}\,\rho\,(\textup{tr}\chi)\chibho  +(\nablasl_3\rho)\,\slashed{\mathcal{D}}_2^{\star}\overset{\text{\scalebox{.6}{$(1)$}}}{\mathfrak{\slashed{\mathfrak{f}}}}_{4}+(\nablasl_4\rho)\,\slashed{\mathcal{D}}_2^{\star}\overset{\text{\scalebox{.6}{$(1)$}}}{\mathfrak{\slashed{\mathfrak{f}}}}_{3}  \, , \\ 
\overset{\text{\scalebox{.6}{$(1)$}}}{\underline{P}}&=\slashed{\mathcal{D}}_2^{\star}\slashed{\mathcal{D}}_1^{\star}(-\rhoo,-\sigmao) -\frac{3}{4}\,\rho\,(\textup{tr}\chib)\chiho-\frac{3}{4}\,\rho\,(\textup{tr}\chi)\chibho  +(\nablasl_3\rho)\,\slashed{\mathcal{D}}_2^{\star}\overset{\text{\scalebox{.6}{$(1)$}}}{\mathfrak{\slashed{\mathfrak{f}}}}_{4}+(\nablasl_4\rho)\,\slashed{\mathcal{D}}_2^{\star}\overset{\text{\scalebox{.6}{$(1)$}}}{\mathfrak{\slashed{\mathfrak{f}}}}_{3} 
\end{align*}
hold.
\end{prop}

\medskip

\begin{remark} \label{rmk_derived_gauge_inv_quantities_rels}
Consider an initial-data normalised solution $\mathfrak{S}_N$ to the system of linearised equations.~Then,
\begin{align*}
\overset{\text{\scalebox{.6}{$(1)$}}}{P}&= \slashed{\mathcal{D}}_2^{\star}\slashed{\mathcal{D}}_1^{\star}(-\rhoo,\sigmao) -\frac{3}{4}\,\rho\,(\textup{tr}\chib)\chiho-\frac{3}{4}\,\rho\,(\textup{tr}\chi)\chibho  +(\nablasl_4\rho)\,\slashed{\mathcal{D}}_2^{\star}\overset{\text{\scalebox{.6}{$(1)$}}}{\mathfrak{\slashed{\mathfrak{f}}}}_{3} \, , \\ 
\overset{\text{\scalebox{.6}{$(1)$}}}{\underline{P}}&=\slashed{\mathcal{D}}_2^{\star}\slashed{\mathcal{D}}_1^{\star}(-\rhoo,-\sigmao) -\frac{3}{4}\,\rho\,(\textup{tr}\chib)\chiho-\frac{3}{4}\,\rho\,(\textup{tr}\chi)\chibho  +(\nablasl_4\rho)\,\slashed{\mathcal{D}}_2^{\star}\overset{\text{\scalebox{.6}{$(1)$}}}{\mathfrak{\slashed{\mathfrak{f}}}}_{3}  \, .
\end{align*}
Note that the last term in both identities does not appear in the corresponding identities (183) and (184) in \cite{DHR}.~However, we have
\begin{align*}
\overset{\text{\scalebox{.6}{$(1)$}}}{P}&= \slashed{\mathcal{D}}_2^{\star}\slashed{\mathcal{D}}_1^{\star}(-\rhoo,\sigmao) -\frac{3}{4}\,\rho\,(\textup{tr}\chib)\chiho    \, , &
\overset{\text{\scalebox{.6}{$(1)$}}}{\underline{P}}&=\slashed{\mathcal{D}}_2^{\star}\slashed{\mathcal{D}}_1^{\star}(-\rhoo,-\sigmao) -\frac{3}{4}\,\rho\,(\textup{tr}\chib)\chiho    
\end{align*}
on $\mathcal{H}^+_{v\geq v_0}$, which do coincide with the identities (183) and (184) in \cite{DHR} restricted to the event horizon.~This fact will be exploited to repeat in our setting the analysis for the horizon fluxes of Section 13.2 in \cite{DHR}.
\end{remark}

\medskip

The following proposition states integrated energy decay for the gauge invariant linearised quantities.

\medskip

\begin{prop}[Integrated decay for the gauge invariant linearised quantities] \label{iled_gauge_inv_quantities}
Consider a solution $\mathfrak{S}$ to the linearised system of equations.~Assume that the initial energy fluxes \eqref{flux_1} and \eqref{flux_4} are finite.~Then, for any $\delta>0$,
\begin{itemize}
\item Theorem 1 of \cite{DHR} holds for $\overset{\text{\scalebox{.6}{$(1)$}}}{\Psi}$ and $\overset{\text{\scalebox{.6}{$(1)$}}}{\underline{\Psi}}$.~In particular, we have the integrated energy decay estimate
\begin{equation} \label{iled_P}
\mathbb{I}^{n,T,\nablasl}_{\mathcal{I},\delta}[\overset{\text{\scalebox{.6}{$(1)$}}}{\Psi}]+\mathbb{I}^{n,T,\nablasl}_{\textup{deg}}[\overset{\text{\scalebox{.6}{$(1)$}}}{\Psi}] \lesssim \mathbb{F}^{n,T,\nablasl}_{0}[\overset{\text{\scalebox{.6}{$(1)$}}}{\Psi}] 
\end{equation}
for any $n\in\mathbb{N}_0$.~An analogous integrated energy decay estimate holds for $\overset{\text{\scalebox{.6}{$(1)$}}}{\underline{\Psi}}$.
\item Theorem 2 of \cite{DHR} holds for $\overset{\text{\scalebox{.6}{$(1)$}}}{\Psi}$, $\overset{\text{\scalebox{.6}{$(1)$}}}{\psi}$, $\alphao$ and $\overset{\text{\scalebox{.6}{$(1)$}}}{\underline{\Psi}}$, $\overset{\text{\scalebox{.6}{$(1)$}}}{\underline{\psi}}$, $\alphabo$.~In particular, we have
\begin{equation} \label{iled_alpha_psi}
\mathbb{I}^{n,T,\nablasl}_{\textup{master}}[\overset{\text{\scalebox{.6}{$(1)$}}}{\Psi},\mathfrak{D}\overset{\text{\scalebox{.6}{$(1)$}}}{\psi},\mathfrak{D}\alphao]  \lesssim \mathbb{F}^{n,T,\nablasl}_{0}[\overset{\text{\scalebox{.6}{$(1)$}}}{\Psi},\mathfrak{D}\overset{\text{\scalebox{.6}{$(1)$}}}{\psi},\mathfrak{D}\alphao] 
\end{equation}
for any $n\in\mathbb{N}_0$.~An analogous integrated energy decay estimate holds for $\overset{\text{\scalebox{.6}{$(1)$}}}{\underline{\Psi}}$, $ \overset{\text{\scalebox{.6}{$(1)$}}}{\underline{\psi}}$, $\alphabo$.
\end{itemize}
\end{prop}

\medskip

The two following statements are corollaries of Proposition \ref{iled_gauge_inv_quantities} and collect the integrated energy decay estimates for the gauge invariant linearised quantities which are necessary for our later gauge dependent analysis.

\medskip

\begin{corollary}
Consider a solution $\mathfrak{S}$ to the linearised system of equations.~Assume that the initial energy flux \eqref{flux_1} is finite.~Then, for any $\delta>0$, the estimate \eqref{iled_P} implies the integrated energy decay estimates
\begin{gather}
\int_{u_0}^{\infty}d\bar{u}\int_{v_0}^{\infty}d\bar{v}\int_{\mathbb{S}^2_{\bar{u},\bar{v}}}\sin\theta\,d\theta\,d\phi \,\frac{\Omega^2}{r^{1+\delta}}\, | \mathcal{A}^{[n]}(r^5\overset{\text{\scalebox{.6}{$(1)$}}}{P})|^2  \lesssim \mathbb{F}^{n,T,\nablasl}_{0}[\overset{\text{\scalebox{.6}{$(1)$}}}{\Psi}] \, ,  \\
\int_{u_0}^{\infty}d\bar{u}\int_{v_0}^{\infty}d\bar{v}\int_{\mathbb{S}^2_{\bar{u},\bar{v}}}\sin\theta\,d\theta\,d\phi \,\frac{\Omega^2}{r^{1+\delta}}\left( 1-\frac{3M}{r}\right)^2 | \mathcal{A}^{[n+1]}(r^5\overset{\text{\scalebox{.6}{$(1)$}}}{P})|^2  \lesssim \mathbb{F}^{n,T,\nablasl}_{0}[\overset{\text{\scalebox{.6}{$(1)$}}}{\Psi}] 
\end{gather}
for any $n\in\mathbb{N}_0$.~Analogous integrated energy decay estimates hold for $\overset{\text{\scalebox{.6}{$(1)$}}}{\underline{P}}$, $\overset{\text{\scalebox{.6}{$(1)$}}}{\underline{\Psi}}$.~In particular, the initial energy flux $\mathbb{F}^{2,T,\nablasl}_{0}[\overset{\text{\scalebox{.6}{$(1)$}}}{\Psi}]$ controls $\mathcal{A}^{[2]}(r^5\overset{\text{\scalebox{.6}{$(1)$}}}{P})$ without degeneration (at $r=3M$) and $\mathcal{A}^{[3]}(r^5\overset{\text{\scalebox{.6}{$(1)$}}}{P})$ with degeneration (at $r=3M$).
\end{corollary}

\medskip

\begin{corollary}
Consider a solution $\mathfrak{S}$ to the linearised system of equations.~Assume that the initial energy fluxes \eqref{flux_2}-\eqref{flux_4b} are finite.~Then, for any $\delta>0$, we have the integrated energy decay estimates
\begin{gather}
\int_{u_0}^{\infty}d\bar{u}\int_{v_0}^{\infty}d\bar{v}\int_{\mathbb{S}^2_{\bar{u},\bar{v}}}\sin\theta\,d\theta\,d\phi \,\frac{\Omega^2}{r^{1+\delta}}\, | r^4\overset{\text{\scalebox{.6}{$(1)$}}}{\psi} |^2  \lesssim \mathbb{F}_{0}[\overset{\text{\scalebox{.6}{$(1)$}}}{\Psi},\overset{\text{\scalebox{.6}{$(1)$}}}{\psi}] \, , \\
\int_{u_0}^{\infty}d\bar{u}\int_{v_0}^{\infty}d\bar{v}\int_{\mathbb{S}^2_{\bar{u},\bar{v}}}\sin\theta\,d\theta\,d\phi \,\frac{\Omega^2}{r^{1+\delta}}\, | r^3\overset{\text{\scalebox{.6}{$(1)$}}}{\underline{\psi}} |^2  \lesssim \mathbb{F}_{0}[\overset{\text{\scalebox{.6}{$(1)$}}}{\underline{\Psi}},\overset{\text{\scalebox{.6}{$(1)$}}}{\underline{\psi}}]
\end{gather}
and
\begin{gather}
\int_{u_0}^{\infty}d\bar{u}\int_{v_0}^{\infty}d\bar{v}\int_{\mathbb{S}^2_{\bar{u},\bar{v}}}\sin\theta\,d\theta\,d\phi \,\frac{\Omega^2}{r^{1+\delta}}\, | r^3\alphao |^2  \lesssim \mathbb{F}_{0}[\overset{\text{\scalebox{.6}{$(1)$}}}{\Psi},\overset{\text{\scalebox{.6}{$(1)$}}}{\psi},\alphao] \, , \\
\int_{u_0}^{\infty}d\bar{u}\int_{v_0}^{\infty}d\bar{v}\int_{\mathbb{S}^2_{\bar{u},\bar{v}}}\sin\theta\,d\theta\,d\phi \,\frac{\Omega^2}{r^{1+\delta}}\, | r\alphabo |^2  \lesssim \mathbb{F}_{0}[\overset{\text{\scalebox{.6}{$(1)$}}}{\underline{\Psi}},\overset{\text{\scalebox{.6}{$(1)$}}}{\underline{\psi}},\alphabo] \, .
\end{gather}
The integrated energy decay estimate \eqref{iled_alpha_psi} implies the integrated energy decay estimates
\begin{gather}
\int_{u_0}^{\infty}d\bar{u}\int_{v_0}^{\infty}d\bar{v}\int_{\mathbb{S}^2_{\bar{u},\bar{v}}}\sin\theta\,d\theta\,d\phi \,\frac{\Omega^2}{r^{1+\delta}}\, (| \mathcal{A}^{[n]}(r^4(\mathfrak{D}\overset{\text{\scalebox{.6}{$(1)$}}}{\psi}))|^2 +| \mathcal{A}^{[n]}( r^3(\mathfrak{D}\alphao))|^2) \lesssim \mathbb{F}^{n,T,\nablasl}_{0}[\overset{\text{\scalebox{.6}{$(1)$}}}{\Psi},\mathfrak{D}\overset{\text{\scalebox{.6}{$(1)$}}}{\psi},\mathfrak{D}\alphao] \, , \\
\int_{u_0}^{\infty}d\bar{u}\int_{v_0}^{\infty}d\bar{v}\int_{\mathbb{S}^2_{\bar{u},\bar{v}}}\sin\theta\,d\theta\,d\phi \,\frac{\Omega^2}{r^{1+\delta}}\, (| \mathcal{A}^{[n]}(r^3(\mathfrak{D}\overset{\text{\scalebox{.6}{$(1)$}}}{\underline{\psi}}))|^2 +| \mathcal{A}^{[n]}( r(\mathfrak{D}\alphabo))|^2) \lesssim \mathbb{F}^{n,T,\nablasl}_{0}[\overset{\text{\scalebox{.6}{$(1)$}}}{\underline{\Psi}},\mathfrak{D}\overset{\text{\scalebox{.6}{$(1)$}}}{\underline{\psi}},\mathfrak{D}\alphabo] 
\end{gather}
and the degenerate integrated energy decay estimates
\begin{gather}
\int_{u_0}^{\infty}d\bar{u}\int_{v_0}^{\infty}d\bar{v}\int_{\mathbb{S}^2_{\bar{u},\bar{v}}}\sin\theta\,d\theta\,d\phi \,\frac{\Omega^2}{r^{1+\delta}}\left(1-\frac{3M}{r}\right)^2 | \mathcal{A}^{[n+1]}(r^4(\mathfrak{D}\overset{\text{\scalebox{.6}{$(1)$}}}{\psi}))|^2  \lesssim \mathbb{F}^{n,T,\nablasl}_{0}[\overset{\text{\scalebox{.6}{$(1)$}}}{\Psi},\mathfrak{D}\overset{\text{\scalebox{.6}{$(1)$}}}{\psi},\mathfrak{D}\alphao] \, , \\
\int_{u_0}^{\infty}d\bar{u}\int_{v_0}^{\infty}d\bar{v}\int_{\mathbb{S}^2_{\bar{u},\bar{v}}}\sin\theta\,d\theta\,d\phi \,\frac{\Omega^2}{r^{1+\delta}}\left(1-\frac{3M}{r}\right)^2 | \mathcal{A}^{[n+1]}(r^3(\mathfrak{D}\overset{\text{\scalebox{.6}{$(1)$}}}{\underline{\psi}}))|^2  \lesssim \mathbb{F}^{n,T,\nablasl}_{0}[\overset{\text{\scalebox{.6}{$(1)$}}}{\underline{\Psi}},\mathfrak{D}\overset{\text{\scalebox{.6}{$(1)$}}}{\underline{\psi}},\mathfrak{D}\alphabo] \, .
\end{gather}
In particular, the initial energy flux $\mathbb{F}^{2,T,\nablasl}_{0}[\overset{\text{\scalebox{.6}{$(1)$}}}{\Psi},\mathfrak{D}\overset{\text{\scalebox{.6}{$(1)$}}}{\psi},\mathfrak{D}\alphao]$ controls $\mathcal{A}^{[3]}(r^4\overset{\text{\scalebox{.6}{$(1)$}}}{\psi})$ without degeneration (at $r=3M$) and $\mathcal{A}^{[4]}(r^4\overset{\text{\scalebox{.6}{$(1)$}}}{\psi})$ with degeneration (at $r=3M$).~An analogous statement holds for the initial energy flux $\mathbb{F}^{2,T,\nablasl}_{0}[\overset{\text{\scalebox{.6}{$(1)$}}}{\underline{\Psi}},\mathfrak{D}\overset{\text{\scalebox{.6}{$(1)$}}}{\underline{\psi}},\mathfrak{D}\alphabo]$.
\end{corollary}

\medskip

We state the following top-order integrated energy decay estimates for angular derivatives of $\alphao$ and $\alphabo$, corresponding to Proposition 12.3.2 of \cite{DHR} with $i=5$.

\medskip

\begin{prop}[Top-order integrated decay for $\alphao$ and $\alphabo$] \label{iled_alpha_alphab}
Consider a solution $\mathfrak{S}$ to the linearised system of equations.~Assume that the initial energy fluxes \eqref{flux_4} and \eqref{flux_4b} are finite.~Then, for any $\delta>0$, $v\geq v_0$ and $u\geq u_0$, we have the integrated energy decay estimates
\begin{align*}
\int_{v_0}^v d\bar{v} \int_{u_0}^{u} d\bar{u}\int_{\mathbb{S}^2_{\bar{u},\bar{v}}} \sin\theta \,d\theta \, d\phi \,\frac{\Omega^2}{r^{1+\delta}} \left( 1-\frac{3M}{r}\right)^2  |\mathcal{A}^{[5]}(r^3\alphao)| ^2 &\lesssim \mathbb{F}^{2,T,\nablasl}_{0}[\overset{\text{\scalebox{.6}{$(1)$}}}{\Psi},\mathfrak{D}\overset{\text{\scalebox{.6}{$(1)$}}}{\psi},\mathfrak{D}\alphao] \, , \\
\int_{v_0}^v d\bar{v} \int_{u_0}^{u} d\bar{u}\int_{\mathbb{S}^2_{\bar{u},\bar{v}}} \sin\theta \,d\theta \, d\phi \,\frac{\Omega^2}{r^{1+\delta}} \left( 1-\frac{3M}{r}\right)^2  |\mathcal{A}^{[5]}(r \alphabo)| ^2  &\lesssim \mathbb{F}^{2,T,\nablasl}_{0}[\overset{\text{\scalebox{.6}{$(1)$}}}{\underline{\Psi}},\mathfrak{D}\overset{\text{\scalebox{.6}{$(1)$}}}{\underline{\psi}},\mathfrak{D}\alphabo] \, .
\end{align*}
\end{prop}

\section{Analysis of the gauge dependent linearised quantities} \label{sec_analysis_gauge_dependent}

In this section, we present the analysis of the gauge dependent linearised quantities in the system.

\subsection{Horizon energy fluxes} \label{sec_horizon_fluxes}

The following propositions state the necessary estimates for the horizon energy fluxes.

\medskip

\begin{prop} \label{prop_flux_horizon_shears}
Let $\mathfrak{S}_N$ be an initial-data normalised solution.~Assume that the initial energy flux \eqref{flux_2} is finite.~Then,
\begin{equation} \label{flux_horizon_shears_zero_order}
\int_{\mathcal{H}^+(v_0,\infty)}dv \sin\theta \,d\theta\, d\phi\,\left(\frac{9}{4}\,\rho^2|\chiho|^2+|\slashed{\mathcal{D}}_2^{\star}\,\slashed{\textup{div}}\chiho|+|\slashed{\mathcal{D}}_2^{\star}\betao|^2 +\frac{6M}{r^3}\, |\betao|^2 \right)  \lesssim \mathbb{F}_0[\overset{\text{\scalebox{.6}{$(1)$}}}{\Psi},\overset{\text{\scalebox{.6}{$(1)$}}}{\psi}]
\end{equation} 
and
\begin{equation} \label{flux_horizon_shears_higher_order}
\int_{\mathcal{H}^+(v_0,\infty)}dv \sin\theta \,d\theta\, d\phi\,( |\nablasl_3\chiho|^2+|\slashed{\mathcal{D}}_2^{\star}\,\slashed{\textup{div}}\nablasl_3\chiho|+|\slashed{\mathcal{D}}_2^{\star}\etao|^2 + |\slashed{\textup{div}}\,\slashed{\mathcal{D}}_2^{\star}\etao|^2+ |\slashed{\mathcal{D}}_2^{\star}\,\slashed{\textup{div}}\,\slashed{\mathcal{D}}_2^{\star}\etao|^2)  \lesssim \mathbb{F}_0[\overset{\text{\scalebox{.6}{$(1)$}}}{\Psi},\overset{\text{\scalebox{.6}{$(1)$}}}{\psi}] \, .  
\end{equation}
\end{prop}

\medskip

\begin{proof}
In view of the initial-data normalisation identities of Definition \ref{def_initial_gauge_norm}, the proof is completely analogous to that of Propositions 13.2.1 and 13.2.2 in \cite{DHR}.
\end{proof}

\medskip

\begin{prop} \label{prop_flux_horizon_metric}
Let $\mathfrak{S}_N$ be an initial-data normalised solution.~Assume that the initial energy flux \eqref{flux_2} is finite.~Then,
\begin{equation}  \label{flux_horizon_metric}
\int_{\mathcal{H}^+(v_0,\infty)}dv \sin\theta \,d\theta\, d\phi\, ( |\slashed{\textup{div}}\,\slashed{\mathcal{D}}^{\star}_2\nablasl\slashed{\textup{div}}\,\slashed{\textup{div}}\overset{\text{\scalebox{.6}{$(1)$}}}{\widehat{\slashed{g}}} |^2+|\slashed{\textup{div}}\,\slashed{\mathcal{D}}^{\star}_2{}^{\star}\nablasl\slashed{\textup{curl}}\,\slashed{\textup{div}}\overset{\text{\scalebox{.6}{$(1)$}}}{\widehat{\slashed{g}}} |^2)  \lesssim \mathbb{F}_0[\overset{\text{\scalebox{.6}{$(1)$}}}{\Psi},\overset{\text{\scalebox{.6}{$(1)$}}}{\psi}]  \, .
\end{equation} 
\end{prop}

\medskip

\begin{proof}
From the estimate \eqref{flux_horizon_shears_zero_order} and Proposition \ref{iled_gauge_inv_quantities}, we control the horizon flux
\begin{equation*} 
\int_{\mathcal{H}^+(v_0,\infty)}dv \sin\theta \,d\theta\, d\phi\,  |\slashed{\text{div}}\,\slashed{\mathcal{D}}^{\star}_2\slashed{\mathcal{D}}^{\star}_1(-\rhoo,\sigmao) |^2   \lesssim \mathbb{F}_0[\overset{\text{\scalebox{.6}{$(1)$}}}{\Psi},\overset{\text{\scalebox{.6}{$(1)$}}}{\psi}]  
\end{equation*}  
(see the analogous estimate (335) of Proposition 13.2.3 in \cite{DHR}).~We recall that $(\text{tr}\overset{\text{\scalebox{.6}{$(1)$}}}{\slashed{g}})|_{\ell\geq 1}=0$ on $\mathcal{H}^+$, and thus the equation \eqref{gauss_eqn} on $\mathcal{H}^+$ yields
\begin{equation*}
\slashed{\text{div}}\,\slashed{\mathcal{D}}^{\star}_2\nablasl\slashed{\text{div}}\,\slashed{\text{div}}\overset{\text{\scalebox{.6}{$(1)$}}}{\widehat{\slashed{g}}}=-2\,\slashed{\text{div}}\,\slashed{\mathcal{D}}^{\star}_2\nablasl \rhoo \, , 
\end{equation*}
which allows to control the first term in \eqref{flux_horizon_metric}.~For the second term in \eqref{flux_horizon_metric}, we exploit the identity 
\begin{equation*}
\slashed{\text{div}}\,\slashed{\mathcal{D}}^{\star}_2{}^{\star}\nablasl\slashed{\text{curl}}\,\slashed{\text{div}}\overset{\text{\scalebox{.6}{$(1)$}}}{\widehat{\slashed{g}}}= 2\,\slashed{\text{div}}\,\slashed{\mathcal{D}}^{\star}_2{}^{\star}\nablasl \sigmao
\end{equation*}
holding on $\mathcal{H}^+$ for any initial-data normalised solution $\mathfrak{S}_N$.
\end{proof}

\medskip

\begin{remark} \label{rmk_higher_order_fluxes}
Higher order horizon fluxes of the form of Proposition 13.2.3 in \cite{DHR} can also be obtained.
\end{remark}

\medskip

\begin{remark}
At this stage, one can already establish polynomial decay in $v$ along the event horizon $\mathcal{H}^+$ for some of the horizon fluxes.
\end{remark}

\subsection{Integrated energy decay for the linearised outgoing shear} \label{sec_iled_chih}

In this section, we state integrated energy decay for the linearised outgoing shear.~The proof can be obtained by following the proof of Propositions 13.3.1 and 13.3.3 in \cite{DHR}.

\medskip

\begin{prop}[Integrated decay for the linearised outgoing shear] \label{iled_chih}
Let $\mathfrak{S}_N$ be an initial-data normalised solution.~Assume that the initial energy flux \eqref{flux_3} and energy \eqref{def_initial_energy_norm} are finite.~Then, for any $\delta>0$, $v\geq v_0$ and $u\geq u_0$, we have
\begin{align}
\int_{v_0}^v d\bar{v} \int_{u_0}^u d\bar{u} \int_{\mathbb{S}^2_{\bar{u},\bar{v}}} \sin\theta\, d\theta \, d\phi \, \frac{\Omega^2}{r^{1+\delta}}\left(|\nablasl_3^2(r^2\chiho)|^2+|\nablasl_3(r^2\chiho)|^2+|r^2\chiho|^2\right) \lesssim & \, \lVert \nablasl_3^2 \chiho \rVert_{L^2(\underline{C}_{v_0})}^2 \\ &+ \mathbb{F}_0[\overset{\text{\scalebox{.6}{$(1)$}}}{\Psi},\overset{\text{\scalebox{.6}{$(1)$}}}{\psi},\alphao] \nonumber
\end{align}
and 
\begin{equation}
\int_{v_0}^v d\bar{v} \int_{u_0}^u d\bar{u} \int_{\mathbb{S}^2_{\bar{u},\bar{v}}}\sin\theta\, d\theta \, d\phi \, \frac{\Omega^2}{r^{1+\delta}}\left(|\nablasl_3^2(\mathcal{A}^{[3]}(r^2\chiho))|^2+|\nablasl_3(\mathcal{A}^{[3]}(r^2\chiho))|^2+|\mathcal{A}^{[3]}(r^2\chiho)|^2\right) \lesssim \mathbb{E}_0 \, .
\end{equation}
\end{prop}

\subsection{Integrated energy decay for the linearised ingoing shear} \label{sec_iled_chibh}

In this section, we state and prove integrated energy decay for the linearised ingoing shear.

\medskip

\begin{prop}[Integrated decay for the linearised ingoing shear] \label{iled_chibh}
Let $\mathfrak{S}_N$ be an initial-data normalised solution.~Assume that the initial energy flux \eqref{flux_3} and energy \eqref{def_initial_energy_norm} are finite.~Then, for any $\delta>0$, $v\geq v_0$ and $u\geq u_0$, we have
\begin{equation} \label{iled_chibh_zero_order}
\int_{v_0}^v d\bar{v} \int_{u_0}^{u} d\bar{u}\int_{\mathbb{S}^2_{\bar{u},\bar{v}}} \sin\theta \,d\theta \, d\phi \,\frac{\Omega^2}{r^{1+\delta}} \, |r\chibho| ^2 \lesssim \lVert \nablasl_3^2 \chiho \rVert_{L^2(\underline{C}_{v_0})}^2+\lVert \chibho \rVert_{L^2(\underline{C}_{v_0})}^2 + \mathbb{F}_0[\overset{\text{\scalebox{.6}{$(1)$}}}{\Psi},\overset{\text{\scalebox{.6}{$(1)$}}}{\psi},\alphao]
\end{equation}
and
\begin{equation}  \label{iled_chibh_higher_order}
\int_{v_0}^v d\bar{v} \int_{u_0}^{u} d\bar{u}\int_{\mathbb{S}^2_{\bar{u},\bar{v}}} \sin\theta \,d\theta \, d\phi \,\frac{\Omega^2}{r^{1+\delta}} \, |\mathcal{A}^{[3]}(r\chibho)| ^2 \lesssim \mathbb{E}_0 \, .
\end{equation}
\end{prop}

\medskip

\begin{proof}
Let $r_2=10M$.~The red-shift estimate 
\begin{equation} \label{ineq_chibh_1}
\int_{v_0}^v d\bar{v} \int_{u_0(r_2,\bar{v})}^u d\bar{u}\int_{\mathbb{S}^2_{\bar{u},\bar{v}}} \sin\theta \,d\theta \, d\phi \,\Omega^2 | \chibho| ^2 \lesssim  \lVert \nablasl_3^2 \chiho \rVert_{L^2(\underline{C}_{v_0})}^2+\lVert \chibho \rVert_{L^2(\underline{C}_{v_0})}^2 + \mathbb{F}_0[\overset{\text{\scalebox{.6}{$(1)$}}}{\Psi},\overset{\text{\scalebox{.6}{$(1)$}}}{\psi},\alphao]      
\end{equation}
is immediate from equation \eqref{4_chibho} written in the form
\begin{equation*}
\nablasl_4(r\chibho)+\omegah\,(r\chibho)= \chiho \, . 
\end{equation*}
To establish \eqref{ineq_chibh_1}, one contracts the equation with $r\chibho$, applies the Cauchy-Schwartz inequality, integrates over $[u_0(r_2,v),u]\times[v_0,v]\times \mathbb{S}^2_{u,v}$ and, using Proposition \ref{iled_chih}, obtains the estimate
\begin{align*}
\sup_{\bar{v}\in (v_0,v)} & \int_{u_0(r_2,\bar{v})}^u d\bar{u}\int_{\mathbb{S}^2_{\bar{u},\bar{v}}} \sin\theta \,d\theta \, d\phi \,\Omega^2 |r\chibho| ^2 \\ &+ \int_{v_0}^v d\bar{v} \int_{u_0(r_2,\bar{v})}^u d\bar{u}\int_{\mathbb{S}^2_{\bar{u},\bar{v}}} \sin\theta \,d\theta \, d\phi \,\Omega^2 | r\chibho| ^2 \lesssim  \lVert \nablasl_3^2 \chiho \rVert_{L^2(\underline{C}_{v_0})}^2 +\lVert \chibho \rVert_{L^2(\underline{C}_{v_0})}^2 + \mathbb{F}_0[\overset{\text{\scalebox{.6}{$(1)$}}}{\Psi},\overset{\text{\scalebox{.6}{$(1)$}}}{\psi},\alphao]  \, .   
\end{align*}
The first term on the left hand side and the weights in $r$ can be dropped.

\medskip

We now wish to extend \eqref{ineq_chibh_1} to a global (in $r$) estimate on $\mathcal{R}_{u_0,v_0}$.~We write equation \eqref{4_chibho} as
\begin{equation*}
\nablasl_4(\Omega^2 r\chibho)=\frac{\Omega^2}{r^2}\,(r^2\chiho) \, .
\end{equation*}
Let $r_1=5M$ and $\xi=\xi(r)$ a cut-off function such that $\xi\equiv 0$ for $2M\leq r\leq r_1$ and $\xi=1$ for $r\geq r_2$.~We contract the equation with $\Omega^2 r\chibho$, apply the Cauchy-Schwartz inequality ($\epsilon>0$), take an overall multiplication by $r^{-\delta}\xi$ and obtain
\begin{equation*}
\frac{1}{2}\,\nablasl_4(r^{-\delta}\xi\cdot | \Omega^2 r\chibho| ^2)-r^{-\delta}(\nablasl_4\xi)\cdot |\Omega^2 r\chibho| ^2+\delta \, \frac{ \Omega^2 }{r^{1+\delta}} \,\xi  \cdot |\Omega^2 r\chibho|^2  - \epsilon\,\frac{\Omega^2}{2\, r^{2+\delta}}\,\xi\cdot | \Omega^2 r\chibho| ^2  \leq \frac{\Omega^2}{2\,\epsilon\, r^{2+\delta}} \,\xi\cdot |  r^2\chiho| ^2 \, . 
\end{equation*}
For any $\delta>0$, the constant $\epsilon$ can be chosen sufficiently small so that the fourth term on the left hand side can be absorbed by the third one.~The second term is only supported on $r_1\leq r \leq r_2$ and, upon integration, can be controlled by the estimate \eqref{ineq_chibh_1}.~We integrate over $[u_0,u(r_1,v)]\times[v_0,v]\times \mathbb{S}^2_{u,v}$ and, using the fact that $\Omega^2$ is bounded below away from $\mathcal{H}^+$ and again the estimate \eqref{ineq_chibh_1} to bound the fixed-$r$ boundary term arising, we obtain the estimate
\begin{align}
& \sup_{\bar{v}\in (v_0,v)}  \int_{u_0}^{u(r_1,\bar{v})} d\bar{u} \int_{\mathbb{S}^2_{\bar{u},\bar{v}}} \sin\theta \,d\theta \, d\phi \,\Omega^2 r^{-\delta}\cdot  |  r\chibho| ^2  \label{ineq_chibh_2}    \\ & + \int_{v_0}^v d\bar{v} \int_{u_0}^{u(r_1,\bar{v})} d\bar{u}\int_{\mathbb{S}^2_{\bar{u},\bar{v}}} \sin\theta \,d\theta \, d\phi \,\Omega^2 r^{-1-\delta}\cdot |  r\chibho| ^2 \lesssim \lVert \nablasl_3^2 \chiho \rVert_{L^2(\underline{C}_{v_0})}^2+\lVert \chibho \rVert_{L^2(\underline{C}_{v_0})}^2  \label{ineq_chibh_1} 
 + \mathbb{F}_0[\overset{\text{\scalebox{.6}{$(1)$}}}{\Psi},\overset{\text{\scalebox{.6}{$(1)$}}}{\psi},\alphao] \, .   \nonumber
\end{align}
Combining the estimates \eqref{ineq_chibh_1} and \eqref{ineq_chibh_2} yields the final estimate \eqref{iled_chibh_zero_order}.

\medskip

To obtain the higher order estimate \eqref{iled_chibh_higher_order}, one simply commutes equation \eqref{4_chibho} with the angular operator $\mathcal{A}^{[3]}$ and repeats the argument above.
\end{proof}

\subsection{Integrated energy decay for the remaining linearised quantities} \label{sec_iled_remaining_quantities}

In this section, we state and prove integrated energy decay for all the remaining linearised quantities in the system.~Interestingly, our scheme never involves integrating an ingoing transport equation for a gauge dependent linearised quantity in the ingoing direction.~We remark that all the linearised quantities that we estimate are supported on $\ell\geq 2$-spherical modes.

\medskip

\begin{prop}[Integrated decay for $\betao$ and $\betabo$] \label{iled_beta_betab}
Let $\mathfrak{S}_N$ be an initial-data normalised solution.~Assume that the initial energy \eqref{def_initial_energy_norm} is finite.~Then, for any $\delta>0$, $v\geq v_0$ and $u\geq u_0$, we have
\begin{equation}
\int_{v_0}^v d\bar{v} \int_{u_0}^{u} d\bar{u}\int_{\mathbb{S}^2_{\bar{u},\bar{v}}} \sin\theta \,d\theta \, d\phi \,\frac{\Omega^2}{r^{1+\delta}}\, ( |\mathcal{A}^{[3]}r\slashed{\mathcal{D}}_2^{\star}(r^3\betao)| ^2 +|\mathcal{A}^{[3]}r\slashed{\mathcal{D}}_2^{\star}(r^2\betabo)| ^2 )\lesssim \mathbb{E}_0 \, .
\end{equation}
\end{prop}

\medskip

\begin{proof}
The proposition follows from the identities 
\begin{align*}
\mathcal{A}^{[3]}(r^4\overset{\text{\scalebox{.6}{$(1)$}}}{\psi}) &= \mathcal{A}^{[3]}r\slashed{\mathcal{D}}_2^{\star}(r^3\betao)-\frac{3M}{r}\,\mathcal{A}^{[3]}(r^2\chiho) \, , &  
\mathcal{A}^{[3]}(r^3\overset{\text{\scalebox{.6}{$(1)$}}}{\underline{\psi}}) &= \mathcal{A}^{[3]}r\slashed{\mathcal{D}}_2^{\star}(r^2\betabo)+\frac{3M}{r}\,\mathcal{A}^{[3]}(r\chibho)
\end{align*}
and the combination of Propositions \ref{iled_chih} and \ref{iled_chibh} with Proposition \ref{iled_gauge_inv_quantities}.
\end{proof}

\medskip

\begin{prop}[Integrated decay for $\etao$] \label{iled_eta}
Let $\mathfrak{S}_N$ be an initial-data normalised solution.~Assume that the initial energy \eqref{def_initial_energy_norm} is finite.~Then, for any $\delta>0$, $v\geq v_0$ and $u\geq u_0$, we have
\begin{equation}
\int_{v_0}^v d\bar{v} \int_{u_0}^{u} d\bar{u}\int_{\mathbb{S}^2_{\bar{u},\bar{v}}} \sin\theta \,d\theta \, d\phi \,\frac{\Omega^2}{r^{1+\delta}}\, |\mathcal{A}^{[3]}r\slashed{\mathcal{D}}_2^{\star}(r\etao)| ^2 \lesssim \mathbb{E}_0 \, .
\end{equation}
\end{prop}

\medskip

\begin{proof}
One exploits the $\mathcal{A}^{[3]}$-commuted equation \eqref{3_chiho} and Propositions \ref{iled_chih} and \ref{iled_chibh}.
\end{proof}

\medskip

\begin{prop}[Integrated decay for $\overset{\text{\scalebox{.6}{$(1)$}}}{\mathfrak{\slashed{\mathfrak{f}}}}_{3}$] \label{iled_fA3}
Let $\mathfrak{S}_N$ be an initial-data normalised solution.~Assume that the initial energy \eqref{def_initial_energy_norm} is finite.~Then, for any $\delta>0$, $v\geq v_0$ and $u\geq u_0$, we have
\begin{equation}
\int_{v_0}^v d\bar{v} \int_{u_0}^{u} d\bar{u}\int_{\mathbb{S}^2_{\bar{u},\bar{v}}} \sin\theta \,d\theta \, d\phi \,\frac{\Omega^2}{r^{1+\delta}}\, |\mathcal{A}^{[3]}r \slashed{\mathcal{D}}_2^{\star} \overset{\text{\scalebox{.6}{$(1)$}}}{\mathfrak{\slashed{\mathfrak{f}}}}_{3} | ^2 \lesssim \mathbb{E}_0 \, .
\end{equation}
\end{prop}

\medskip

\begin{proof}
We commute equation \eqref{4_fA3} with $\mathcal{A}^{[3]}r\slashed{\mathcal{D}}_2^{\star}$ and use the initial-data normalisation identity
\begin{equation}  \label{commuted_gauge_rel}
\mathcal{A}^{[3]}r\slashed{\mathcal{D}}_2^{\star} \overset{\text{\scalebox{.6}{$(1)$}}}{\mathfrak{\slashed{\mathfrak{f}}}}_{3}=\frac{r^2}{2M}(\mathcal{A}^{[3]}r \slashed{\mathcal{D}}_2^{\star}\etao-\mathcal{A}^{[3]}r\slashed{\mathcal{D}}_2^{\star} \zetao)
\end{equation}
to obtain the red-shifted transport equation
\begin{equation*}
\nablasl_4 (\mathcal{A}^{[3]}r\slashed{\mathcal{D}}_2^{\star} \overset{\text{\scalebox{.6}{$(1)$}}}{\mathfrak{\slashed{\mathfrak{f}}}}_{3})+\frac{1}{2}\,(\text{tr}\chi) (\mathcal{A}^{[3]}r \slashed{\mathcal{D}}_2^{\star} \overset{\text{\scalebox{.6}{$(1)$}}}{\mathfrak{\slashed{\mathfrak{f}}}}_{3})+\omegah\,(\mathcal{A}^{[3]}r \slashed{\mathcal{D}}_2^{\star}  \overset{\text{\scalebox{.6}{$(1)$}}}{\mathfrak{\slashed{\mathfrak{f}}}}_{3})=\frac{1}{r}\,\mathcal{A}^{[3]}r \slashed{\mathcal{D}}_2^{\star} (r\etao ) \, .
\end{equation*}
The proof then follows the scheme of the proof of Proposition \ref{iled_chibh} and exploits Proposition \ref{iled_eta}.~The proof of the red-shift estimate is completely analogous.~For integrated decay in the large-$r$ region, one writes the equation as
\begin{equation*}
\nablasl_4 (\Omega^2\,\mathcal{A}^{[3]}r\slashed{\mathcal{D}}_2^{\star} \overset{\text{\scalebox{.6}{$(1)$}}}{\mathfrak{\slashed{\mathfrak{f}}}}_{3})+\frac{\Omega^2}{r}\, (\Omega^2\,\mathcal{A}^{[3]}r \slashed{\mathcal{D}}_2^{\star} \overset{\text{\scalebox{.6}{$(1)$}}}{\mathfrak{\slashed{\mathfrak{f}}}}_{3})=\frac{1}{r}\, \Omega^2\,\mathcal{A}^{[3]}r \slashed{\mathcal{D}}_2^{\star} (r\etao ) 
\end{equation*}
and follows the same procedure adopted to prove the estimate \eqref{ineq_chibh_2}.
\end{proof}

\medskip

\begin{prop}[Integrated decay for $\zetao$] \label{iled_zeta}
Let $\mathfrak{S}_N$ be an initial-data normalised solution.~Assume that the initial energy \eqref{def_initial_energy_norm} is finite.~Then, for any $\delta>0$, $v\geq v_0$ and $u\geq u_0$, we have
\begin{equation}
\int_{v_0}^v d\bar{v} \int_{u_0}^{u} d\bar{u}\int_{\mathbb{S}^2_{\bar{u},\bar{v}}} \sin\theta \,d\theta \, d\phi \,\frac{\Omega^2}{r^{1+\delta}} \,  |\mathcal{A}^{[3]}r\slashed{\mathcal{D}}_2^{\star}(r\zetao)| ^2 \lesssim \mathbb{E}_0 \, .
\end{equation}
\end{prop}

\medskip

\begin{proof}
One uses the $\mathcal{A}^{[3]}r\slashed{\mathcal{D}}_2^{\star}$-commuted red-shifted equation \eqref{4_zetao} combined with Propositions \ref{iled_beta_betab} and \ref{iled_fA3}.~The scheme of the proof is analogous to that of Proposition \ref{iled_chibh}.
\end{proof}

\medskip

\begin{remark}
Exploiting the identity \eqref{commuted_gauge_rel} and Proposition \ref{iled_fA3}, one can immediately show that integrated energy decay for (angular derivatives of) the difference $\etao-\zetao$ gains a power of $r$ compared to integrated energy decay for (angular derivatives of) $\etao$ (Proposition \ref{iled_eta}) and (angular derivatives of) $\zetao$ (Proposition \ref{iled_zeta}) separately.
\end{remark}

\medskip

\begin{prop}[Top-order integrated decay for $\rhoo$ and $\sigmao$] \label{iled_rho_sigma}
Let $\mathfrak{S}_N$ be an initial-data normalised solution.~Assume that the initial energy \eqref{def_initial_energy_norm} is finite.~Then, for any $\delta>0$, $v\geq v_0$ and $u\geq u_0$, we have
\begin{equation}
\int_{v_0}^v d\bar{v} \int_{u_0}^{u} d\bar{u}\int_{\mathbb{S}^2_{\bar{u},\bar{v}}} \sin\theta \,d\theta \, d\phi \,\frac{\Omega^2}{r^{1+\delta}} \left( 1-\frac{3M}{r}\right)^2  |\mathcal{A}^{[3]}r^2\slashed{\mathcal{D}}_2^{\star}\slashed{\mathcal{D}}_1^{\star}(r^3\rhoo,r^3\sigmao)| ^2 \lesssim \mathbb{E}_0 \, .
\end{equation}
\end{prop}

\medskip

\begin{proof}
The estimate can be proven by direct application of the identity 
\begin{equation*}
\mathcal{A}^{[3]}(r^5\overset{\text{\scalebox{.6}{$(1)$}}}{P})= \mathcal{A}^{[3]} r^2\slashed{\mathcal{D}}_2^{\star}\slashed{\mathcal{D}}_1^{\star}(-r^3\rhoo,r^3\sigmao) -\frac{3M}{r}\,\mathcal{A}^{[3]}(r^2\chiho)+3 M\, \Omega^2\,(r\chibho)  +6M\,\Omega^2\,(r\slashed{\mathcal{D}}_2^{\star}\overset{\text{\scalebox{.6}{$(1)$}}}{\mathfrak{\slashed{\mathfrak{f}}}}_{3})
\end{equation*}
and Propositions \ref{iled_gauge_inv_quantities}, \ref{iled_chih}, \ref{iled_chibh} and \ref{iled_fA3}.~We recall that the initial energy flux $\mathbb{F}^{2,T,\nablasl}_{0}[\overset{\text{\scalebox{.6}{$(1)$}}}{\Psi},\mathfrak{D}\overset{\text{\scalebox{.6}{$(1)$}}}{\psi},\mathfrak{D}\alphao]$ only controls $\mathcal{A}^{[3]}(r^5\overset{\text{\scalebox{.6}{$(1)$}}}{P})$ with degeneration at trapping ($r=3M$).
\end{proof}

\medskip

\begin{remark}
In contrast with Proposition \ref{iled_beta_betab}, the (degenerate) integrated decay estimate from Proposition \ref{iled_rho_sigma} already controls the linearised curvature components $\rhoo$ and $\sigmao$ at top order, without losing derivatives.~We will establish an analogous top-order estimate for $\betao$ and $\betabo$ in Proposition \ref{iled_beta_betab_bis}.
\end{remark}

\medskip

\begin{prop}[Integrated decay for $(\trchio)$] \label{iled_trchi}
Let $\mathfrak{S}_N$ be an initial-data normalised solution.~Assume that the initial energy \eqref{def_initial_energy_norm} is finite.~Then, for any $\delta>0$, $v\geq v_0$ and $u\geq u_0$, we have
\begin{equation}
\int_{v_0}^v d\bar{v} \int_{u_0}^{u} d\bar{u}\int_{\mathbb{S}^2_{\bar{u},\bar{v}}} \sin\theta \,d\theta \, d\phi \,\frac{\Omega^2}{r^{1+\delta}} \, | \mathcal{A}^{[3]}r^2\slashed{\mathcal{D}}_2^{\star}\nablasl (r^2\Omega^{-4}(\overset{\text{\scalebox{.6}{$(1)$}}}{\textup{tr}\chi}) ) |^2 \lesssim \mathbb{E}_0 \, .
\end{equation}
\end{prop}

\medskip

\begin{proof}
We write equation \eqref{4_trchio} as the red-shifted equation
\begin{equation*}
\nablasl_4 (\Omega^{-4}\,(\trchio))+ (\text{tr}\chi)(\Omega^{-4}\,(\trchio))+\omegah \, (\Omega^{-4}\,(\trchio)) =0
\end{equation*}
and recall Proposition \ref{prop_vanish_Omega4_trchio}.~One can then commute with $\mathcal{A}^{[3]}r^2\slashed{\mathcal{D}}_2^{\star}\nablasl$ and apply the scheme of the proof of Proposition \ref{iled_chibh}.
\end{proof}

\medskip

\begin{remark} \label{rmk_trchi_position_scheme}
Proposition \ref{iled_trchi} can, in principle, be proven at the very start of our scheme to control the gauge dependent linearised quantities.~However, the proposition is only needed at this stage of the scheme, when one wants to prove the integrated energy decay estimate of Proposition \ref{iled_trchib_atrchib}.   
\end{remark}

\medskip

\begin{prop}[Integrated decay for $(\trchibo)$ and $(\overset{\text{\scalebox{.6}{$(1)$}}}{\slashed{\varepsilon}\cdot\chib})$] \label{iled_trchib_atrchib}
Let $\mathfrak{S}_N$ be an initial-data normalised solution.~Assume that the initial energy \eqref{def_initial_energy_norm} is finite.~Then, for any $\delta>0$, $v\geq v_0$ and $u\geq u_0$, we have
\begin{equation}
\int_{v_0}^v d\bar{v} \int_{u_0}^{u} d\bar{u}\int_{\mathbb{S}^2_{\bar{u},\bar{v}}} \sin\theta \,d\theta \, d\phi \,\frac{\Omega^2}{r^{1+\delta}} \left( 1-\frac{3M}{r}\right)^2 ( |\mathcal{A}^{[3]}r^2\slashed{\mathcal{D}}_2^{\star}\nablasl(r(\overset{\text{\scalebox{.6}{$(1)$}}}{\textup{tr}\chib}))| ^2+|\mathcal{A}^{[3]}r^2\slashed{\mathcal{D}}_2^{\star}\nablasl(r(\overset{\text{\scalebox{.6}{$(1)$}}}{\slashed{\varepsilon}\cdot\chib}))| ^2) \lesssim \mathbb{E}_0 \, .
\end{equation}
\end{prop}

\medskip

\begin{proof}
We use the $\mathcal{A}^{[3]}r^2\slashed{\mathcal{D}}_2^{\star}\nablasl$-commuted red-shifted equations \eqref{4_trchibo} and \eqref{4_atrchibo} and Propositions \ref{iled_trchi} and \ref{iled_rho_sigma}.~We note that, in particular, the degeneration at trapping ($r=3M$) is introduced by Proposition \ref{iled_rho_sigma}.
\end{proof}

\medskip

\begin{prop}[Integrated decay for $\ybo$] \label{iled_yb}
Let $\mathfrak{S}_N$ be an initial-data normalised solution.~Assume that the initial energy \eqref{def_initial_energy_norm} is finite.~Then, for any $\delta>0$, $v\geq v_0$ and $u\geq u_0$, we have
\begin{equation}
\int_{v_0}^v d\bar{v} \int_{u_0}^{u} d\bar{u}\int_{\mathbb{S}^2_{\bar{u},\bar{v}}} \sin\theta \,d\theta \, d\phi \,\frac{\Omega^2}{r^{1+\delta}} \, |\mathcal{A}^{[3]}r\slashed{\mathcal{D}}_2^{\star}\ybo| ^2  \lesssim \mathbb{E}_0 \, .
\end{equation}
\end{prop}

\medskip

\begin{proof}
We use the $\mathcal{A}^{[3]}r\slashed{\mathcal{D}}_2^{\star}$-commuted red-shifted equation \eqref{4_ybo} and Propositions \ref{iled_eta} and \ref{iled_beta_betab}.
\end{proof}

\medskip

\begin{prop}[Integrated decay for $\overset{\text{\scalebox{.6}{$(1)$}}}{\mathfrak{\underline{\mathfrak{f}}}}$] \label{iled_fA}
Let $\mathfrak{S}_N$ be an initial-data normalised solution.~Assume that the initial energy \eqref{def_initial_energy_norm} is finite.~Then, for any $\delta>0$, $v\geq v_0$ and $u\geq u_0$, we have
\begin{equation}
\int_{v_0}^v d\bar{v} \int_{u_0}^{u} d\bar{u}\int_{\mathbb{S}^2_{\bar{u},\bar{v}}} \sin\theta \,d\theta \, d\phi \,\frac{\Omega^2}{r^{1+\delta}} \,  | \mathcal{A}^{[3]}r\slashed{\mathcal{D}}_2^{\star}  (r^{-1}\overset{\text{\scalebox{.6}{$(1)$}}}{\mathfrak{\underline{\mathfrak{f}}}} ) |^2 \lesssim \mathbb{E}_0 \, .
\end{equation}
\end{prop}

\medskip

\begin{proof}
We use the $\mathcal{A}^{[3]}r\slashed{\mathcal{D}}_2^{\star} $-commuted red-shifted equation \eqref{4_fA} and Proposition \ref{iled_eta}.
\end{proof}

\medskip

\begin{prop}[Integrated decay for $\omegabo$] \label{iled_omegab}
Let $\mathfrak{S}_N$ be an initial-data normalised solution.~Assume that the initial energy \eqref{def_initial_energy_norm} is finite.~Then, for any $\delta>0$, $v\geq v_0$ and $u\geq u_0$, we have
\begin{equation}
\int_{v_0}^v d\bar{v} \int_{u_0}^{u} d\bar{u}\int_{\mathbb{S}^2_{\bar{u},\bar{v}}} \sin\theta \,d\theta \, d\phi \,\frac{\Omega^2}{r^{1+\delta}}\, |r\,\slashed{\textup{div}}\,r^2\slashed{\mathcal{D}}_2^{\star}\nablasl\omegabo| ^2  \lesssim \mathbb{E}_0 \, .
\end{equation}
\end{prop}

\medskip

\begin{proof}
We exploit the $r\,\slashed{\text{div}}\,r^2\slashed{\mathcal{D}}_2^{\star}\nablasl$-commuted red-shifted equation \eqref{4_omegabo}.~We note the auxiliary identity 
\begin{align}
\Omega^2\,(r\,\slashed{\text{div}}\,r^2\,\slashed{\mathcal{D}}_2^{\star}\nablasl\overset{\text{\scalebox{.6}{$(1)$}}}{\mathfrak{\underline{\mathfrak{f}}}}_3)=& \,-\Omega^2\,r\,( r\,\slashed{\text{div}}\,r \slashed{\mathcal{D}}_2^{\star}\ybo)+ 2\,\mathcal{A}^{[3]}(r\chibho)-r\,\slashed{\text{div}}\,r^2\slashed{\mathcal{D}}_2^{\star}\nablasl(r(\trchibo)) \label{aux_iled_omegab}  \\ & + r\,\slashed{\text{div}}\,r^2\slashed{\mathcal{D}}_2^{\star}{}^{\star}\nablasl(r(\overset{\text{\scalebox{.6}{$(1)$}}}{\slashed{\varepsilon}\cdot\chib})) - r\,\slashed{\text{div}}\,r\slashed{\mathcal{D}}_2^{\star}(r\etao)-2\,r\,\slashed{\text{div}}\,r\slashed{\mathcal{D}}_2^{\star}(r^2\betabo)   \nonumber
\end{align}
to replace $(\nablasl_4\omegah)(r\,\slashed{\text{div}}\,r^2\,\slashed{\mathcal{D}}_2^{\star}\nablasl\overset{\text{\scalebox{.6}{$(1)$}}}{\mathfrak{\underline{\mathfrak{f}}}}_3)$ on the right hand side.~The identity \eqref{aux_iled_omegab} is obtained by taking the $r\,\slashed{\text{div}}\,r^2\slashed{\mathcal{D}}_2^{\star}$-derivative of equation \eqref{codazzi_chibho} and then commuting it by a $\nablasl_3$-derivative.
\end{proof}

\medskip

\begin{prop}[Integrated decay for $\overset{\text{\scalebox{.6}{$(1)$}}}{\mathfrak{\underline{\mathfrak{f}}}}_3$] \label{iled_f3}
Let $\mathfrak{S}_N$ be an initial-data normalised solution.~Assume that the initial energy \eqref{def_initial_energy_norm} is finite.~Then, for any $\delta>0$, $v\geq v_0$ and $u\geq u_0$, we have
\begin{equation}
\int_{v_0}^v d\bar{v} \int_{u_0}^{u} d\bar{u}\int_{\mathbb{S}^2_{\bar{u},\bar{v}}} \sin\theta \,d\theta \, d\phi \,\frac{\Omega^2}{r^{1+\delta}} \, | r\,\slashed{\textup{div}}\,r^2\slashed{\mathcal{D}}_2^{\star}\nablasl  (r^{-1}\overset{\text{\scalebox{.6}{$(1)$}}}{\mathfrak{\underline{\mathfrak{f}}}}_3 ) |^2 \lesssim \mathbb{E}_0 \, .
\end{equation}
\end{prop}

\medskip

\begin{proof}
We exploit the $r\,\slashed{\text{div}}\,r^2\slashed{\mathcal{D}}_2^{\star}\nablasl$-commuted red-shifted equation \eqref{4_f3} and Proposition \ref{iled_omegab}.
\end{proof}

\medskip

\begin{prop}[Integrated decay for $(\textup{tr}\overset{\text{\scalebox{.6}{$(1)$}}}{\slashed{g}})$] \label{iled_trg}
Let $\mathfrak{S}_N$ be an initial-data normalised solution.~Assume that the initial energy \eqref{def_initial_energy_norm} is finite.~Then, for any $\delta>0$, $v\geq v_0$ and $u\geq u_0$, we have
\begin{equation}
\int_{v_0}^v d\bar{v} \int_{u_0}^{u} d\bar{u}\int_{\mathbb{S}^2_{\bar{u},\bar{v}}} \sin\theta \,d\theta \, d\phi \,\frac{\Omega^2}{r^{1+\delta}} \, | \mathcal{A}^{[3]}r^2\slashed{\mathcal{D}}_2^{\star}\nablasl (\Omega^{-2}(\textup{tr}\overset{\text{\scalebox{.6}{$(1)$}}}{\slashed{g}}) )|^2 \lesssim \mathbb{E}_0 \, .
\end{equation}
\end{prop}

\medskip

\begin{proof}
We exploit the $\mathcal{A}^{[3]}r^2\slashed{\mathcal{D}}_2^{\star}\nablasl$-commuted red-shifted equation 
\begin{equation*}
\nablasl_4(\mathcal{A}^{[3]}r^2\slashed{\mathcal{D}}_2^{\star}\nablasl (\Omega^{-2}(\textup{tr}\overset{\text{\scalebox{.6}{$(1)$}}}{\slashed{g}})))+\omegah\,(\mathcal{A}^{[3]}r^2\slashed{\mathcal{D}}_2^{\star}\nablasl(\Omega^{-2}(\textup{tr}\overset{\text{\scalebox{.6}{$(1)$}}}{\slashed{g}})))=\frac{2\Omega^2}{r^2}\,\mathcal{A}^{[3]}r^2\slashed{\mathcal{D}}_2^{\star}\nablasl(r^2\,\Omega^{-4}(\trchio)) 
\end{equation*}
obtained from equation \eqref{4_trg} and Proposition \ref{iled_trchi}.~We note that (angular derivatives of) $r\nablasl(\Omega^{-2}(\textup{tr}\overset{\text{\scalebox{.6}{$(1)$}}}{\slashed{g}}))$ are uniformly bounded on $\underline{C}_{v_0}$ for smooth seed initial data by the initial-data normalisation $\nablasl(\textup{tr}\overset{\text{\scalebox{.6}{$(1)$}}}{\slashed{g}})=0$ on $\mathbb{S}^2_{\infty,v_0}$.
\end{proof}

\medskip

\begin{prop}[Integrated decay for $\overset{\text{\scalebox{.6}{$(1)$}}}{\widehat{\slashed{g}}}$] \label{iled_ghat}
Let $\mathfrak{S}_N$ be an initial-data normalised solution.~Assume that the initial energy \eqref{def_initial_energy_norm} is finite.~Then, for any $\delta>0$, $v\geq v_0$ and $u\geq u_0$, we have
\begin{equation}
\int_{v_0}^v d\bar{v} \int_{u_0}^{u} d\bar{u}\int_{\mathbb{S}^2_{\bar{u},\bar{v}}} \sin\theta \,d\theta \, d\phi \,\frac{\Omega^2}{r^{1+\delta}} \, | \mathcal{A}^{[3]}\overset{\text{\scalebox{.6}{$(1)$}}}{\widehat{\slashed{g}}} | ^2 \lesssim \mathbb{E}_0 \, .
\end{equation}
\end{prop}

\medskip

\begin{proof}
We exploit the $\nablasl_3\mathcal{A}^{[3]}$-commuted red-shifted equation
\begin{equation}
\nablasl_4(\nablasl_3(\mathcal{A}^{[3]}\overset{\text{\scalebox{.6}{$(1)$}}}{\widehat{\slashed{g}}}))+\omegah\,(\nablasl_3(\mathcal{A}^{[3]}\overset{\text{\scalebox{.6}{$(1)$}}}{\widehat{\slashed{g}}}))=\frac{2}{r^{2}}\, \nablasl_3(\mathcal{A}^{[3]}(r^2\chiho))+\frac{4}{r^{3}}\, (\mathcal{A}^{[3]}(r^2\chiho))
\end{equation}
obtained from commuting equation \eqref{4_ghat} and Proposition \ref{iled_chih} to obtain the integrated energy decay estimate 
\begin{equation} \label{iled_nabla_3_ghat}
\int_{v_0}^v d\bar{v} \int_{u_0}^{u} d\bar{u}\int_{\mathbb{S}^2_{\bar{u},\bar{v}}} \sin\theta \,d\theta \, d\phi \,\frac{\Omega^2}{r^{1+\delta}} \, | \nablasl_3(\mathcal{A}^{[3]}\overset{\text{\scalebox{.6}{$(1)$}}}{\widehat{\slashed{g}}}) | ^2 \lesssim \mathbb{E}_0 \, .
\end{equation}
Let $r_1=10M$.~By applying an argument as in Lemma 13.3.1 of \cite{DHR}, one obtains the estimate
\begin{align*}
\int_{v_0}^v d\bar{v} \int_{u(r_1,\bar{v})}^{u} d\bar{u}\int_{\mathbb{S}^2_{\bar{u},\bar{v}}} \sin\theta \,d\theta \, d\phi \, \Omega^2 \, | \mathcal{A}^{[3]}\overset{\text{\scalebox{.6}{$(1)$}}}{\widehat{\slashed{g}}} | ^2 \lesssim & \,  \int_{v_0}^v d\bar{v}  \int_{\mathbb{S}^2_{\infty,\bar{v}}} \sin\theta \,d\theta \, d\phi \,   |  \mathcal{A}^{[3]}\overset{\text{\scalebox{.6}{$(1)$}}}{\widehat{\slashed{g}}} | ^2 \\ &+\int_{v_0}^v d\bar{v} \int_{u(r_1,\bar{v})}^{u} d\bar{u}\int_{\mathbb{S}^2_{\bar{u},\bar{v}}} \sin\theta \,d\theta \, d\phi \, \Omega^2 \, | \nablasl_3(\mathcal{A}^{[3]}\overset{\text{\scalebox{.6}{$(1)$}}}{\widehat{\slashed{g}}}) | ^2 \, .
\end{align*} 
We now exploit the estimate \eqref{iled_nabla_3_ghat} and Proposition \ref{prop_flux_horizon_metric} to bound the right hand side and obtain
\begin{equation} \label{iled_ghat_close_horizon}
\int_{v_0}^v d\bar{v} \int_{u(r_1,\bar{v})}^{u} d\bar{u}\int_{\mathbb{S}^2_{\bar{u},\bar{v}}} \sin\theta \,d\theta \, d\phi \, \Omega^2 \, | \mathcal{A}^{[3]}\overset{\text{\scalebox{.6}{$(1)$}}}{\widehat{\slashed{g}}} | ^2 \lesssim \mathbb{E}_0 \, .
\end{equation}
To make the estimate \eqref{iled_ghat_close_horizon} global, one exploits the equation
\begin{equation*}
\nablasl_4(\mathcal{A}^{[3]}\overset{\text{\scalebox{.6}{$(1)$}}}{\widehat{\slashed{g}}})=\frac{2}{r^2}\,\mathcal{A}^{[3]}(r^2\chiho)
\end{equation*}
and Proposition \ref{iled_chih}.
\end{proof}

\medskip

\begin{remark} \label{rmk_alt_est_nabla_3_ghat}
The estimate \eqref{iled_nabla_3_ghat} could have alternatively been obtained by exploiting the equation \eqref{3_ghat} and Propositions \ref{iled_chibh} and \ref{iled_fA}.
\end{remark}

\subsection{Top-order estimates for the linearised curvature components} \label{sec_top_order_estimates_curvature}

In this section, we derive top-order (degenerate) integrated energy decay for the linearised curvature components $\betao$ and $\betabo$.~These estimates do not lose derivatives.~Combined with Propositions \ref{iled_gauge_inv_quantities} and \ref{iled_alpha_alphab}, this section completes the top-order estimates for all linearised curvature components. 

\medskip

We start with two propositions which state integrated energy decay for four angular derivatives of the linearised outgoing and ingoing shears.

\medskip

\begin{prop}[Refined integrated decay for the linearised outgoing shear] \label{iled_chiho_bis}
Let $\mathfrak{S}_N$ be an initial-data normalised solution.~Assume that the initial energy \eqref{def_initial_energy_norm} is finite.~Then, for any $\delta>0$, $v\geq v_0$ and $u\geq u_0$, we have
\begin{equation}
\int_{v_0}^v d\bar{v} \int_{u_0}^{u} d\bar{u}\int_{\mathbb{S}^2_{\bar{u},\bar{v}}} \sin\theta \,d\theta \, d\phi \,\frac{\Omega^2}{r^{1+\delta}} \,  |\mathcal{A}^{[2]}r^2\slashed{\mathcal{D}}_2^{\star}\slashed{\textup{div}} (r \chiho)| ^2 \lesssim \mathbb{E}_0 \, .
\end{equation}
\end{prop}

\medskip

\begin{proof}
We take the $\mathcal{A}^{[2]}r^2\slashed{\mathcal{D}}_2^{\star}$-derivative of equation \eqref{codazzi_chiho} and exploit Propositions \ref{iled_beta_betab}, \ref{iled_fA3}, \ref{iled_zeta} and \ref{iled_trchi}.
\end{proof}

\medskip

\begin{remark}
The integrated energy decay estimate of Proposition \ref{iled_chiho_bis} gains one derivative and loses one power of $r$ if compared to the estimate of Proposition \eqref{iled_chih} for three angular derivatives of the linearised outgoing shear.~Proposition \ref{iled_chiho_bis} suffices to prove a top-order integrated energy decay estimate for $\betao$ which does not lose in $r$-weight compared to the estimate of Proposition \ref{iled_beta_betab}.
\end{remark}

\medskip

\begin{prop}[Refined integrated decay for the linearised ingoing shear] \label{iled_chibho_refined} 
Let $\mathfrak{S}_N$ be an initial-data normalised solution.~Assume that the initial energy \eqref{def_initial_energy_norm} is finite.~Then, for any $\delta>0$, $v\geq v_0$ and $u\geq u_0$, we have
\begin{equation}
\int_{v_0}^v d\bar{v} \int_{u_0}^{u} d\bar{u}\int_{\mathbb{S}^2_{\bar{u},\bar{v}}} \sin\theta \,d\theta \, d\phi \,\frac{\Omega^2}{r^{1+\delta}} \, |\mathcal{A}^{[2]}r^2\slashed{\mathcal{D}}_2^{\star}\slashed{\textup{div}} (r \chibho)| ^2 \lesssim \mathbb{E}_0 \, .
\end{equation}
\end{prop}

\medskip

\begin{proof}
We take the $\mathcal{A}^{[2]}r^2\slashed{\mathcal{D}}_2^{\star}$-derivative of equation \eqref{codazzi_chibho} and exploit Propositions \ref{iled_beta_betab}, \ref{iled_fA3}, \ref{iled_zeta} and \ref{iled_trchib_atrchib}.
\end{proof}

\medskip

The following is the main proposition of the section.

\medskip

\begin{prop}[Top-order integrated decay for $\betao$ and $\betabo$] \label{iled_beta_betab_bis}
Let $\mathfrak{S}_N$ be an initial-data normalised solution.~Assume that the initial energy \eqref{def_initial_energy_norm} is finite.~Then, for any $\delta>0$, $v\geq v_0$ and $u\geq u_0$, we have
\begin{equation}
\int_{v_0}^v d\bar{v} \int_{u_0}^{u} d\bar{u}\int_{\mathbb{S}^2_{\bar{u},\bar{v}}} \sin\theta \,d\theta \, d\phi \,\frac{\Omega^2}{r^{1+\delta}}\left( 1-\frac{3M}{r}\right)^2 ( |\mathcal{A}^{[4]}r\slashed{\mathcal{D}}_2^{\star}(r^3\betao)| ^2 +|\mathcal{A}^{[4]}r\slashed{\mathcal{D}}_2^{\star}(r^2\betabo)| ^2 )\lesssim \mathbb{E}_0 \, .
\end{equation}
\end{prop}

\medskip

\begin{proof}
The proof follows the same argument as the proof of Proposition \ref{iled_beta_betab}, but one now takes the $\mathcal{A}^{[4]}$-derivative of the gauge invariant linearised quantities.~We recall that the initial energy flux $\mathbb{F}^{2,T,\nablasl}_{0}[\overset{\text{\scalebox{.6}{$(1)$}}}{\Psi},\mathfrak{D}\overset{\text{\scalebox{.6}{$(1)$}}}{\psi},\mathfrak{D}\alphao]$ only controls $\mathcal{A}^{[4]}(r^4\overset{\text{\scalebox{.6}{$(1)$}}}{\psi})$ with degeneration at trapping ($r=3M$), the same holding for the flux $\mathbb{F}^{2,T,\nablasl}_{0}[\overset{\text{\scalebox{.6}{$(1)$}}}{\underline{\Psi}},\mathfrak{D}\overset{\text{\scalebox{.6}{$(1)$}}}{\underline{\psi}},\mathfrak{D}\alphabo]$ relative to $\mathcal{A}^{[4]}(r^3\overset{\text{\scalebox{.6}{$(1)$}}}{\underline{\psi}})$.
\end{proof}

\section{The proof of the main theorem}  \label{sec_proof_main_th}

The proof of Theorem \ref{th_main_theorem} follows from Propositions \ref{iled_alpha_alphab}, \ref{iled_beta_betab_bis}, \ref{iled_rho_sigma}, \ref{iled_chih}, \ref{iled_chibh}, \ref{iled_trchi}, \ref{iled_trchib_atrchib}, \ref{iled_omegab}, \ref{iled_yb}, \ref{iled_eta}, \ref{iled_zeta}, \ref{iled_fA3}, \ref{iled_fA}, \ref{iled_f3}, \ref{iled_trg} and \ref{iled_ghat} (with the propositions ordered according to how the estimates appear in the statement of the theorem).

\appendix

\section{Well-posedness of the system of linearised gravity} \label{appendix_well_posedness}

We present the proof of Proposition \ref{prop_well_posedness}.

\medskip

\begin{proof}
Given smooth seed initial data, we start by showing that all the linearised quantities can be determined on the initial hypersurface $\mathcal{S}_{u_0,v_0}$.~On $\underline{C}_{v_0}$, one uniquely determines (in order)
\begin{align*}
&(\overset{\text{\scalebox{.6}{$(1)$}}}{\slashed{\varepsilon}\cdot\chi}) \, , \, (\overset{\text{\scalebox{.6}{$(1)$}}}{\slashed{\varepsilon}\cdot\chib}) & &\text{using the equations \eqref{A_fA4} and \eqref{A_fA3},}\\
&\ybo & &\text{using the equation \eqref{3_fA3},}\\
&\chibho & &\text{using the equation \eqref{3_ghat},}\\
&\alphabo & &\text{using the equation \eqref{3_chibho},}\\
&\betabo & &\text{using the equation \eqref{codazzi_chibho} on $\mathbb{S}^2_{\infty,v_0}$ and equation \eqref{3_betabo},}\\
&\sigmao & &\text{using the equation \eqref{curl_zetao} on $\mathbb{S}^2_{\infty,v_0}$ and equation \eqref{3_sigmao},}\\
&(\trchibo) & &\text{using the equation \eqref{3_trchibo},}\\
&(\text{tr}\overset{\text{\scalebox{.6}{$(1)$}}}{\slashed{g}})& &\text{using the equation \eqref{3_trgo},}\\
&\overset{\text{\scalebox{.6}{$(1)$}}}{\widetilde{\slashed{K}}} & &\text{using the expression \eqref{aux_expression_K},}\\
&\rhoo & &\text{using the equation \eqref{gauss_eqn} on $\mathbb{S}^2_{\infty,v_0}$ and equation \eqref{3_rhoo},}\\
&\zetao \, , \, \etao & & \text{using the equations \eqref{3_fA4} and \eqref{3_zetao},}\\
&(\trchio) &  &\text{using the equation \eqref{3_trchio}.}
\end{align*}
We note that all the linearised equations intrinsic to the ingoing null cone $\underline{C}_{v_0}$ were used, with the exception of the equations \eqref{3_atrchibo} and \eqref{3_atrchio} on $\underline{C}_{v_0}$ and the equations \eqref{curl_zetao}, \eqref{codazzi_chibho} and \eqref{gauss_eqn} on $\underline{C}_{v_0}\setminus \mathbb{S}^2_{\infty,v_0}$.~One can check that the equations \eqref{3_atrchibo} and \eqref{3_atrchio} are obtained by taking a $\nablasl_3$-derivative of the equations \eqref{A_fA4} and \eqref{A_fA3}, the ($\slashed{\text{curl}}$-commuted) equation \eqref{3_zetao} is obtained by taking a $\nablasl_3$-derivative of the equation \eqref{curl_zetao} and that the constraint equations \eqref{codazzi_chibho} and \eqref{gauss_eqn} are propagated from the initial horizon sphere $\mathbb{S}^2_{\infty,v_0}$ to the whole initial ingoing null cone $\underline{C}_{v_0}$.

\medskip

Using, in order, the equations \eqref{4_ghat}, \eqref{4_chiho}, \eqref{4_fb4}, \eqref{4_fA4} and \eqref{4_atrchio}, one uniquely determines
\begin{align*}
&\chiho \, , \, \alphao,\overset{\text{\scalebox{.6}{$(1)$}}}{\mathfrak{\underline{\mathfrak{f}}}}_4 \, , \, \overset{\text{\scalebox{.6}{$(1)$}}}{\mathfrak{\slashed{\mathfrak{f}}}}_{4} \, , \, (\overset{\text{\scalebox{.6}{$(1)$}}}{\slashed{\varepsilon}\cdot\chi}) &  &\text{on $C_{u_0}$.}
\end{align*}
Using the equation \eqref{3_chiho}, one uniquely determines
\begin{align*}
&\chiho &  &\text{on $\underline{C}_{v_0}$.}
\end{align*}
The equation \eqref{codazzi_chiho} then gives $\betao$ on $S^2_{u_0,v_0}$, which, using (in order) the equations \eqref{4_betao}, \eqref{4_sigmao} and \eqref{4_rhoo}, yields
\begin{align*}
&\betao \, , \, \sigmao \, , \, \rhoo &  &\text{on $C_{u_0}$.}
\end{align*}
Using the equations \eqref{3_betao} and \eqref{3_alphao}, one uniquely determines
\begin{align*}
&\betao \, , \, \alphao &  &\text{on $\underline{C}_{v_0}$.}
\end{align*}
This completes the computation of all the linearised quantities in the system on $\underline{C}_{v_0}$.

\medskip

One can check that all the remaining linearised quantities can be uniquely determined on $C_{u_0}$.~To determine all the linearised quantities in the system on $\mathcal{R}_{u_0,v_0}$, one starts from the decoupled linearised quantities $\alphao$ and $\alphabo$ solving the spin $\pm 2$ Teukolsky equations.~One can then compute all the gauge dependent linearised quantities on $\mathcal{R}_{u_0,v_0}$ and check that the linearised constraint equations are propagated.
\end{proof}

\section{Early estimates for the linearised induced metric and frame coefficients}  \label{sec_early_estimates}

As an alternative to what done in Section \ref{sec_analysis_gauge_dependent}, one can exploit integrated energy decay for the gauge invariant linearised quantities (Proposition \ref{iled_gauge_inv_quantities}) and the linearised outgoing shear (Proposition \ref{iled_chih}) to obtain integrated energy decay for the linearised induced metric and frame coefficients (with the exception of Proposition \ref{iled_f3}) \emph{without relying on any control over the linearised ingoing shear}.

\medskip

The following is the alternative route that one would have to take after proving Proposition \ref{iled_chih}:
\begin{itemize}
\item By repeating the proof of integrated energy decay for (angular derivatives of) the linearised induced metric
\begin{align} \label{aux_metric_coefficients}
&(\textup{tr}\overset{\text{\scalebox{.6}{$(1)$}}}{\slashed{g}}) \, , &  &\overset{\text{\scalebox{.6}{$(1)$}}}{\widehat{\slashed{g}}}
\end{align}
in Propositions \ref{iled_trg} and \ref{iled_ghat}, one realises that the proof only relies on Propositions \ref{iled_gauge_inv_quantities} and \ref{iled_chih} and, in the case of $(\textup{tr}\overset{\text{\scalebox{.6}{$(1)$}}}{\slashed{g}})$, on Proposition \ref{iled_trchi}.~We note that Proposition \ref{iled_trchi} can be proven independently from the rest of the linearised system of equations, see Remark \ref{rmk_trchi_position_scheme}.~See also the related Remark \ref{rmk_alt_est_nabla_3_ghat}.

\begin{remark}
The proof of integrated energy decay for $\overset{\text{\scalebox{.6}{$(1)$}}}{\widehat{\slashed{g}}}$ in Proposition \ref{iled_ghat} does \emph{not} use Proposition \ref{iled_trchi} for $(\overset{\text{\scalebox{.6}{$(1)$}}}{\textup{tr}\chi})$.
\end{remark}

\begin{remark}
Integrated energy decay for the linearised quantities \eqref{aux_metric_coefficients} can be obtained \emph{before} and independently from integrated energy decay for the linearised frame coefficients.
\end{remark}

\item By using the identities
\begin{align*}
\mathcal{A}^{[i]}(r^4\overset{\text{\scalebox{.6}{$(1)$}}}{\psi}) &= \mathcal{A}^{[i]}r\slashed{\mathcal{D}}_2^{\star}(r^3\betao)-\frac{3M}{r}\,\mathcal{A}^{[i]}(r^2\chiho) \, , \\
\mathcal{A}^{[i]}\nablasl_3(r^4\overset{\text{\scalebox{.6}{$(1)$}}}{\psi}) &= \mathcal{A}^{[i]}r\slashed{\mathcal{D}}_2^{\star}\nablasl_3(r^3\betao)-\frac{3M}{r^2}\,\mathcal{A}^{[i]}(r^2\chiho)-\frac{3M}{r}\,\mathcal{A}^{[i]}\nablasl_3(r^2\chiho)
\end{align*} 
and exploiting Propositions \ref{iled_gauge_inv_quantities} and \ref{iled_chih}, one can prove the integrated energy decay estimate for $\betao$ stated in Proposition \ref{iled_beta_betab} and the estimates
\begin{align*}
\int_{v_0}^v d\bar{v} \int_{u_0}^{u} d\bar{u}\int_{\mathbb{S}^2_{\bar{u},\bar{v}}} \sin\theta \,d\theta \, d\phi \,\frac{\Omega^2}{r^{1+\delta}}\,  |\nablasl_3(\mathcal{A}^{[2]}r\slashed{\mathcal{D}}_2^{\star}(r^3\betao))| ^2  \lesssim \mathbb{E}_0 \, , \\
\int_{v_0}^v d\bar{v} \int_{u_0}^{u} d\bar{u}\int_{\mathbb{S}^2_{\bar{u},\bar{v}}} \sin\theta \,d\theta \, d\phi \,\frac{\Omega^2}{r^{1+\delta}}\,\left(  1-\frac{3M}{r}\right)^2  |\nablasl_3(\mathcal{A}^{[3]}r\slashed{\mathcal{D}}_2^{\star}(r^3\betao))| ^2  \lesssim \mathbb{E}_0 \, ,
\end{align*}
where we note the degenerate (at trapping, $r=3M$) factor in the second estimate.

\item By commuting once with $\nablasl_3$ (and angular derivatives) the equation \eqref{4_etao}, one obtains the red-shifted equation
\begin{equation}  \label{aux_iled_eta_alternative}
\nablasl_4 (\nablasl_3\mathcal{A}^{[i]}r\slashed{\mathcal{D}}_2^{\star}(r\etao))+\omegah\,(\nablasl_3\mathcal{A}^{[i]}r\slashed{\mathcal{D}}_2^{\star}(r\etao))     =  - \frac{1}{r^2}\nablasl_3\mathcal{A}^{[i]}r\slashed{\mathcal{D}}_2^{\star}(r^3\betao)-\frac{2}{r^3}\mathcal{A}^{[i]}r\slashed{\mathcal{D}}_2^{\star}(r^3\betao) \, .
\end{equation}
Equation \eqref{aux_iled_eta_alternative} can be used to prove the estimates
\begin{align}
\int_{v_0}^v d\bar{v} \int_{u_0}^{u} d\bar{u}\int_{\mathbb{S}^2_{\bar{u},\bar{v}}} \sin\theta \,d\theta \, d\phi \,\frac{\Omega^2}{r^{1+\delta}}\,  |\nablasl_3\mathcal{A}^{[2]}r\slashed{\mathcal{D}}_2^{\star}(r\etao)| ^2  &\lesssim  \mathbb{E}_0  \, ,  \nonumber\\
\int_{v_0}^v d\bar{v} \int_{u_0}^{u} d\bar{u}\int_{\mathbb{S}^2_{\bar{u},\bar{v}}} \sin\theta \,d\theta \, d\phi \,\frac{\Omega^2}{r^{1+\delta}}\,\left(  1-\frac{3M}{r}\right)^2  |\nablasl_3\mathcal{A}^{[3]}r\slashed{\mathcal{D}}_2^{\star}(r\etao)| ^2  &\lesssim  \mathbb{E}_0 \, . \label{aux_iled_eta_alternative_2} 
\end{align} 
We note that
\begin{equation*}
\lVert \nablasl_3\mathcal{A}^{[i]}\slashed{\mathcal{D}}_2^{\star}\etao  \rVert_{L^2(\underline{C}_{v_0})} \lesssim \mathbb{E}_0
\end{equation*}
for $i\leq 3$.

Let $r_1=5M/2$.~As noted in the proof of Proposition \ref{iled_ghat}, one can use an estimate like \eqref{aux_iled_eta_alternative_2} to control the following integral close to the event horizon
\begin{equation}
\int_{v_0}^v d\bar{v} \int_{u_0(r_1,\bar{v})}^{u} d\bar{u}\int_{\mathbb{S}^2_{\bar{u},\bar{v}}} \sin\theta \,d\theta \, d\phi \,\Omega^2 \,  |\mathcal{A}^{[3]}r\slashed{\mathcal{D}}_2^{\star}(r\etao)| ^2  \lesssim \mathbb{E}_0 \, . \label{aux_iled_eta_alternative_3}
\end{equation}
The constant $r_1$ is chosen so that one can remove the degenerate (at trapping, $r=3M$) factor from \eqref{aux_iled_eta_alternative_2}.~We also note that to derive the estimate \eqref{aux_iled_eta_alternative_3}, one relies on control over the horizon energy flux 
\begin{equation}
\int_{\mathcal{H}^+(v_0,v)} d\bar{v}\, \sin\theta \,d\theta \, d\phi \,\Omega^2 \,  |\mathcal{A}^{[3]}r\slashed{\mathcal{D}}_2^{\star}(r\etao)| ^2  \lesssim \mathbb{E}_0 \, ,
\end{equation}
which one can deduce from Proposition \ref{prop_flux_horizon_shears} (see Remark \ref{rmk_higher_order_fluxes}).~The estimate \eqref{aux_iled_eta_alternative_3} can be made global by exploiting the equation \eqref{4_etao}, thus achieving Proposition \ref{iled_eta}.

We emphasise that, in contrast with what done in our scheme of Section \ref{sec_analysis_gauge_dependent} (i.e.~in the proof of Proposition \ref{iled_eta}), we achieve Proposition \ref{iled_eta} avoiding the use of equation \eqref{3_chiho}, which involves (and requires control over) the linearised ingoing shear.

\item One can now repeat the proof of Propositions \ref{iled_fA3} and \ref{iled_fA} for the linearised frame coefficients
\begin{align*}
&\overset{\text{\scalebox{.6}{$(1)$}}}{\mathfrak{\slashed{\mathfrak{f}}}}_{3} \, , &  &\overset{\text{\scalebox{.6}{$(1)$}}}{\mathfrak{\underline{\mathfrak{f}}}} 
\end{align*}
respectively.
\end{itemize}

This alternative route may find interesting applications to the analysis of the system of linearised gravity of \cite{benomio_kerr_system} in the general $|a|<M$ case, in which the outgoing transport equation for the linearised ingoing shear couples with the linearised induced metric and frame coefficients on the right hand side.

\bibliography{Stream_schwarzschild_b} 
\bibliographystyle{hsiam}

\end{document}